\documentclass[11pt]{article}
\pdfoutput=1
\usepackage[utf8]{inputenc}
\usepackage[numbers,sort&compress]{natbib}
\usepackage{hyperref}
\usepackage[margin=2.5cm]{geometry}
\usepackage{tocloft}
\usepackage[skip=0.2cm, indent=0pt]{parskip}
\usepackage{amsmath,amssymb,mathtools,amsthm,latexsym,amsxtra}
\usepackage{bm,bbm,dsfont,mathrsfs}
\usepackage{tensor}
\usepackage{stackrel}
\usepackage{youngtab}
\usepackage{simplewick}
\usepackage{slashed}
\usepackage{braket}
\usepackage{frcursive}
\usepackage{relsize}
\usepackage{dsfont}

\usepackage{graphicx,epsfig,epstopdf,psfrag}
\usepackage{tikz,pgfplots}
\usetikzlibrary{intersections,pgfplots.fillbetween,decorations.pathmorphing}
\pgfplotsset{compat=1.18}

\usepackage{array,booktabs,multirow,tablefootnote}

\usepackage{subcaption}

\usepackage[nosort]{cite}

\usepackage{xcolor}
\usepackage[normalem]{ulem} 
\usepackage{soul}
\setstcolor{red}
\usepackage{cancel}
\usepackage{scalerel}
\usepackage{makeidx}
\usepackage{appendix}
\usepackage{float}
\usepackage{verbatim}
\usepackage{url}

\DeclareUnicodeCharacter{2212}{-}

\hypersetup{
    colorlinks,
    citecolor=blue,
    filecolor=blue,
    linkcolor=blue,
    urlcolor=blue
}
\usepackage{cleveref}

\usepackage{pgfornament,multicol}

\def\e{{\epsilon}}

\newcommand{\bi}{\begin{itemize}}
\newcommand{\ei}{\end{itemize}}
\newcommand{\bea}{\begin{eqnarray}}
\newcommand{\eea}{\end{eqnarray}}
\newcommand{\be}{\begin{equation}}
\newcommand{\ee}{\end{equation}}
\newcommand{\dd}{\text{d}}

\newcommand{\cA}{\mathcal{A}}
\newcommand{\cB}{\mathcal{B}}
\newcommand{\cC}{\mathcal{C}}
\newcommand{\cD}{\mathcal{D}}

\newcommand{\cG}{\mathcal{G}}

\newcommand{\cL}{\mathcal{L}}

\newcommand{\cM}{\mathcal{M}}

\newcommand{\cN}{\mathcal{N}}
\newcommand{\cO}{\mathcal{O}}

\newcommand{\cT}{\mathcal{T}}
\newcommand{\cU}{\mathcal{U}}
\newcommand{\cV}{\mathcal{V}}

\newcommand{\cZ}{\mathcal{Z}}

\newcommand{\bR}{\mathbb{R}}
\newcommand{\bT}{\mathbb{T}}

\newcommand{\bZ}{\mathbb{Z}}
\newcommand{\tn}[1]{\textnormal{#1}}
\newcommand{\nf}{\tn{N}_{\tn{f}}}

\def\XXint#1#2#3{{\setbox0=\hbox{$#1{#2#3}{\int}$}
     \vcenter{\hbox{$#2#3$}}\kern-.5\wd0}}

\def\={\, = \,}

\makeatletter
\DeclareRobustCommand{\cev}[1]{%
  \mathpalette\do@cev{#1}%
}
\newcommand{\do@cev}[2]{%
  \fix@cev{#1}{+}%
  \reflectbox{$\m@th#1\vec{\reflectbox{$\fix@cev{#1}{-}\m@th#1#2\fix@cev{#1}{+}$}}$}%
  \fix@cev{#1}{-}%
}
\newcommand{\fix@cev}[2]{%
  \ifx#1\displaystyle
    \mkern#23mu
  \else
    \ifx#1\textstyle
      \mkern#23mu
    \else
      \ifx#1\scriptstyle
        \mkern#22mu
      \else
        \mkern#22mu
      \fi
    \fi
  \fi
}
\makeatother

\numberwithin{equation}{subsection}
\allowdisplaybreaks
\makeatletter
\AddToHook{cmd/appendix/before}{\def\cref@section@alias{appendix}\def\cref@subsection@alias{appendix}}
\makeatother

\begin{document}

\pagenumbering{Alph}
\begin{titlepage}
\thispagestyle{empty}

\begin{flushright}
\end{flushright}

\vskip3cm

\begin{center}

{\Huge \textsc \bf de Sitter Vacua \& pUniverses}

\vskip1cm

\begin{center}

Jeremias Aguilera-Damia,$^\dagger$\footnote{jeremiasad@fqa.ub.edu}
Dionysios Anninos,$^{\ddag\square}$\footnote{dionysios.anninos@kcl.ac.uk}
Tarek Anous,$^\Delta$\footnote{t.anous@qmul.ac.uk} 
Johnny Gleeson,$^{\Delta}$\footnote{j.b.gleeson@qmul.ac.uk} 
\\ and Alan Rios Fukelman$^\Delta$\footnote{a.riosfukelman@qmul.ac.uk}

\end{center}

\vskip0.3cm

{ \small \it
$^{\dagger}$Departament de F\'{\i}sica Qu\`antica i Astrof\'{\i}sica and Institut de Ci\`encies del Cosmos,
Universitat de Barcelona, Mart\'{\i} Franqu\`es 1, 08028 Barcelona, Spain
}
\vskip0.1cm
{\small \it
$^{\ddag}$Department of Mathematics, King's College London, 
The Strand, London WC2R 2LS, U.K.
}
\vskip 0.1cm
{ \small \it{$^\square$Instituut voor Theoretische Fysica, KU Leuven, Celestijnenlaan 200D, B-3001 Leuven, Belgium}}
\vskip0.1cm
{\small \it
$^{\Delta}$School of Mathematical Sciences, Queen Mary University of London,
Mile End Road \\ London E1 4NS, United Kingdom
}

\end{center}

\vskip1cm

\begin{abstract}

\noindent We analyze a simple extension of the Schwinger model, which we refer to as the $p$-Schwinger model, on a de Sitter background. In this theory, the charged massless fermions carry non-unit integer charge $p$. In Minkowski space, the $p$-Schwinger model has discrete zero- and one-form global symmetries that are spontaneously broken, yielding $p$ degenerate ground states.  We demonstrate that these features persist upon placing the $p$-Schwinger model on a global de Sitter background, establishing that such discrete global symmetries can be spontaneously broken for quantum field theories in de Sitter space. In particular, the theory is endowed with $p$ distinct, but locally-indistinguishable, de Sitter-invariant states, the de Sitter vacua, satisfying the Hadamard property. We couple a variant of the $p$-Schwinger model with $\nf$ flavors to quantum gravity with $\Lambda>0$, and demonstrate the existence of a semiclassical de Sitter saddle at large $\nf$. In the gravitational theory, the $p$ de Sitter-invariant vacua are speculatively interpreted as microstates of the de Sitter horizon in the low-energy effective field theory. 
\end{abstract}

\end{titlepage}

\pagenumbering{arabic}
\tableofcontents

\section{Introduction}

This paper is concerned with the fate of certain spontaneously-broken discrete symmetries in de Sitter space. We will follow a constructive approach, building on \citep{Jayewardena:1988td,Anninos:2024fty,Smith:2026dae}. These works offer a complete solution to the ordinary Schwinger model \citep{Schwinger:1962tp}, namely, two dimensional quantum electrodynamics with massless charged fermions, on both a Euclidean and Lorentzian de Sitter background. The Schwinger model on a de Sitter background includes a host of interesting features: correlation functions that grow logarithmically in time  at any loop order, non-perturbative corrections stemming from gauge theoretic instantons, an anomalously-broken axial symmetry, and both local and non-local gauge-invariant operators. These are phenomena of general relevance to de Sitter quantum field theory, in any number of dimensions. The theory can be studied directly in terms of the fermionic fields coupled to the Abelian gauge field, but also via bosonization, which permits a more Gaussian approach. One thing that is missing in the ordinary Schwinger model is the presence of global symmetries, and consequently any spontaneous symmetry breaking patterns that might accompany them. 

The aim of this paper is to analyze a simple but qualitatively interesting extension of the ordinary Schwinger model. In the case of interest, the fermions have a non-unit integer charge $p$. We refer to this model as the $p$-Schwinger model. In Minkowski space, the $p$-Schwinger model is known to exhibit an interesting pattern of symmetry breaking \citep{Misumi:2019dwq,Witten:1978bc,Komargodski:2020mxz,Armoni:2018bga,Seiberg:2010qd}. The $U(1)$-axial symmetry is anomalous, but because of the non-minimal charge $p$ of the fermions, a $\mathbb{Z}_p^{(0)}$ subgroup survives. This $\mathbb{Z}_p^{(0)}$ zero-form symmetry in turn, is spontaneously broken entirely. In addition, the theory has a $\mathbb{Z}_p^{(1)}$ one-form global symmetry that is itself also entirely spontaneously broken.\footnote{Zero-form symmetries act on ordinary local operators, whilst one-form symmetries act on line operators. For a pedagogical review see \citep{Iqbal:2025dsh}.} An illustrative way to understand the Minkowksian theory is that, in the deep infrared, the theory, though gapped, flows to a non-trivial topological phase given by the Abelian BF theory at level $p$. In its turn, the Abelian BF theory is a particularly simple theory to analyze \citep{Bergeron:1994ym, Blau:1993hj,Kapustin:2014gua,seiberg2015}, and its symmetry breaking structure mirrors that of the $p$-Schwinger model to the note. It exhibits, in particular, $p$-independent vacuum states. 

Before moving to the case of interest, we further mention that the spontaneous symmetry breaking pattern of the $p$-Schwinger model on Minkowski space is particularly robust. For instance, it persists upon placing the theory on a compact spatial slice, as well as at all temperatures by prohibiting any thermal tunneling processes of finite energy \citep{Komargodski:2020mxz}. Moreover, only space-filling line-operators, rather than ordinary local operators, can mediate between different one-form charged sectors.  As such, a sector carrying definite $\mathbb{Z}_p^{(1)}$ one-form  charge is often referred to as a Universe in the literature. Part of the reason behind the robustness of the symmetry breaking pattern is that the two $\mathbb{Z}_p$ symmetries participate in a mixed $\mathbb{Z}_p$-valued 't Hooft anomaly which implies their generators fail to commute \citep{Komargodski:2020mxz, Kapustin:2014gua}. The mixed 't Hooft anomaly implies, for the gapped theory, that there cannot be a unique vacuum state. Rather, we have a spontaneously-broken symmetry structure yielding degenerate vacua.

We now turn to the case of interest, which is the $p$-Schwinger model on a global de Sitter background. Due to the compactness of the de Sitter Cauchy slices, and the finite volume of the Euclidean de Sitter solution, which is a two-sphere, it is unclear whether we can immediately port the results of the Minkowski model to de Sitter space. For instance, the deep infrared regime is not a completely well-defined concept for a theory where we cannot take vacuum correlations to arbitrarily large spatial separation (in de Sitter units). Moreover, the absence of time-translation invariance causes novel effects in the vacuum, such as the spontaneous creation of particles. To make matters even more perplexing, physical observers are confined within a single static patch and do not have access to the global structure of physical operators. Perhaps then, and also in light of some of the infrared enhancements observed for the ordinary Schwinger model, those global symmetries that are spontaneously broken in the Minkowskian model are restored upon placing the theory on de Sitter space. Such a conclusion would align with several discussions about spontaneous symmetry breaking in de Sitter space, though we should caution that those discussions are typically relevant to continuous rather than discrete symmetries. As we shall see, however, the conclusion does not hold for the theories at hand.

In what follows, we will not rely on general arguments. Instead, we will proceed to exactly solve the $p$-Schwinger model in de Sitter space. We will  find that, at least from the global perspective, the de Sitter theory exhibits spontaneous symmetry breaking of both $\mathbb{Z}_p$ global symmetries. As we shall see, one (schematic) way to anticipate this result is that the symmetries are generated by topological operators in the Minkowskian theory, and it can be reasonably hoped that these operators, appropriately regulated, persist on a curved space. We will indeed carefully construct these operators for the de Sitter case. Somewhat strikingly, as a consequence of the symmetry breaking pattern, the $p$-Schwinger model exhibits $p$ independent de Sitter-invariant Hadamard states that are locally indistinguishable.\footnote{It is worth mentioning that a similar scenario is expected to arise in more complicated non-Abelian two-dimensional gauge theories, such as Adjoint QCD. These theories hold intricate structures of zero-form and one-form symmetries, comprising various mixed anomalies with analogous implications for the low energy dynamics in Minkowski space \citep{Komargodski:2020mxz,Damia:2024kyt}. However, at the present time these theories do not admit an exact solution via path integral methods.  } We emphasize the Hadamard structure to distinguish these states from the more familiar de Sitter-invariant $\alpha$-vacua, which exhibit non-Hadamard antipodal singularities (see for instance \citep{Mottola:1984ar,Allen:1985ux,Bousso:2001mw} and more recently \citep{Miller:2025jbz}). To the expert, these ideas may seem natural or straightforward extensions of the Minkowskian case, however we found it important to present the $p$-Schwinger model on a de Sitter background in full detail, given the rich properties already uncovered for the ordinary one. Further to this, an important part of our motivation, which we initiate in the last section, is to couple these theories to two-dimensional gravity with $\Lambda>0$, so as to analyze the gravitational fate of these field theoretic structures. 

Before proceeding to the outline of the paper, we also mention that a similar situation can occur in four-dimensional quantum field theory on a Minkowski background. Specifically, in section 4 of \citep{Cherman:2020cvw} a four-dimensional model mirroring the mixed anomaly structure of the $p$-Schwinger model is considered. The model now has a global zero-form $\mathbb{Z}_{2\nf\, p}$ symmetry and a global three-form $\mathbb{Z}_p$ symmetry, where $p$ is the discrete coupling for the three-form, and $\nf$ is the number of fermionic flavors in the model. The two global symmetries again participate in a mixed 't Hooft anomaly. As such, the model of  \citep{Cherman:2020cvw} exhibits robust Universes in the same sense as the two-dimensional $p$-Schwinger model. As an additional example,  take a compact scalar field $\Phi\cong\Phi+2\pi$, viewed as the effective field theory of a superfluid. The theory encodes a three-form current $J_{\text{w}}^{\mu\nu\rho} =  \epsilon^{\mu\nu\rho\sigma}\partial_\sigma \Phi$, which is topologically conserved. By considering the Hodge-dual current, we conclude that this theory also has topological line operators.  We can gauge the winding symmetry with a three-form $U(1)$ gauge field, $A_{\mu\nu\rho}$,  coupled to the current as $p \, J_{\text{w}}^{\mu\nu\rho}A_{\mu\nu\rho}$ with $p\in\bZ$. If $p>1$, the theory enjoys a $\bZ_p$ zero-form and $\bZ_p$ three-form global symmetry which participate in a mixed 't Hooft anomaly of the type described above. ({The bosonized version of the charge-$p$ Schwinger model is the two-dimensional version of this construction.})  It is tempting to anticipate that these symmetry breaking structures will persist for such theories on a four-dimensional de Sitter background. We leave the details of this for future work.

\subsection*{Outline of the paper}

In section \ref{sec: BF}, we review the Abelian BF model, its Hilbert space structure, and its symmetry breaking pattern. This is well known material and we claim no new results. Rather, we present the material in a way that is well-adapted for subsequent sections. 

In section \ref{sec:pSchwinger}, we introduce the $p$-Schwinger model on a two-dimensional de Sitter background. We work mostly in Euclidean signature, so really, the model is analyzed on the round two-sphere. As analyzed for the ordinary Schwinger model in \citep{Anninos:2024fty}, this encodes a host of Lorentzian features. To our knowledge, this is the first time this model has been analyzed on either the two-sphere or two-dimensional de Sitter space. The main result of the section is the explicit construction of the topological operators generating the zero- and one-form $\mathbb{Z}_p$ symmetries, and the explicit construction of $p$ independent de Sitter-invariant Hadamard states. We verify the Hadamard property by computing a slew of local correlation functions, non-perturbatively in the gauge coupling $q$. In other words, the section demonstrates that interacting quantum field theories in de Sitter space can have spontaneously broken discrete global symmetries. Much of our analyses can be performed on a general curved space. 

In section \ref{sec:gravity}, we extend the $p$-Schwinger model to a model that includes dynamical gravity with $\Lambda>0$. To do so, we first extend the $p$-Schwinger model to an $SU(\nf)$ flavored version, whose low energy structure includes a gapless level-one $SU(\nf)$ Wess-Zumino-Witten model, in  addition to the sector with the aforementioned discrete symmetry breaking pattern. At large $\nf$, the gravitational model displays a semiclassical de Sitter saddle that experiences small metric fluctuations that are accompanied by the fluctuations of the matter fields. From the perspective of the gravitational model, we interpret the $p$ locally indistinguishable de Sitter-invariant states (the $p$ independent vacua) as microstates of the de Sitter horizon.

Conventions for our path integral and operator regularization schemes are detailed in the two appendices. 

\section{BF Theory on \texorpdfstring{dS$_2$}{dS2}}\label{sec: BF}

We start by considering BF theory on a rigid two-dimensional de Sitter background, eventually working our way up to the Hartle-Hawking construction of the Hilbert space. Nothing we say here about BF theory will be new, but our presentation will focus on certain aspects that will be useful to us in later sections.

Let us start with the global metric on two-dimensional de Sitter space: 
\begin{equation}\label{eq:deSittermetricglobal}
\frac{\dd s^2}{\ell^2}=-\dd\tau^2+\cosh^2\tau\,\dd \varphi^2~,\qquad  \tau\in\mathbb{R} ,\quad \varphi\cong\varphi+2\pi~,
\end{equation}
where $\ell$ is the de Sitter length scale. This geometry is conformally equivalent to a cylinder. To see this, simply reparametrize $\tau\rightarrow\log\left(-\cot\frac{T}{2}\right)$, yielding:
\begin{equation}\label{eq:deSittermetric}
\frac{\dd s^2}{\ell^2}=\frac{-\dd T^2+\dd \varphi^2}{\sin^2 T}~,\qquad  T\in (-\pi, 0),\quad \varphi\cong\varphi+2\pi~.
\end{equation}
We will orient our spacetime manifold such that $\epsilon_{T\varphi}=-\epsilon_{\varphi T}=\sqrt{-g}$.

On top of this spacetime we place an Abelian BF theory with the following action:\footnote{The extra factor of $i$ is kept as a bookkeeping device since the Lorentzian path integral is weighted by $e^{i S^{\textnormal{BF}}_{ L}}$.}
\begin{equation} \label{BF:ActionLor}
    i S^{\textnormal{BF}}_{ L}= -\frac{ip}{4\pi} \int_{\mathcal{M}} \dd^2 x \sqrt{-g}\,   B(\mathbf{x})  \epsilon^{\mu \nu} F_{\mu \nu}(\mathbf{x}) \, ,
\end{equation}
where $p \in \mathbb{Z}$ is the coupling constant. The matter content includes a compact scalar field $B(\mathbf{x})$, of period $2\pi$, and a $U(1)$ gauge field $A_\mu$ with corresponding field strength $F_{\mu \nu}\equiv \partial_\mu A_\nu - \partial_\nu A_\mu$. The redundancies in field-space are captured by the following equivalences: 
\begin{equation}
B(\mathbf{x})\cong B(\mathbf{x})+2\pi, \qquad A_\mu\cong A_\mu+i h(\mathbf{x})^{-1}\partial_\mu h(\mathbf{x})
\end{equation} 
with $h=e^{i\alpha(\mathbf{x})}\in U(1)$, meaning the gauge parameter $\alpha(\mathbf{x})$ is also a compact scalar of period $2\pi$. 

This action yields a well-known {topological quantum field theory} \citep{Bergeron:1994ym, Blau:1993hj, Kapustin:2014gua,seiberg2015}, which is independent of the background metric. As we will see in the following section, quantization will only be sensitive to the fact that this spacetime is topologically equivalent to a cylinder.

\subsection{Quantum operators}

 To quantize the theory, let us work in temporal gauge $A_T=0$, while remembering to impose the constraint obtained by varying \eqref{BF:ActionLor} with respect to $A_T$, namely $\partial_\varphi B=0$~. In this gauge, the only surviving mode of $B(\mathbf{x})$ is its spatial zero-mode $b$:
\begin{equation}\label{eq:Bzeromode}
b\equiv \frac{1}{2\pi}\oint\dd\varphi\, B~, 
\end{equation}
with $b\cong b+2\pi$. 
The temporal gauge admits residual gauge transformations of the form $A_\varphi\rightarrow A_\varphi-\partial_\varphi\alpha$, under which the zero-mode: 
\begin{equation}
a\equiv\oint\dd\varphi A_\varphi 
\end{equation}
shifts by $a\rightarrow a-[\alpha(2\pi)-\alpha(0)]$. Since $h= e^{i\alpha(\mathbf{x})}\in U(1)$, we require $[\alpha(2\pi)-\alpha(0)]\in 2\pi\bZ$, which further implies that $a$ has a compact target space: $a\cong a+2\pi$. Hence, in gauge-fixed form, our theory reduces to that of two compact, quantum-mechanical degrees of freedom with action: 
\begin{equation}\label{eq:bfgaugefixed}
S_L^{\rm BF}=\frac{p}{2\pi}\int \dd T\, b\,\dot{a}~.
\end{equation}
The above action indicates two things: first, that $a$ and $b$ are canonically conjugate, meaning we can quantize the theory by imposing the following commutation relation: 
\begin{equation}\label{eq:canonicalcomm}
[a,b]=\frac{2\pi i}{p}~.
\end{equation}
Second, \eqref{eq:bfgaugefixed} shows that the Hamiltonian of BF theory vanishes---the theory has no dynamics. 
Since our fields are compact, the globally well-defined operators are:  
\begin{equation}
\hat{U}_m\equiv e^{i m b}~, \qquad \hat{L}_n\equiv e^{i n a}~,\qquad m,n \in \mathbb{Z}~,
\end{equation}
and, by using the Baker-Campbell-Hausdorff formula, along with \eqref{eq:canonicalcomm}, we find the following operator algebra \citep{Bergeron:1994ym, Blau:1993hj,Kapustin:2014gua,seiberg2015}: 
\begin{equation}
\begin{split}
    &\hat{U}_m\, \hat{L}_n = e^{\frac{2\pi i }{p} nm}   \hat{L}_n\, \hat{U}_m \, , \\ 
    \hat{U}_m \hat{U}_{m'} &= \hat{U}_{m+m'} \, , \qquad \hat{L}_n \hat{L}_{n'} = \hat{L}_{n+n'} \, .
\end{split}
\label{BF:ClockAlgebra}
\end{equation}
Let us rearrange this algebra in a suggestive manner: 
\begin{equation}\label{eq:chargeofU}
\hat{L}_n\,\hat{U}_m\, \hat{L}_n^\dagger  =e^{-\frac{2\pi i }{p} nm}\,\hat{U}_m~, \qquad\hat{L}_n^\dagger\equiv \hat{L}_{-n}~.
\end{equation}
This expression can be understood diagrammatically as follows:\footnote{In this diagram $L_n^\dagger$ is represented as an $L_n$ with its curve orientation reversed, and operators acting from right to left in \eqref{eq:chargeofU} are ordered acting earliest to latest. Since BF theory is topological, we only need to keep track of the topology of spacetime: $\mathbb{R}\times S^1$, and we can move operators freely in spacetime so long as we keep track of how they link with each other. Finally we draw these diagrams with time pointing downward since this aligns with our `right handed' orientation $\epsilon_{T\varphi}=+\sqrt{-g}$.} 
\begin{equation}\label{eq:chargemeasure}
\begin{gathered}\includegraphics[height=3.85cm]{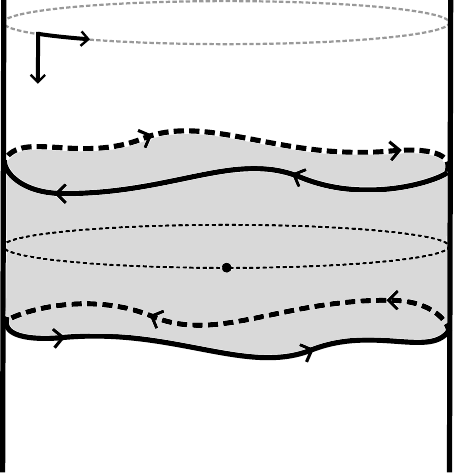}\put(-82,99){{\scriptsize $\varphi$}}\put(-99,82){{\scriptsize $T$}}\put(-25,81){{\footnotesize $\hat{L}_n$}}\put(-62,43){{\footnotesize $\hat{U}_m$}}
    \put(-25,19){{\footnotesize $\hat{L}_n$}}\end{gathered}=\begin{gathered}\includegraphics[height=3.85cm]{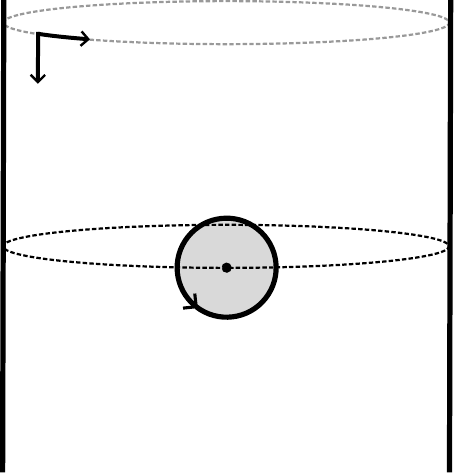}\put(-82,99){{\scriptsize $\varphi$}}\put(-99,82){{\scriptsize $T$}}\put(-62,43){{\footnotesize $\hat{U}_m$}}\put(-46,29){{\footnotesize $\hat{L}_n$}}\end{gathered}=e^{-\frac{2\pi i }{p} nm}\times\begin{gathered}\includegraphics[height=3.85cm]{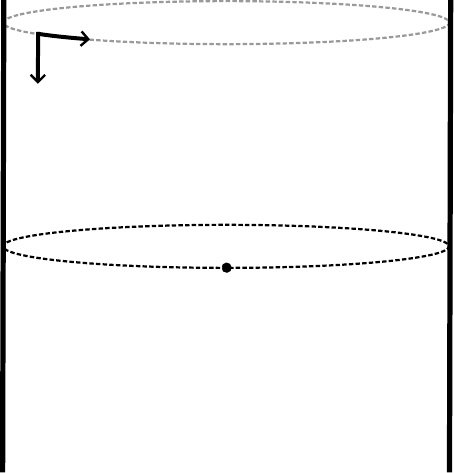}\put(-82,99){{\scriptsize $\varphi$}}\put(-99,82){{\scriptsize $T$}}\put(-62,43){{\footnotesize $\hat{U}_m$}}\end{gathered}
\end{equation}
in other words, if an $\hat{L}_n$ wraps a $\hat{U}_m$, it measures its \emph{charge}. The origin of this charge becomes clear when we go back to the definition of the operator $\hat{U}_m\equiv e^{imb}$. Under shifts of $b\rightarrow b-\frac{2\pi n}{p}$, the operator $\hat{U}_m$ gets multiplied by a $\mathbb{Z}_p$ phase, which is precisely what is being measured by $\hat{L}_n$. What we have uncovered is that this model is invariant under a discrete shift symmetry of the underlying operator $B(\mathbf{x})$. We remind the reader that in its gauge invariant form, $\hat{U}_m(\mathbf{x}) \equiv e^{i m B(\mathbf{x})}$ is in fact a local operator, despite how it may appear in this section.

This model actually has a dual $\mathbb{Z}_p$ symmetry. To see this, let us alternatively rearrange \eqref{BF:ClockAlgebra} as follows 
\begin{equation}
\hat{U}_m\, \hat{L}_n\hat{U}_m^\dagger  =e^{\frac{2\pi i }{p} nm}\hat{L}_n~, \qquad\hat{U}_m^\dagger\equiv \hat{U}_{-m}~,
\end{equation}
which, in diagram form, looks like: 
\begin{equation}
\begin{gathered}\includegraphics[height=3.85cm]{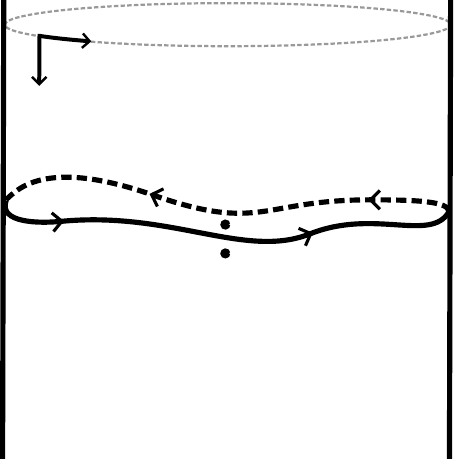}\put(-84,98){{\scriptsize $\varphi$}}\put(-101,81){{\scriptsize $T$}}\put(-60,37){{\footnotesize $\hat{U}_m$}}\put(-60,65){{\footnotesize $\hat{U}_m^\dagger$}}\put(-18,45){{\footnotesize $\hat{L}_n$}}\end{gathered}=e^{\frac{2\pi i }{p} nm}\times\begin{gathered}\includegraphics[height=3.85cm]{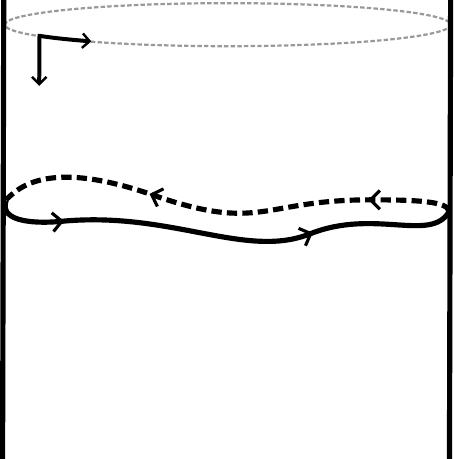}\put(-84,98){{\scriptsize $\varphi$}}\put(-101,81){{\scriptsize $T$}}\put(-18,45){{\footnotesize $\hat{L}_n$}}\end{gathered}~,
\end{equation}
and we conclude, in similar fashion, that when an $\hat{L}_n$ operator is `wrapped' by a pair of $\hat{U}_m$'s, we recover the charge of the $\hat{L}_n$ operator. Hence we have landed on the commonly-used language in the literature that $\hat{L}_n$ and $\hat{U}_m$ are each other's respective \emph{charge operators}. The $\mathbb{Z}_p$ symmetry being measured in this case is a discrete shift in the operator $a$, which we can trace back to a discrete shift in the holonomy of $A_\varphi$. 

As a final remark we note that the algebra \eqref{BF:ClockAlgebra} is invariant under shifts of $n$ and $m$ by multiples of $p$. Hence we deduce the equivalence: 
\begin{equation}
\hat{L}_n\cong \hat{L}_{n+p}, \qquad \hat{U}_m\cong \hat{U}_{m+p}~, \qquad\forall\, m,n\in \mathbb{Z}  
\end{equation}
and notably 
\begin{equation}
\hat{L}_p\cong \hat{U}_p \cong \mathds{1}~. 
\end{equation}
This means that we only have a finite number of operators
\begin{equation}
\hat{L}_n,~\hat{U}_m~,\qquad m,~n\in 0,1,\dots,p-1~,
\end{equation}
and by extension, a finite number of states. 

In summary, and to be slightly more precise in our discussion of the symmetries, we have just reviewed the well-known fact that BF theory, with coupling constant $p$, has a pair of $\mathbb{Z}_p$ symmetries, one being a zero-form symmetry, and the other being a one-form symmetry.  
The topological line operators $\hat L_n$ implement the $\mathbb{Z}_p^{(0)}$ 0-form symmetry while being charged under the $\mathbb{Z}_p^{(1)}$ 1-form symmetry.
Conversely,  the topological local operators $\hat U_m$ are charged under the $\mathbb{Z}_p^{(0)}$ 0-form symmetry, while implementing the 1-form symmetry. In the next section, we will review the fact that these symmetries are \emph{spontaneously broken}.

\subsection{Hilbert space (or vacua versus pUniverses)}\label{sub:hilbert_space}

As our diagrams in the previous section show, the $\hat{U}_m$ operators are both topological and local; by the equations of motion, it does not actually matter where in spacetime these operators are placed. On the other hand, the $\hat{L}_n$ are a set of topological \emph{line} operators, which act on all of space. To define the Hilbert space, it will be convenient to diagonalize one set of these operators. 

\subsubsection*{The vacua}
Let us work in the basis that diagonalizes the $\hat{L}_n$ operators. These are called \emph{vacua}: 
\begin{equation}\label{eq: BF fixed holonomy basis}
    \mathcal{H}_{S^1} = \textnormal{span}\left\{~ {\ket{m}} ~\lvert \quad m=0,1,\dots, p-1  \right\} \, , \qquad \hat{L}_n{\ket{m}} = e^{-\frac{2\pi i}{p}nm}{\ket{m}}~,
\end{equation}
which, by \eqref{BF:ClockAlgebra}, are constructed as follows:
\begin{equation}
    {\ket{m}} \equiv \hat{U}_m{\ket{0}} \, ,
\end{equation}
and  satisfy: 
\begin{equation}\label{eq:Zp0Hilbertoverlaps}
{\braket{n|m}}=\begin{cases}1~, & m=n \text{ mod } p\\
0~, & \text{otherwise} \end{cases}~,\qquad\hat{U}_k {\ket{m}}={\ket{k+m}}~,\qquad {\ket{m}}\cong {\ket{m+p}}~.
\end{equation}
In other words, we have a basis of $p$ distinct states, each of which is charged under the $\mathbb{Z}_p^{(0)}$ symmetry. Since the $\bZ_p^{(1)}$ operators obtain a vev in this basis, we also see that the $\mathbb{Z}_p^{(1)}$ symmetry is spontaneously broken.  However, note that it is possible to mediate transitions to other states of different charge by acting with the local operators $\hat{U}_m$. 

\subsubsection*{The pUniverses}
Alternatively, we could have  chosen to work in the basis that diagonalizes the $\hat{U}_m$ operators, which are called universes in the literature \citep{Komargodski:2020mxz}, but we will instead opt for the name \textbf{pUniverses}: 
\begin{equation}\label{eq: shift sym basis}
    \mathcal{H}_{S^1} = \textnormal{span}\left\{~ \widetilde{\ket{n}} ~\lvert \quad n=0,1,\dots, p-1  \right\} \, , \qquad \hat{U}_m\widetilde{\ket{n}} = e^{\frac{2\pi i}{p}nm}\widetilde{\ket{n}}~.
\end{equation}
Using the algebra \eqref{BF:ClockAlgebra}, it is straightforward to check that these states can be constructed as follows: 
\begin{equation}
\widetilde{\ket{n}}\equiv \hat{L}_n\widetilde{\ket{0}}~, 
\end{equation}
and satisfy:
\begin{equation}\label{eq:shift}
\widetilde{\braket{m|n}}=\begin{cases}1~, & m=n \text{ mod } p\\
0~, & \text{otherwise} \end{cases}~,\qquad\hat{L}_k\widetilde{\ket{n}}=\widetilde{\ket{k+n}}~,\qquad \widetilde{\ket{n}}\cong \widetilde{\ket{n+p}}~.
\end{equation}
As before, we have a basis of states charged under $\bZ_p^{(1)}$, and the $\mathbb{Z}_p^{(0)}$ charge operators obtain a vev in each of these states; hence we conclude that the $\mathbb{Z}_p^{(0)}$ symmetry is spontaneously broken. However, unlike the previous basis, by \eqref{eq:shift} we conclude that there exists no local operator that mediates between the various states charged under the $\mathbb{Z}_p^{(1)}$ (the only such operators act on all of space) and hence no local operations can restore the symmetry. This spontaneous symmetry breaking is much more robust.

\subsubsection*{A note on nomenclature}
As stated above, the states $\ket{n}$, which carry definite $\bZ^{(0)}_p$ charge, are referred to as the \textbf{vacua} of the theory. Using this language in BF theory is a bit of an overkill. Since the Hamiltonian vanishes, all states are technically vacua. However, the nomenclature is in place to distinguish these states from the set of \textbf{pUniverses} $\widetilde{\ket{m}}$ charged under $\bZ^{(1)}_p$. The distinguishing feature between the vacua and the pUniverses is that one can mediate between the vacua by acting with local operators, whereas moving between the different pUniverses requires acting on all of space---an inaccessible operation for local observers, so the pUniverses are completely isolated from one another. Since the coupling constant $p$ tells us that there are $p$ inequivalent universes, we have opted to call them pUniverses rather than universes, as a mnemonic.  

Since both the vacua and the pUniverses form complete set on our finite-dimensional Hilbert space, it must be possible to express one set in terms of the other. A short calculation yields:  
\begin{equation}\label{eq:changeofbasiszp0zp1}
\widetilde{\ket{m}}=\frac{1}{\sqrt{p}}\sum_{j=0}^{p-1}e^{-\frac{2\pi i}{p}jm}\ket{j}~,\qquad{\ket{n}}=\frac{1}{\sqrt{p}}\sum_{j=0}^{p-1}e^{+\frac{2\pi i}{p}jn}\widetilde{\ket{j}},
\end{equation}
with the important consequence that 
\begin{equation}
\widetilde{\ket{0}}\neq\ket{0}~,
\end{equation}
unless $p=1$.

\subsection{Hartle-Hawking construction of vacua and pUniverses}

\subsubsection{Summary: The Hartle-Hawking state carries definite \texorpdfstring{$\mathbb{Z}_p^{(0)}$}{Zp0} charge}

The discussion so far has been standard and can be found in many references \citep{Bergeron:1994ym, Blau:1993hj,Kapustin:2014gua,seiberg2015}. To connect with what is to come, we will now repeat the above construction of the BF Hilbert space and operator algebra using the Hartle-Hawking picture.\footnote{Traditionally, the Hartle-Hawking state is associated to a certain wavefunction in a theory of gravity \citep{Hartle:1983ai}. Here, there is no gravity, but the construction of the state resorts to Euclidean signature in a similar way so we stick to the Hartle-Hawking terminology.} Since we are ultimately interested in de Sitter physics, let us spend a few moments to recall that the Euclidean continuation of the de Sitter metric results in the round $S^2$. To see this, consider the analytic continuation of the metric \eqref{eq:deSittermetric} by taking $T=-i X-\frac{\pi}{2}$:
\begin{equation}
\frac{\dd s^2}{\ell^2}=\frac{\dd X^2+\dd \varphi^2}{\cosh^2 X}~,\qquad  X\in \mathbb{R},\quad \varphi\cong\varphi+2\pi~.
\end{equation}
While these coordinates may not seem immediately familiar, they represent the fact that the sphere (with its poles removed) is conformally equivalent to the Euclidean cylinder, as was the case with its Lorentzian counterpart \eqref{eq:deSittermetric}. However, the sphere is topologically distinct from the cylinder, with important consequences. The more familiar spherical coordinates on the $S^2$ can be obtained by parameterizing  $X=\log\tan\frac{\vartheta}{2}$, upon which the metric is expressed as: 
\begin{equation}
\frac{\dd s^2}{\ell^2}=\dd\vartheta^2+\sin^2\vartheta\,\dd\varphi^2~,\qquad  \vartheta\in (0,\pi),\quad \varphi\cong\varphi+2\pi~.
\end{equation}
We could have obtained this metric more directly by starting from the Lorentzian metric \eqref{eq:deSittermetricglobal} and continuing $\tau\rightarrow -i\left(\vartheta-\frac{\pi}{2}\right)$. What this reveals is that Euclidean time runs forward from the north pole at $\vartheta=0$ to the south pole at $\vartheta=\pi$. Following the analytic continuations carefully means orienting our coordinates such that $\epsilon_{\vartheta\varphi}=-\epsilon_{\varphi\vartheta}=\sqrt{g}$.

In the standard Hartle-Hawking picture, the quantum field theoretic Euclidean path integral over the entire two-sphere (with or without operator insertions at the poles) computes an overlap between states. For example, the bare (appropriately gauge-fixed) path integral computes for us: 
\begin{equation}\label{eq:BDoverlapZdef}
{\braket{0|0}}\propto \begin{gathered}\includegraphics[height=2.3cm]{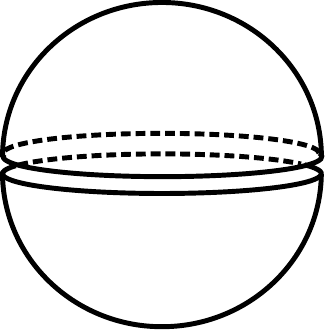}\end{gathered}= \int \frac{DB DA_\mu}{\textnormal{vol}\,\mathcal{G}} \,   e^{-S^{\textnormal{BF} }_E}\equiv \cZ_{\rm BF}~,
\end{equation}
where we have used the `$\propto$' symbol to remind the reader that the path integral $\mathcal{Z}$ computes the un-normalized overlap between the Hartle-Hawking state and its Hermitian conjugate. The division by $\textnormal{vol}\,\mathcal{G}$ in the definition of the path integral ensures that we don't overcount gauge orbits. Defined appropriately, the state $\ket{0}$ should then be understood as one of the vacua of \eqref{eq: BF fixed holonomy basis}, meaning it is  \emph{carries definite $\mathbb{Z}_p^{(0)}$ charge}. 

Since Euclidean time runs from the north pole to the south pole, the ket $\ket{0}$ is represented by the portion of the path integral computed over the northern hemisphere, while the bra $\bra{0}$ is represented by the remaining portion in the southern hemisphere. We will give a path integral derivation that the Hartle-Hawking state carries $\mathbb{Z}_p^{(0)}$ charge below.

In order to give a path integral representation of the Hilbert space, we need to reintroduce positional dependence to our topological operators. To this end we define: 
\begin{equation}\label{eq: BF top ops}
\hat{U}_n(\mathbf{x})\equiv e^{i n B(\mathbf{x})}~, \qquad \hat{L}_m[\cC]\equiv e^{i m\oint_\cC A_\mu \dd x^\mu}~.
\end{equation}
A correct treatment of the path integral will ensure that correlation functions of these operators only depend topologically on their insertion locations/paths. 

Besides the norm of the state $\ket{0}$, in \eqref{eq:zp0overlapcal} we will show that the remaining overlaps in this basis are computed in a similar fashion, i.e. by inserting local operator insertions into the bare path integral :
\begin{align}
{\braket{m|n}}&= \begin{gathered}\includegraphics[height=2.3cm]{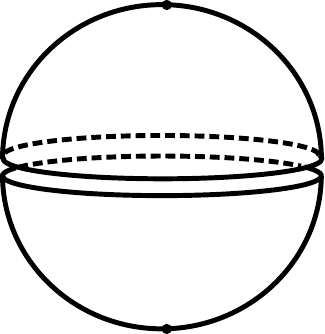}\put(-40,72){{\footnotesize $\hat{U}_n$}($\mathbf{x}$)}\put(-40,-12){{\footnotesize $\hat{U}_m^\dagger$}($\mathbf{y}$)}\end{gathered}\qquad\qquad~,\nonumber\\&= \frac{1}{\cZ_{\rm BF}}\int \frac{DB DA_\mu}{\textnormal{vol}\mathcal{G}} \,e^{-im B(\mathbf{y})}e^{in B(\mathbf{x})}   e^{-S^{\textnormal{BF} }_E}=\begin{cases}1~, & m=n \text{ mod } p \\
0~, & \text{otherwise} \end{cases}~,\label{eq:Zp0overlaps}
\end{align}
where the ket $\ket{n}$ is again drawn as arising from the path integral over the northern hemisphere while the bra $\bra{m}$ is drawn as coming from the path integral performed over the southern hemisphere. 

Let us add a few details. In Euclidean signature, the path integral is weighted by the exponential of minus the \emph{Euclidean} BF action on the $S^2$:\footnote{ Note the sign difference with respect to \eqref{BF:ActionLor}. This difference is simply due to the fact that raising indices in Lorentzian signature introduces a sign, while doing so in Euclidean signature does not.}
\begin{equation}
    e^{-S^{\textnormal{BF}}_{ E}}= \exp\left[+\frac{i p}{4\pi} \int_{S^2} \dd^2 x \sqrt{g}\,   B(\mathbf{x})  \epsilon^{\mu \nu} F_{\mu \nu}(\mathbf{x}) \right]\, ,
    \label{BF:Action}
\end{equation}
 and the field strength is normalized to satisfy the Dirac quantization condition:
\be\label{eq: Dirac quant}
-\frac{1}{4\pi}\int_{S^2}\dd^2 x\sqrt{g} \epsilon^{\mu\nu}F_{\mu\nu}=k\in \mathbb{Z} \, .
\ee
Since the field $B(\mathbf{x})$ is a compact scalar, $B(\mathbf{x}) \cong B(\mathbf{x}) + 2\pi$, in a background of flux $k$, any shift of the form
\begin{equation}\label{eq:zpbf}
B(\mathbf{x})\rightarrow B(\mathbf{x})+\frac{2\pi n}{p}~,
\end{equation}
for $n=0,1,\dots,p-1$, shifts the action by 
\begin{equation}\label{eq:pathintegralinvBF}
S^{\textnormal{BF}}_{ E}\rightarrow S^{\textnormal{BF}}_{ E}+2\pi i k n~,
\end{equation}
meaning the path integral remains unchanged. Thus we have uncovered the Euclidean origin of the underlying $\mathbb{Z}_p^{(0)}$ symmetry described in the previous section. 

Finally we can use the above analysis to outline the Hartle-Hawking construction of the states charged under the $\mathbb{Z}_p^{(1)}$ one-form symmetry---the pUniverses. This requires making use of \eqref{eq:changeofbasiszp0zp1}. Taking \eqref{eq:changeofbasiszp0zp1} at face value suggests that the norm of the state $\widetilde{\ket{0}}$ is computed as follows:
\begin{equation}
\widetilde{\braket{0|0}}\propto \begin{gathered}\includegraphics[height=2.3cm]{Images/Overlaps2.pdf}\put(-63,77){{\footnotesize $\frac{1}{\sqrt{p}}\sum\limits_{r=0}^{p-1}\hat{U}_r$}($\mathbf{x}$)}\put(-63,-17){{\footnotesize $\frac{1}{\sqrt{p}}\sum\limits_{s=0}^{p-1}\hat{U}_s^\dagger$}($\mathbf{y}$)}\end{gathered}= \frac{1}{p}\int \frac{DB DA_\mu}{\textnormal{vol}\mathcal{G}} \,  \sum_{r,s=0}^{p-1}\,e^{-is B(\mathbf{y})}e^{ir B(\mathbf{x})}  e^{-S^{\textnormal{BF} }_E}\equiv \widetilde{\mathcal{Z}}_{\rm BF}~.
\end{equation}
We will verify this picture shortly. The immediate takeaway is that the Hartle-Hawking construction seems more naturally designed to produce vacua rather than pUniverses. 

The remaining overlaps are obtained by inserting topological line operators into the Euclidean path integral:
\begin{equation}\label{eq:1formstateoverlap}
\widetilde{\braket{m|n}}= \begin{gathered}\includegraphics[height=3cm]{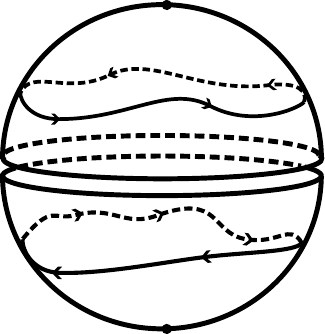}\put(-63,97){{\footnotesize $\frac{1}{\sqrt{p}}\sum\limits_{r=0}^{p-1}\hat{U}_r$}($\mathbf{x}$)}\put(-63,-17){{\footnotesize $\frac{1}{\sqrt{p}}\sum\limits_{s=0}^{p-1}\hat{U}_s^\dagger$}($\mathbf{y}$)}\put(-5,69){\footnotesize $\hat{L}_n[\mathcal{C}]$}\put(-5,9){\footnotesize $\hat{L}_m[\mathcal{C}']$}\end{gathered}~\qquad=\begin{cases}1~, & m=n \text{ mod } p \\
0~, & \text{otherwise} \end{cases}.\end{equation}
We will compute this via the following expression: 
\begin{equation}\widetilde{\braket{m|n}}= \frac{1}{p\,\widetilde{\mathcal{Z}}_{\rm BF}}\int \frac{DB DA_\mu}{\textnormal{vol}\mathcal{G}} \,  \sum_{r,s=0}^{p-1}\,e^{-is B(\mathbf{y})}e^{im\oint_{\mathcal{C}'}A_\nu\dd z^\nu}e^{in\oint_\mathcal{C} A_\mu \dd w^\mu}e^{ir B(\mathbf{x})}  e^{-S^{\textnormal{BF} }_E}~,
\end{equation}
whose calculation we outline around \eqref{eq:zp1overlapcal}.
In summary, we must first prepare the 1-form symmetry vacuum, then act with a topological line to produce a particular state of definite 1-form charge. 
From \eqref{eq:1formstateoverlap}, we see that the local operator insertions preparing the $\widetilde{\ket{0}}$ pUniverse provide an obstruction to shrinking the topological lines, ensuring they have a physical effect.

Actually, as we will see, we can use \eqref{eq:chargemeasure} to shrink the topological line operators and pick up the required phases: 
\begin{equation}\label{eq:shrink}
\widetilde{\braket{m|n}}= \qquad\begin{gathered}\includegraphics[height=3cm]{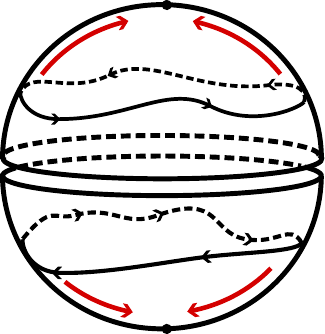}\put(-63,97){{\footnotesize $\frac{1}{\sqrt{p}}\sum\limits_{r=0}^{p-1}\hat{U}_r$}($\mathbf{x}$)}\put(-63,-17){{\footnotesize $\frac{1}{\sqrt{p}}\sum\limits_{s=0}^{p-1}\hat{U}_s^\dagger$}($\mathbf{y}$)}\put(-5,69){\footnotesize $\hat{L}_n[\mathcal{C}]$}\put(-5,9){\footnotesize $\hat{L}_m[\mathcal{C}']$}\end{gathered}\qquad=\qquad\begin{gathered}\includegraphics[height=3cm]{Images/Overlaps2.pdf}\put(-63,97){{\footnotesize $\frac{1}{\sqrt{p}}\sum\limits_{r=0}^{p-1}e^{-\frac{2\pi i}{p}nr}\,\hat{U}_r$}($\mathbf{x}$)}\put(-63,-17){{\footnotesize $\frac{1}{\sqrt{p}}\sum\limits_{s=0}^{p-1}e^{+\frac{2\pi i}{p}ms}\,\hat{U}_s^\dagger$}($\mathbf{y}$)}\end{gathered}~.
\end{equation}
as needed by \eqref{eq:changeofbasiszp0zp1}.

In the next section we will explain how to calculate these overlaps precisely using the path integral.

\subsubsection{Calculations}

We will spare the reader the details of computing \emph{all} of the above overlaps. Since this theory is linear, we can verify everything by computing certain simple configurations and adding them together in superpositions. 

To proceed we introduce yet another set of so-called `stereographic' coordinates on the $S^2$: 
\begin{equation}\label{spherestereo}
\dd s^2= \frac{4 \ell^4\dd\mathbf{x}\cdot \dd \mathbf{x}}{(\ell^2+ \mathbf{x}\cdot \mathbf{x})^2}\equiv \Omega(\mathbf{x})^2\dd\mathbf{x}\cdot \dd \mathbf{x}~,
\end{equation}
where $\ell$ is the radius of the sphere, and $\mathbf{x}\in\mathbb{R}^2$. The origin of these coordinates is located at the north pole of the $S^2$. These coordinates demonstrate that the $S^2$ is Weyl-flat with conformal factor $\Omega(\mathbf{x})=\frac{2 \ell^2}{(\ell^2+ \mathbf{x}\cdot \mathbf{x})}$. The equator of the $S^2$ lies along $\mathbf{x}\cdot\mathbf{x}=\ell^2$. The second chart for the manifold, centered around the south pole, can be obtained by a coordinate inversion: 
\begin{equation}
\mathbf{x}\rightarrow \frac{\ell^2}{\mathbf{x}'\cdot\mathbf{x}'}\mathbf{x}'~.
\end{equation}

Since we're working on the $S^2$, the gauge-field configurations split up into integrally quantized topological sectors \eqref{eq: Dirac quant}. We can use this fact to parametrize the full set of gauge field configurations within a fixed, charge-$k$, topological sector as follows:
\begin{equation} \label{Ak}
A^{(k)}_\mu(\mathbf{x}) = k C_\mu(\mathbf{x}) +\epsilon_{\mu\nu} \partial^\nu\Phi(\mathbf{x}) + {i} h(\mathbf{x})^{-1} \partial_\mu h(\mathbf{x})~.
\end{equation}
The first term represents a $k$-instanton background, and is expressed as $k$ times a one-instanton background \citep{Anninos:2024fty,Jayewardena:1988td}:
\begin{equation}\label{eq:instantongaugefield}
C_\mu\equiv-\frac{1}{{}2}\epsilon_{\mu\nu}\partial^\nu\log\Omega(\mathbf{x})~.
\end{equation} 
The field $\Phi(\mathbf{x})$ is a scalar that parametrizes transverse fluctuations in Lorenz gauge. Since shifts of $\Phi(\mathbf{x})$ by a constant do not affect the field configuration, $\Phi$'s zero-mode will not be included in the path integral. Lastly $h(\mathbf{x}) \in U(1)$ represents the pure gauge contribution.

\subsubsection*{First calculation: The Hartle-Hawking norm}
Let us first explain how to compute $\mathcal{Z}$, c.f. equation \eqref{eq:BDoverlapZdef}, the normalization of the Hartle-Hawking state.  From the definition we have: 
\begin{equation}
\cZ_{\rm BF}=\sum_{k=-\infty}^{\infty}\int \frac{DB DA^{(k)}_\mu}{\textnormal{vol}\,\mathcal{G}} \,   \exp\left[{+\frac{i p}{4\pi} \int_{S^2} \dd^2 x \sqrt{g}\,   B(\mathbf{x})  \epsilon^{\mu \nu} F^{(k)}_{\mu \nu}(\mathbf{x})}\right]~,
\end{equation}
where $F_{\mu\nu}^{(k)}$ is the field strength associated to the field configuration \eqref{Ak}. In the coordinates \eqref{spherestereo}: 
\begin{equation}\label{eq:fieldstrengthinstanton}
F_{\mu\nu}^{(k)}=-\epsilon_{\mu\nu}\left(\frac{k}{2\ell^2}+\nabla^2\Phi\right)~.
\end{equation}
Additionally, we will split the path integral over $B$ between its zero-mode and nonzero-mode contributions: 
\begin{equation}
b\equiv \frac{1}{4\pi\ell^2} \int_{S^2} \dd^2 x \sqrt{g}\,   B(\mathbf{x})~,\qquad\qquad B'(\mathbf{x})\equiv B(\mathbf{x})-b~,
\end{equation}
where $b\cong b+2\pi$ is the same as in \eqref{eq:Bzeromode} and $B'(\mathbf{x})$ has no zero-mode. 

It is impossible to meaningfully compute any path integral without first specifying a regulator. Our choice of heat kernel regulator is summarized in \cref{app:bosonregularization}, to which we refer the reader for more details. This choice naturally introduces an  ultraviolet cutoff scale $\Lambda_{\rm UV}=\ell_{\rm UV}^{-1}$ defined in \eqref{eq:cutoffboson}, which will appear in the equations below. 

With our choice of regulator in hand, we are tasked with computing:\footnote{We refer the reader to the discussion around \eqref{eq:BFmeasure} to understand precisely why the factor of $\ell_{\rm UV}$ appears in the above measure.}
\begin{equation}\label{eq:Zdefintermediate}
\cZ^\epsilon_{\rm BF}=\sum_{k=-\infty}^{\infty}\frac{\ell}{\ell_{\rm UV}}\int_{0}^{2\pi}{\dd b}\int \frac{\, DB' D\Phi Dh}{\textnormal{vol}\,\mathcal{G}}J_{\Phi,h} \,   \exp\left[-ip\, k\,b  -\frac{i p}{2\pi}\int_{S^2} \dd^2 x \sqrt{g}\,   B'(\mathbf{x})\nabla^2\Phi(\mathbf{x})  \right]~.
\end{equation}
The Jacobian measure factor $J_{\Phi,h}$ in the above path integral \citep{Klebanov:2011td,Giombi:2015haa} arises, formally, from the change of variables in \eqref{Ak}, and is given  by 
\begin{equation}\label{jac}
J_{\Phi,h} = {\left|{\det}_\epsilon'\, \partial_\mu | \times | {\det}_\epsilon'\, \epsilon_{\mu\nu} \partial^\nu \right|} \equiv {\det}_\epsilon' \left(-\nabla^2\right)~,
\end{equation}
where the primes $'$ denote that these determinants have the $L=0$ mode removed and the subscript $\epsilon$ implies we are regulating this determinant using \eqref{eq:heatkernelharishboson}. 
This regulated Jacobian factor is a pure number independent of the fields. 

It is now straightforward to perform the Euclidean path integral over the field configurations of $B(\mathbf{x})$, split between $b$ and $B'(\mathbf{x})$: 
\begin{equation}\label{eq:Zdefintermediate2}
\cZ^\epsilon_{\rm BF}=\frac{\ell}{\ell_{\rm UV}}\,{\det}_\epsilon' \left(-\nabla^2\right)\sum_{k=-\infty}^{\infty}2\pi\delta_{kp,0}\int \frac{ Dh}{\textnormal{vol}\,\mathcal{G}} \int D\Phi\,   \delta\left(-\frac{p}{2\pi} \nabla^2\Phi(\mathbf{x})\right)~,
\end{equation}
hence the integral over $b$ has collapsed the instanton sum to the $k=0$ sector.

The computation of the middle term requires some care and, within our regularization scheme, can be found below in \cref{app:bosonregularization} (see equation \eqref{eq:volgmeasure}), yielding:
\begin{equation}\label{eq: Dh int}
\int \frac{ Dh}{\textnormal{vol}\,\mathcal{G}}= \frac{\ell_{\rm UV}}{2\pi\ell}~.
\end{equation}
Lastly, to perform the integral over $\Phi$, let us change variables:
\begin{equation}
\xi(\mathbf{x})\equiv -\frac{p}{2\pi} \nabla^2\Phi(\mathbf{x})\qquad\rightarrow\qquad D\Phi= \det{}^{-1} \left(\frac{\delta \xi}{\delta \Phi}\right) D\xi=\frac{D\xi}{\det{}'\left(-\frac{p}{2\pi}\nabla^2\right)}~.
\end{equation}
Combining everything, we find: 
\begin{align}
\cZ^\epsilon_{\rm BF}&=\frac{2\pi\ell}{\ell_{\rm UV}}\frac{\ell_{\rm UV}}{2\pi\ell}\frac{{\det}_\epsilon' \left(-\nabla^2\right)}{\det_\epsilon'\left(-\frac{p}{2\pi}\nabla^2\right)}\int D\xi\,   \delta\left(\xi(\mathbf{x})\right)~,\nonumber\\&=\frac{{\det}_\epsilon' \left(-\nabla^2\right)}{\det_\epsilon'\left(-\frac{p}{2\pi}\nabla^2\right)}~ .\label{eq:ZBFcalc}
\end{align}
 To proceed, we will now use the fact that, formally, for any positive operator $-D^2$ and a dimensionless constant $w$:\footnote{This equality should be understood up to divergences which can be removed by local counterterms. The nature of the UV finiteness in (\ref{eq:ZBFcalc}) and, on more general two-manifolds, was uncovered and clarified to us in discussions with Stathis Vitouladitis, see also \citep{asj,asj_TQFT}.} 
\begin{equation}\label{eq:zeromoderescaling}
{\det}'\left(-\frac{D^2}{w^2}\right)={w}^{-2}\,{\det}'\left(-D^2\right)~,
\end{equation}
which can be understood from the zero-mode removal. In other words, were we computing $\det$ rather than $\det'$, the rescaling would be undetectable since it can be absorbed by a field redefinition. Hence we have
\begin{equation}
{\det}_\epsilon'\left(-\frac{p}{2\pi}\nabla^2\right)=\frac{p}{2\pi}{\det}_\epsilon'\left(-\nabla^2\right)~.
\end{equation}
These formal manipulations give us: 
\begin{equation}
\boxed{\cZ^\epsilon_{\rm BF}=\frac{2\pi}{p}}~,
\end{equation}
which can be compared with the partition function of Abelian Chern-Simons theory in three dimensions \citep{Kitaev:2005dm,Levin:2006zz}.

\subsubsection*{Second calculation: The \texorpdfstring{$\mathbb{Z}_p^{(0)}$}{Zp0} `vacuum-to-vacuum' overlaps}

With our calculation of $\cZ_{\rm BF}$ complete, our next task is to compute overlaps between states of different 0-form charge, of the kind described in \eqref{eq:Zp0overlaps}:
\begin{equation}\label{eq:Zp0BFsetup}
{\braket{m|n}}= \begin{gathered}\includegraphics[height=2.3cm]{Images/Overlaps2.pdf}\put(-40,72){{\footnotesize $\hat{U}_n$}($\mathbf{x}$)}\put(-40,-12){{\footnotesize $\hat{U}_m^\dagger$}($\mathbf{y}$)}\end{gathered}= \frac{1}{\cZ_{\rm BF}}\int \frac{DB DA_\mu}{\textnormal{vol}\mathcal{G}} \,e^{-im B(\mathbf{y})}e^{in B(\mathbf{x})}   e^{-S^{\textnormal{BF} }_E}~.
\end{equation}
Explicitly, we must compute: 
\begin{multline}
{\braket{m|n}}=\frac{1}{\cZ_{\rm BF}}\sum_{k=-\infty}^{\infty}\frac{\ell}{\ell_{\rm UV}}\int_{0}^ {2\pi}\dd b\int \frac{ DB' D\Phi Dh}{\textnormal{vol}\,\mathcal{G}}J_{\Phi,h} \,   \exp\bigg\lbrace-ib\left(p\, k+m-n\right)  \\ +i\int_{S^2} \dd^2 x' \sqrt{g}\, \left[B'(\mathbf{x}') \left(n\frac{\delta\left(\mathbf{x}'-\mathbf{x}\right)}{\sqrt{g}}-m\frac{\delta\left(\mathbf{x}'-\mathbf{y}\right)}{\sqrt{g}}\right) -\frac{p}{2\pi}B'(\mathbf{x}')\nabla^2\Phi(\mathbf{x}')\right]  \bigg\rbrace~,
\end{multline}
 where we have again split $B(\mathbf{x})=b+B'(\mathbf{x})$ and the numerator of the above expression can be compared with \eqref{eq:Zdefintermediate}. 
 
 The next step is subtle and requires a bit of care. If we naively path integrate over $B'$, we would incorrectly conclude that we should impose the constraint: 
\begin{equation}\label{eq:wrongconstraint}
\nabla^2\Phi(\mathbf{x}')=\frac{2\pi}{p}\left(n\frac{\delta\left(\mathbf{x}'-\mathbf{x}\right)}{\sqrt{g}}-m\frac{\delta\left(\mathbf{x}'-\mathbf{y}\right)}{\sqrt{g}}\right)~.
\end{equation}
 However \eqref{eq:wrongconstraint} is inconsistent for $n\neq m$. Since  the $S^2$ is closed, we must have that $\int_{S^2}\nabla^2\Phi=0$. However, integrating both sides of \eqref{eq:wrongconstraint} leads to $0=\frac{2\pi}{p}(n-m)$. To correct the analysis, we simply shift $\Phi$ as follows:
\begin{equation}\label{eq:phishift}
\Phi(\mathbf{x}')\equiv\widetilde{\Phi}(\mathbf{x}')+\Phi_{\rm b}(\mathbf{x}')~,
\end{equation} 
where $\Phi_{\rm b}$ is a fixed background function that satisfies
\begin{equation}\label{eq:chidef}
 \nabla^2\Phi_{\rm b}(\mathbf{x}')=\frac{2\pi}{p}\left(n\frac{\delta\left(\mathbf{x}'-\mathbf{x}\right)}{\sqrt{g}}-m\frac{\delta\left(\mathbf{x}'-\mathbf{y}\right)}{\sqrt{g}}-\frac{n-m}{4\pi\ell^2}\right)~.
\end{equation}
Unlike \eqref{eq:wrongconstraint}, the above equation is consistent with $\int_{S^2}\nabla^2\Phi_{\rm b}=0$~. As before we demand that $\widetilde{\Phi}$ have no zero-mode:
\begin{equation}
\int_{S^2}\dd^2x\sqrt{g}\,\widetilde{\Phi}=0~.
\end{equation}
Since this field redefinition is a simple shift, its Jacobian is trivial and our path integral is now:
\begin{multline}
{\braket{m|n}}=\frac{1}{\cZ_{\rm BF}}\sum_{k=-\infty}^{\infty}\frac{\ell}{\ell_{\rm UV}}\int_{0}^ {2\pi}\dd b\int \frac{ DB' D\widetilde{\Phi} Dh}{\textnormal{vol}\,\mathcal{G}}J_{\Phi,h} \,   \exp\bigg\lbrace-ib\left(p\, k+m-n\right)  \\ +i\int_{S^2} \dd^2 x' \sqrt{g}\, \left[ -\frac{p}{2\pi}B'(\mathbf{x}')\nabla^2\widetilde{\Phi}(\mathbf{x}')\right]  \bigg\rbrace~,
\end{multline}
which the reader should compare with \eqref{eq:Zdefintermediate}.

Having given the computation of $\mathcal{Z}$ in excruciating detail, we can now afford to skip a few steps in the remainder of the derivation. Integrating out $B$ we find:
\begin{align}
{\braket{m|n}}&=\frac{1}{\cZ_{\rm BF}}\sum_{k=-\infty}^{\infty}2\pi\delta_{n,m+kp}\frac{\ell}{\ell_{\rm UV}}\int \frac{Dh}{\textnormal{vol}\,\mathcal{G}}J_{\Phi,h}   \int D\widetilde{\Phi} \,\delta\left(-\frac{p}{2\pi} \nabla^2\widetilde{\Phi}(\mathbf{x}')\right)~,\nonumber\\
&=\sum_{k=-\infty}^{\infty}\delta_{n,m+kp}~,
\end{align}
and in going from the first to the second line we used \eqref{eq:Zdefintermediate2} to cancel the factor of $\cZ_{\rm BF}$ between the numerator and denominator. This leads to a very simple result: 
\begin{equation}\label{eq:zp0overlapcal}
\boxed{\braket{m|n}=\begin{cases}1~, & m=n \text{ mod } p\\
0~, & \text{otherwise} \end{cases}}~,
\end{equation}
consistent with the operator picture given in \eqref{eq:Zp0Hilbertoverlaps}. In summary, we have confirmed that the vacua $\ket{n}$ can be constructed via a Euclidean path integral by inserting the topological local operator $\hat{U}_n(\mathbf{x})$ anywhere in the upper hemisphere. 

\subsubsection*{Third calculation: Topological line inserted between \texorpdfstring{$\mathbb{Z}_p^{(0)}$}{Zp0} vacua}

The next calculation we will perform is the expectation value of a topological line operator sandwiched between states of definite $\mathbb{Z}_p^{(0)}$ charge: 
\begin{equation}\label{eq:lineopexpectation}
\bra{m}\hat{L}_j[\mathcal{C}]\ket{n}= \begin{gathered}\includegraphics[height=3cm]{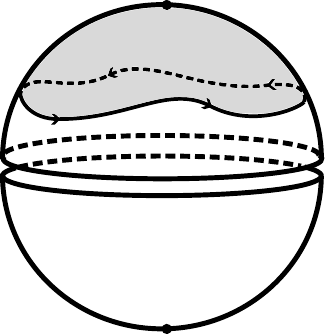}\put(-53,92){{\footnotesize $\hat{U}_n$}($\mathbf{x}$)}\put(-53,-12){{\footnotesize $\hat{U}_m^\dagger$}($\mathbf{y}$)}\put(-5,67){\footnotesize $\hat{L}_j[\cC]$}\end{gathered}=\frac{1}{\cZ_{\rm BF}}\int \frac{DB DA_\mu}{\textnormal{vol}\mathcal{G}} \,e^{-im B(\mathbf{y})}e^{i j\oint_\cC A_\mu \dd x^\mu}e^{in B(\mathbf{x})}   e^{-S^{\textnormal{BF} }_E}~,\end{equation}
where $\cC$ is the curve traced out by the line operator $\hat{L}_j[\mathcal{C}]$. We will label the smallest area subtended by the curve $\cC$ as $\cD$, denoted by the grey region in the above diagram. We will use $\cA_\cD$ to denote its area. Using Stokes' theorem and \eqref{eq:fieldstrengthinstanton}, we find: 
\begin{equation}\label{eq:StokestheoremWilson}
    \oint_\cC A^{(k)}_\mu \dd x^\mu = \frac{1}{2} \int_{\mathcal{D} }F_{\mu \nu}^{(k)} \dd x^\mu \wedge \dd x^\nu = -\frac{k}{2\ell^2} \mathcal{A}_\mathcal{D} - \int_{\mathcal{D}} \dd^2 x \sqrt{g} \nabla^2 \Phi(\mathbf{x}) \, .
\end{equation}
To proceed, we again split $B=b + B'$ between its zero mode and nonzero modes: 
\begin{multline}
\bra{m}\hat{L}_j[\mathcal{C}]\ket{n}=\frac{1}{\cZ_{\rm BF}}\sum_{k=-\infty}^{\infty}e^{-ijk\frac{\mathcal{A}_\mathcal{D}}{2\ell^2} }\frac{\ell}{\ell_{\rm UV}}\int \frac{Dh}{\textnormal{vol}\,\mathcal{G}}J_{\Phi,h} \int_0^{2\pi} \dd b\, e^{-ib(p\, k+m-n)}\\\int D\Phi e^{-ij\int_{\mathcal{D}} \dd^2 x' \sqrt{g} \nabla^2 \Phi(\mathbf{x}')}   \int DB'\, e^{i\int_{S^2} \dd^2 x' \sqrt{g}\, B'(\mathbf{x}') \left(-\frac{p}{2\pi}\nabla^2\Phi(\mathbf{x}')+n\frac{\delta\left(\mathbf{x}'-\mathbf{x}\right)}{\sqrt{g}}-m\frac{\delta\left(\mathbf{x}'-\mathbf{y}\right)}{\sqrt{g}}\right)  }~,
\end{multline}
which we write as
\begin{multline}\label{eq:lineoperatorintermed}
\bra{m}\hat{L}_j[\mathcal{C}]\ket{n}=\frac{1}{\cZ_{\rm BF}}\sum_{k=-\infty}^{\infty}2\pi \delta_{n,m+kp}e^{-ijk\frac{\mathcal{A}_\mathcal{D}}{2\ell^2} }\frac{\ell}{\ell_{\rm UV}}\int \frac{Dh}{\textnormal{vol}\,\mathcal{G}}J_{\Phi,h} \\\int DB' D\Phi\, e^{i\int_{S^2} \dd^2 x' \sqrt{g}\,\left[ B'(\mathbf{x}') \left(-\frac{p}{2\pi}\nabla^2\Phi(\mathbf{x}')+n\frac{\delta\left(\mathbf{x}'-\mathbf{x}\right)}{\sqrt{g}}-m\frac{\delta\left(\mathbf{x}'-\mathbf{y}\right)}{\sqrt{g}}\right)-j \Theta_\cD(\mathbf{x}')\nabla^2\Phi\right]  } ~.
\end{multline}
In the above equation we have introduced the function $\Theta_\cD$, which is a generalization of the Heaviside function, defined to have the following properties: 
\begin{equation}\label{eq:Heavisidedef}
\Theta_\cD(\mathbf{x})=\begin{cases} 0~, & \mathbf{x} \notin \cD \\
1~, & \mathbf{x}\in \cD \end{cases}~.
\end{equation}

Proceeding as in the previous calculation, the simplest way to correctly deal with the path integration over $B'$ and $\Phi$ is  to shift the integral over these fields as follows:
\begin{equation}\label{eq:bphishift}
B'(\mathbf{x}')\equiv \widetilde{B}'(\mathbf{x}')+\frac{2\pi j}{p}\left(\frac{\cA_\cD}{4\pi\ell^2}-\Theta_\cD(\mathbf{x}')\right)~,\qquad
\Phi(\mathbf{x}')\equiv\widetilde{\Phi}(\mathbf{x}')+\Phi_{\rm b}(\mathbf{x}')~,
\end{equation} 
where $\Phi_{\rm b}$ is defined  in \eqref{eq:chidef}.
As before we demand that $\widetilde{B}'$ and $\widetilde{\Phi}$ have no zero-modes:
\begin{equation}
\int_{S^2}\dd^2x\sqrt{g}\,\widetilde{B}'=\int_{S^2}\dd^2x\sqrt{g}\,\widetilde{\Phi}=0~.
\end{equation}
Since the Jacobian for these field redefinitions is trivial, inserting them into the \eqref{eq:lineoperatorintermed} gives:
\begin{multline}
\bra{m}\hat{L}_j[\mathcal{C}]\ket{n}=\frac{1}{\cZ_{\rm BF}}\sum_{k=-\infty}^{\infty}2\pi \delta_{n,m+kp}e^{-ij\left(k-\frac{n-m}{p}\right)\frac{\mathcal{A}_\mathcal{D}}{2\ell^2} }\frac{\ell}{\ell_{\rm UV}}\int \frac{Dh}{\textnormal{vol}\,\mathcal{G}}J_{\Phi,h} \\\int D\widetilde{B}' D\widetilde{\Phi}\, e^{i\int_{S^2} \dd^2 x' \sqrt{g}\,\left[ \widetilde{B}'(\mathbf{x}') \left(-\frac{p}{2\pi}\nabla^2\widetilde{\Phi}(\mathbf{x}')\right)-\frac{2\pi j}{p} \Theta_\cD(\mathbf{x}')\left(n\frac{\delta\left(\mathbf{x}'-\mathbf{x}\right)}{\sqrt{g}}-m\frac{\delta\left(\mathbf{x}'-\mathbf{y}\right)}{\sqrt{g}}\right)\right]  } ~.
\end{multline}
The path integral over $\widetilde{B}'$, $\widetilde{\Phi}$, and $h$ cancels with the factor of $\cZ_{\rm BF}$ and we are left with: 
\begin{equation}
\bra{m}\hat{L}_j[\mathcal{C}]\ket{n}=\sum_{k=-\infty}^{\infty} \delta_{n,m+kp}e^{-ij\left(k-\frac{n-m}{p}\right)\frac{\mathcal{A}_\mathcal{D}}{2\ell^2} -i\int_{S^2} \dd^2 x' \sqrt{g}\,\frac{2\pi j}{p} \Theta_\cD(\mathbf{x}')\left(n\frac{\delta\left(\mathbf{x}'-\mathbf{x}\right)}{\sqrt{g}}-m\frac{\delta\left(\mathbf{x}'-\mathbf{y}\right)}{\sqrt{g}}\right)} ~.
\end{equation}
Since $\cD$ overlaps with $\mathbf{x}$ as per equation \eqref{eq:lineopexpectation}, we find
\begin{equation}
\bra{m}\hat{L}_j[\mathcal{C}]\ket{n}=\begin{cases}e^{-\frac{2\pi i }{p}n j}~, & m=n \text{ mod } p\\
0~, & \text{otherwise} \end{cases}~,
\end{equation}
as needed. Note that we could have applied Stokes' theorem in the complement region to $\mathcal{D}$ and would have obtained the same result, accounting for orientation. 

Finally, using the steps outlined in the previous calculation, it is straightforward to derive the following result: 
\begin{equation}
\bra{m}\hat{L}_v[\mathcal{C}']\,\hat{L}_j[\mathcal{C}]\ket{n}= \quad\begin{gathered}\includegraphics[height=3cm]{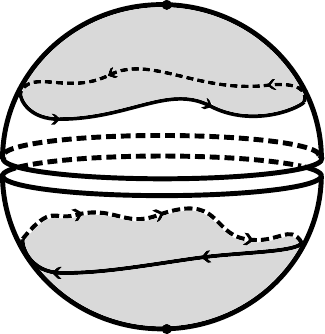}\put(-53,92){{\footnotesize $\hat{U}_n$}($\mathbf{x}$)}\put(-53,-12){{\footnotesize $\hat{U}_m^\dagger$}($\mathbf{y}$)}\put(-5,67){\footnotesize $\hat{L}_j[\cC]$}\put(-5,9){\footnotesize $\hat{L}_v[\cC']$}\end{gathered}\qquad=\begin{cases}e^{\frac{2\pi i }{p}\left(vm-n j\right)}~, & m=n \text{ mod } p\\
0~, & \text{otherwise} \end{cases}~,
\end{equation} 
needed to derive \eqref{eq: shift sym basis}-\eqref{eq:shift} using the Hartle-Hawking construction. That is, we have a path-integral derivation of the following: 
\begin{equation}\label{eq:zp1overlapcal}
\boxed{\widetilde{\braket{m|n}}=\begin{cases}1~, & m=n \text{ mod } p\\
0~, & \text{otherwise} \end{cases}}~,
\end{equation}
meaning that we have provided a Euclidean path-integral construction of the $p$-inequivalent pUniverses charged under $\bZ_p^{(1)}$~.

\subsubsection*{Conclusions}

Let us summarize the results so far. BF theory is a two-dimensional topological theory that has two discrete symmetries: one 0-form symmetry $\mathbb{Z}_p^{(0)}$ and one 1-form symmetry $\mathbb{Z}_p^{(1)}$. These symmetries are spontaneously broken, independent of the spacetime geometry. Via the preceding calculations we provided a Hartle-Hawking construction of the various states charged under either the 0- or the 1-form symmetry given in \cref{sub:hilbert_space}, as well as how to compute overlaps between them, adapted for two-dimensional de Sitter space. 

In the next section we will spruce up this model to one that has actual dynamics. That is, we will consider an honest-to-goodness QFT with local degrees of freedom and non-trivial dimensionful quantities. What we will demand of this theory is that it shares the same 0- and 1-form symmetries as the BF theory, and we will provide a similar Euclidean path-integral construction on the $S^2$ of the various states charged under these symmetries. Since the dynamics can be analytically continued to Lorentzian de Sitter, this will demonstrate the existence of these various states, and hence the spontaneous breaking of the alleged symmetries, in de Sitter, for a fully-fledged quantum field theory. 

\section{Charge-\texorpdfstring{$p$}{p} Schwinger model on \texorpdfstring{$S^2$}{S2}} \label{sec:pSchwinger}

We will study the charge-$p$ Schwinger model (or \emph{$p$-Schwinger model} for short), a theory of a two-component, massless, charged, Dirac spinor $\Psi$ interacting with a compact $U(1)$ gauge field $A_\mu$ in two spacetime dimensions. On a general Euclidean curved background $g_{\mu \nu}$ the model is defined by the following action:
\begin{equation}\label{eq:Sschw}
    S_E^{\textnormal{Schwinger}} =   \int_{\mathcal{M}}  \textnormal{d}^2 x \,\sqrt{g}\left[\bar{\Psi} \gamma^\mu\left(\nabla_\mu  + i  p A_\mu \right) \Psi+\frac{1}{4q^2} F^{\mu \nu} F_{\mu\nu}\right] ~ , 
\end{equation}
where $F_{\mu\nu} = \partial_\mu A_\nu - \partial_\nu A_\mu$ is the $U(1)$ field strength and $q$ is the dimensionful gauge-coupling strength with units $[{\rm length}]^{-1}$. As before, we will restrict to $\cM = S^2$~.

This model differs from the standard Schwinger model in that we are allowing the spinor to carry charge $p \in \mathbb{Z}$, which will allow for richer physics than the $p = 1$ case analyzed for example in \citep{Schwinger:1962tp,Schwinger:1963yda,Lowenstein:1971fc,Jackiw:1984zi,Coleman:1975pw,Coleman:1976uz,Roskies:1980jh}. The $p=1$ model has been studied in curved backgrounds in \citep{Gass:1982iu,Oki:1984tpr,Barcelos-Neto:1986oku,Ferrari:1995uk,Jayewardena:1988td} and more recently in the context of de Sitter quantum field theories in \citep{Anninos:2024fty} whose spinor conventions we adopt.\footnote{For these conventions see appendix A of \citep{Anninos:2024fty}, including the definition of the spinor covariant derivative and Clifford matrices $\gamma^\mu$.}  We will label the highest element of the Clifford algebra, which anticommutes with all the $\gamma^\mu$, as $\gamma_*$.

Crucially, as we discuss below, allowing $p>1$ endows the theory with two global symmetries: a 0-form and a 1-form symmetry, as was the case in the BF theory. We will review this fact now.

\subsection{Symmetries and anomalies}

Before considering the above theory, let us start with a discussion of the theory of a free Dirac fermion. A free Dirac fermion has two separate global symmetries, generated, respectively, by the following conserved vector- and axial-$U(1)$ currents: 
\begin{equation}
j_{\rm V}^\mu\equiv\bar{\Psi}\gamma^\mu \Psi~, \qquad j^\mu_{\rm A}\equiv\bar{\Psi}\gamma^\mu\gamma_* \Psi~,\qquad \nabla_\mu\,j^\mu_{\rm V/A}=0~.\label{eq:fermicurrents}
\end{equation}
In the presence of a nonzero gauge coupling $q$, the vector-$U(1)$ symmetry is promoted to a gauge redundancy, whereby the action \eqref{eq:Sschw} is invariant under the following set of {local} transformations: 
\begin{align}
&\Psi(\mathbf{x})\rightarrow h(\mathbf{x})^p\Psi(\mathbf{x})~, &\bar{\Psi}(\mathbf{x})\rightarrow \bar{\Psi}(\mathbf{x})h(\mathbf{x})^{-p}~,\nonumber\\ & A_\mu\rightarrow A_\mu+i h(\mathbf{x})^{-1}\partial_\mu h(\mathbf{x})^{}~, &h(\mathbf{x})=e^{i\alpha(\mathbf{x})}\in U(1)\label{eq:u1rot}~,
\end{align}
conservation of the $U(1)$-vector current $\nabla_\mu\, j_{\rm V}^\mu=0$ survives this gauging. 

However, the axial-$U(1)$ symmetry famously suffers from a mixed anomaly with the $U(1)$ gauge redundancy, which stems from the fact that, under the \emph{local} rotation 
 \begin{equation}\label{eq:chiralrot}
\Psi(\mathbf{x})\rightarrow e^{i\beta(\mathbf{x})\gamma_*}\Psi(\mathbf{x})~, \qquad\qquad\bar{\Psi}(\mathbf{x})\rightarrow \bar{\Psi}(\mathbf{x})e^{i\beta(\mathbf{x})\gamma_*}~,
\end{equation}
the fermionic measure transforms as follows \citep{Fujikawa:1979ay, Roskies:1980jh,Jackiw:1984zi}: 
\begin{equation}\label{eq:fermiactionrot}
D\bar\Psi D\Psi\rightarrow  {D}\bar\Psi {D}\Psi e^{i\frac{ p}{2\pi}\int \dd^2x\sqrt{g}\, \left( \beta(x)\epsilon^{\mu\nu}F_{\mu\nu}- \frac{i}{p} \beta(x) \nabla^2 \beta(x) \right)}~.
\end{equation} 
Thus, taking account of the anomaly polynomial \eqref{eq:fermiactionrot}, under the rotation \eqref{eq:chiralrot}, the action transforms as 
\begin{align}
    \delta S_E^{\textnormal{Schwinger}} &=i\int \dd^2x\sqrt{g}\,\left\lbrace j_A^{{\mu}}\,\partial_\mu\beta(\mathbf{x})-\frac{ p}{2\pi}\left( \beta(\mathbf{x})\epsilon^{\mu\nu}F_{\mu\nu} - \frac{i}{p} \beta(\mathbf{x}) \nabla^2 \beta(\mathbf{x}) \right)\right\rbrace\label{eq:chiralanomalyaction1} \\
    &= \int_{S^2} \dd^2 x \sqrt{g} \,\beta(\mathbf{x})\left\lbrace i\left[- \nabla_\mu \, j_{\rm A}^{{\mu}}- \frac{ p}{2\pi}  \epsilon^{\mu \nu} F_{\mu \nu}   \right]  - \frac{1}{2\pi} \nabla^2 \beta(\mathbf{x}) \right\rbrace \, ,\label{eq:chiralanomalyaction2}
\end{align}
from which we deduce the failure of axial-current conservation:
\begin{equation}\label{eq:chiralanomequation}
\nabla_\mu \, j_{\rm A}^{{\mu}}=- \frac{ p}{2\pi}  \epsilon^{\mu \nu} F_{\mu \nu} ~.
\end{equation}
Although the Schwinger model fails to be invariant under axial rotations labeled by generic constant $\beta$, for the following particular choices of $\beta$:  
\begin{equation}\label{eq: Z2p values}
    \beta = \frac{2\pi n}{2p} \, , \qquad n = 0,\cdots,2p-1 \, ,
\end{equation}
the action transforms as
\begin{equation}\label{eq:pathintegralinvSchw}
S^{\textnormal{Schwinger}}_{ E}\rightarrow S^{\textnormal{Schwinger}}_{ E}+2\pi i k n~,
\end{equation}
in any instanton background labeled by $k\in \mathbb{Z}$, thanks to the Dirac quantization condition \eqref{eq: Dirac quant}. So as in \eqref{eq:pathintegralinvBF}, we seem to have just shown that the axial-$U(1)$ symmetry is broken down to a discrete $\mathbb{Z}^{(0)}_{2p}$ symmetry, under which the quantum theory remains invariant. 

This is almost the case, but notice that not all values listed in \eqref{eq: Z2p values} correspond to a proper axial symmetry transformation. In fact, for $n=p$, the symmetry transformations \eqref{eq:chiralrot} correspond to $\Psi\rightarrow -\Psi$ and $\bar\Psi\rightarrow -\bar\Psi$ which is just a constant gauge transformation by the phase $e^{\pm i\pi}$. In conclusion, only the values $n=0,\ldots , p-1$ lead to non-trivial chiral symmetry transformations.
In other words, the classical axial-$U(1)$ symmetry of the free fermion is broken to a $\mathbb{Z}^{(0)}_p$ global symmetry in the $p$-Schwinger model as a consequence of the Adler-Bell-Jackiw (ABJ) anomaly \citep{Komargodski:2020mxz,Witten:1978bc}.

\begin{figure}
    \centering
    \includegraphics[width=0.75\linewidth]{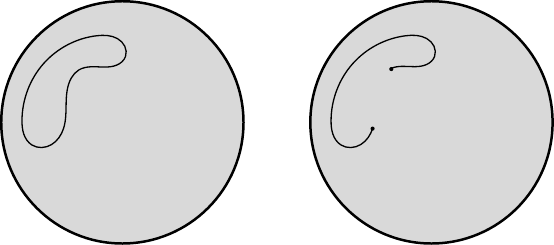}
    \put(-100,95){$\Psi$}
    \put(-105,65){$\bar{\Psi}$}
    \put(-70,110){$\hat{W}_1[\mathcal{C}]$}
    \put(-285,75){$\hat{W}_1[\mathcal{C}]$}
    \caption{Screening of the electric field created by an $j=1$ Wilson line in the $p=1$-Schwinger model.}
    \label{fig:WLBreaking}
\end{figure}
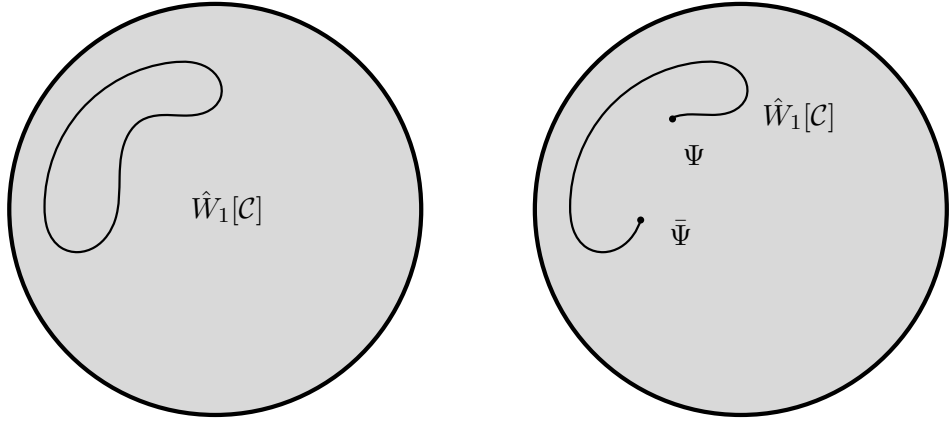

Another remarkable consequence of dynamical electrons having charge $p>1$, is that the electric field flux, which is naturally quantized in units of $q^2$, cannot be completely screened by matter fields. More precisely, consider a gauge invariant source of electric field, i.e. a Wilson line:
\be\label{eq: Wilson line}
\hat{W}_j[\mathcal{C}]=e^{i {j}\oint_{\mathcal{C}} \dd x^\mu A_\mu}
\ee
tracing out a closed curve $\mathcal{C}$ and $j\in \mathbb{Z}$ by compactness of the gauge group. Acting on the vacuum, this line creates an electric field flux tube. Now, in the presence of unit-charged (i.e. $p=1$) matter fields, the electric field can be screened by the creation of electron/positron pairs out of the vacuum. In terms of the Wilson line operator \eqref{eq: Wilson line}, this implies that one may cut open the curve $\mathcal{C}$ by inserting appropriate powers of the matter fields on the endpoints, as in Figure \ref{fig:WLBreaking}. 

But in the charge-$p$ Schwinger model \eqref{eq:Sschw} this is only possible for $j=0$ mod $p$, since single electrons carry too much charge to screen the electric field. Consequently, electric sources of the form \eqref{eq: Wilson line} with $j\neq 0$ mod $p$ are protected, and hence carry a conserved global charge. Due to the fact that this global charge is carried by non-local, one-dimensional operators, this is a 1-form symmetry. Finally, due to the fact that the conservation is modulo $p$, we conclude that the theory possess a $\mathbb{Z}_p^{(1)}$ 1-form global symmetry, acting on the Wilson lines as
\be\label{eq: 1-form action}
\mathbb{Z}_p^{(1)} \, : \,\,\, \hat{W}_j[\mathcal{C}] \, \to \,  e^{\frac{2\pi ij}{p}} \hat{W}_j[\mathcal{C}]
\ee

As it turns out, as a consequence of the chiral anomaly \eqref{eq:fermiactionrot}, the $\mathbb{Z}_p^{(0)}$ chiral symmetry and the $\mathbb{Z}_p^{(1)}$ 1-form symmetry participate in a $\mathbb{Z}_p$-valued mixed 't Hooft anomaly. This anomaly is of the same type as the one described by \eqref{BF:ClockAlgebra} for the BF theory in the previous section. 

From the  perspective of \citep{Gaiotto:2014kfa}, the global symmetries described above imply the existence of a discrete set of topological operators. The $\mathbb{Z}_p^{(0)}$ chiral symmetry will be implemented by topological-line operators $\hat L_n[\cC]$, $n\in 0,1,\dots p-1$, extended over a closed curve $\cC$, while the $\mathbb{Z}_p^{(1)}$ 1-form symmetry will be generated by topological-local operators that we denote by $\hat U_m(\mathbf{x})$, $m \in 0,1,\dots,p-1$, in analogy with the ones already introduced for the BF theory. Although their existence is guaranteed by the presence of these global symmetries, the goal of the remainder of the paper will be to provide concrete expressions for these operators in terms of the fundamental fields, and to compute concrete quantities with them via the path integral.

In the next section we will show that it is possible to construct the corresponding topological operators in the $p$-Schwinger model on $S^2$. Furthermore, as we did in the BF theory, we will show how to use these topological operators to explicitly construct a set of $p$-inequivalent Hartle-Hawking states (vacua and pUniverses) for the \emph{full, interacting model} without resorting to the explicit putative IR structure of the theory. 

Why are we doing this? If we were to consider the $p$-Schwinger model in flat space, by the arguments above, we know the theory must flow to the BF TQFT in the IR, whose vacuum structure we reviewed in \cref{sub:hilbert_space} \citep{Komargodski:2020mxz}. Sadly, we cannot apply this logic when placing the theory on the $S^2$. This can be easily understood upon analytic continuation, where one is led to consider the model on a fixed $2$-dimensional de Sitter space. Famously dS does not have any globally-defined time-like Killing vector and thus there is no global notion of energy. The subtlety is then clear, with no notion of conserved energy how does one consistently integrate out fields and access the IR?

\subsection{Bosonization} \label{sec: boson Schwinger}

We will establish the existence of the topological operators $\hat{U}_n(\mathbf{x})$ and $\hat{L}_m[\cC]$ in the $p$-Schwinger model via Abelian bosonization \citep{Coleman:1974bu,Mandelstam:1975hb}. Let us review this procedure now (see \citep{Frishman:1992mr,Ji:2019ugf} for a more modern treatment). In its simplest presentation, bosonization starts by writing the vector- and axial-currents \eqref{eq:fermicurrents} in terms of a single compact scalar $B\cong B+2\pi$:\footnote{These expressions are compatible with the identity $\gamma_\mu\gamma_*=-i\epsilon_{\mu\nu}\gamma^\nu$~.} 
\begin{equation}\label{jB}
j_{\rm V}^\mu\equiv\bar{\Psi}\gamma^\mu \Psi\rightarrow-\frac{1}{2\pi}\epsilon^{\mu\nu}\partial_\nu B~, \qquad j^\mu_{\rm A}\equiv\bar{\Psi}\gamma^\mu\gamma_* \Psi\rightarrow -\frac{i}{2\pi}\partial^\mu B~,\qquad \bar{\Psi} \gamma^\mu\nabla_\mu  \Psi\rightarrow\frac{1}{8\pi}\partial_\mu \,B\partial^\mu B~.
\end{equation}
According to this replacement rule, the vector current becomes trivially conserved as a consequence of the antisymmetry of the $\epsilon$-tensor. 
The bosonized action of the $p$-Schwinger model follows from these identifications:
\begin{equation}\label{eq: boson action}
    S_E^{\rm BS} = \int_{S^2} \dd^2 x \sqrt{g} \left[  \frac{1}{8\pi} \partial_\mu B\,\partial^\mu B+\frac{1}{4 q^2} F_{\mu \nu}F^{\mu \nu}  - \frac{ip}{4\pi} B\,\epsilon^{\mu \nu} F_{\mu \nu} \right] \, ,
\end{equation}
from which we read off the equations of motion: 
\begin{equation}\label{eq:eomsbosonaction}
\delta B: -\nabla^2 B-ip \,\epsilon^{\mu \nu} F_{\mu \nu}=0~,\qquad \delta A_\nu:  \nabla_\mu\left(\frac{p}{2\pi}\epsilon^{\mu\nu}B+\frac{i}{q^2}F^{\mu\nu}\right)=0~.
\end{equation}
The first of these can be massaged into an expression for the anomalous conservation equation for the axial current, c.f. \eqref{eq:chiralanomequation}: 
\begin{equation}
\nabla_\mu \, j_{\rm A}^{{\mu}}=- \frac{ p}{2\pi}  \epsilon^{\mu \nu} F_{\mu \nu} ~.
\end{equation}

In this language, and since $B$ is identified modulo $2\pi$, the axial current $j_{\rm A}^\mu$ generates a $U(1)$ shift of the scalar field $B(\mathbf{x})$ along its target space, a symmetry which is explicitly broken by the $B$-field's axion-like coupling to the gauge field $A_\mu$. The symmetry-breaking term should immediately be recognized as the Euclidean BF action \eqref{BF:Action}. Indeed, the bosonized $p$-Schwinger model simply supplements the topological BF term with a kinetic term for both $B$ and the gauge field $A_\mu$. 

As reviewed around \eqref{eq:zpbf}, the $U(1)$-shift symmetry isn't completely broken since discrete translations of the form
\begin{equation}\label{eq:zpbosschwing}
B(\mathbf{x})\rightarrow B(\mathbf{x})+\frac{2\pi n}{p}~,
\end{equation} 
leave the theory invariant by Dirac quantization \eqref{eq: Dirac quant}. Thus this bosonic theory has a $\mathbb{Z}_p^{(0)}$ symmetry which acts by discrete shifts of $B$, as in both the BF theory and the fermionic Schwinger action \eqref{eq:Sschw}. And as we reviewed at the end of the last section, this $\bZ_p^{(0)}$ symmetry participates in a $\bZ_p$-valued mixed anomaly with the $\bZ_p^{(1)}$ 1-form symmetry under which the Wilson lines are charged.

Shortly we will show that the theory \eqref{eq: boson action} on the $S^2$ exhibits spontaneous symmetry breaking of both the $\bZ_p^{(0)}$ and the $\bZ_p^{(1)}$ symmetries. We will do this by providing an exact path integral construction of the vacua and pUniverses of this theory, distinguished by their charge under the chiral symmetry $\bZ^{(0)}_p$ or $\bZ^{(1)}_p$, just as we did for the BF theory. Moreover, we will show that all these inequivalent states are explicitly de Sitter-invariant and Hadamard, namely that correlation functions computed in these states will display (only) the standard coincident point singularities prescribed as in flat space.

\subsubsection{Topological operators}

To get a sense of what the topological operators of the bosonized $p$-Schwinger model may be, let us massage the equations of motion  \eqref{eq:eomsbosonaction}. Using the Levi-Civita identities, we arrive at an equivalent representation of the equations of motion derived from \eqref{eq: boson action}: 
\begin{equation}\label{eq:eomsbosonaction2}
\delta B: \epsilon^{\sigma\mu}\nabla_\sigma\left(A_\mu+\frac{i}{2p}\epsilon_{\mu\nu}\partial^\nu B\right)=0~,\qquad \delta A_\nu:  \epsilon^{\nu\mu}\nabla_\mu\left(B+\frac{i\pi}{pq^2}\epsilon^{\rho\sigma}F_{\rho\sigma}\right)=0~.
\end{equation}
These expressions have a simple geometric interpretation. Using the Hodge star operation $\star$, we identify them as $\star\, \dd \star J^{(1)}=0=\star\, \dd \star J^{(2)}$ for a pair of (respectively 1- and 2-form) currents:
\begin{align}
\star J^{(1)}&= \left(A_{\mu}+ \frac{i }{2p}\epsilon_{\mu\nu}\partial^\nu B\right)\dd x^\mu~,\\
\star J^{(2)}&= B+\frac{\pi i}{pq^2}\epsilon^{\mu\nu}F_{\mu\nu}~.\label{eq: J2}
\end{align}

Note that neither of the above currents is globally well-defined, and moreover $\star J^{(1)}$ is also not gauge-invariant, consistent with the absence of continuous global symmetries in this model. Albeit true, we can nevertheless use these currents to construct the  the following globally well-defined and gauge-invariant operators: 
\be\label{eq: bare top ops}
\hat U^{\rm bare}_n(\mathbf{x})= e^{in \star J^{(2)}(\mathbf{x})} \quad , \quad \hat L_m^{\rm bare}[\cC]=e^{i m \oint_\cC \star J^{(1)}} \, , \qquad n,m \in \mathbb{Z}
\ee
with $\partial\cC=\emptyset$. Comparing with \eqref{eq: BF top ops} we conclude that these are the natural generalizations of the topological operators of the BF theory. And because the currents contain explicit factors of $i$, we should specify that, like in BF theory: 
\be\label{eq: daggerdefs}
\hat U^{\dagger{\rm bare}}_n(\mathbf{x})\equiv \hat U^{\rm bare}_{-n}(\mathbf{x})\quad , \quad \hat L_m^{\dagger{\rm bare}}[\cC]\equiv L_{-m}^{\rm bare}[\cC]=L_{m}^{\rm bare}[-\cC]~.\ee

These operators are topological because they are constructed out of conserved currents. Therefore, it is seemingly reasonable to identify them as the generators of the $\bZ_p^{(0)}$ and $\bZ_p^{(1)}$ global symmetry generators. However, even if this identification is conceptually correct, it is unfortunately not sensible because the exponentials of a local- or line-operator in QFT generally suffer from ultraviolet divergences, rendering them ill-defined without specifying a regularization procedure. This is a general fact about QFTs which we surprisingly avoided in the exceptional case of the BF theory. 

\subsubsection*{Regularization of $\hat{U}_n^{\rm bare}(\mathbf{x})$}
Let us focus first on the topological local operators $U_n^{\rm bare}(\mathbf{x})$. We will adopt a regularization scheme whereby we smear the operators prior to exponentiation over a small neighborhood around the insertion point $\mathbf{x}$. To be more precise, let us define a sequence of non-negative real-valued functions $f^\delta_{\mathbf{x}}(\mathbf{w})$ centered around the point $\mathbf{x}$ on $S^2$ labeled by a small parameter $\delta$. We will demand that the functions $f^\delta_{\mathbf{x}}(\mathbf{w})$ have the following properties: 
\begin{equation}\label{eq: f def}
        \int_{S^2} \dd^2 w\sqrt{g} f^\delta_{\mathbf{x}}(\mathbf{w})=1~,\qquad\qquad\int_{S^2} \dd^2 w\sqrt{g} \left(f^\delta_{\mathbf{x}}(\mathbf{w})\right)^2 <\infty\, ,
\end{equation}
for $\delta>0$. Lastly,  we will demand: 
\begin{equation}\label{eq:fporperties}
        \lim_{\delta \to 0} f^\delta_{\mathbf{x}}(\mathbf{w}) =\frac{\delta(\mathbf{w}-\mathbf{x})}{\sqrt{g}} ~.
\end{equation} 
Hence, the function ${f^\delta_\mathbf{x}(\mathbf{w})}$ may be understood as a fattening of the $\frac{\delta(\mathbf{w}-\mathbf{x})}{\sqrt{g}}$ distribution, with a finite width parametrized by $\delta$. 
As we present in Appendix \ref{app: background}, the precise form of the required counterterms is dictated by the proper definition of the global symmetry in terms of background gauge transformations \citep{Gaiotto:2014kfa}. 
Taking this into account, we propose the following regulated topological operator:
\begin{equation}
    \hat{U}^\delta_n(\mathbf{x}) \equiv  \, \mathcal{N}^\delta_n \,\, e^{ in \int_{S^2}\dd^2 z \sqrt{g} \,f_\mathbf{x}^\delta(\mathbf{z})\, *J^{(2)}(\mathbf{z})}  \, ,
    \label{eq: renorm local top}
\end{equation}
with normalization constant:
\begin{equation}
    \mathcal{N}^\delta_n \equiv \exp\left[{- \frac{2\pi^2 n^2}{p^2q^2} \int_{S^2} \dd^2 w\sqrt{g} \left(f^\delta_{\mathbf{x}}(\mathbf{w})\right)^2 }\right] \, .
    \label{RenormConst1}
\end{equation} As we will see, this choice of $\cN_n^\delta$ will ensure that the correlators involving products of $\hat{U}_n^\delta(\mathbf{x})$ will remain finite in the $\delta\rightarrow 0$ limit.

\subsubsection*{Regularization of $\hat{L}_m^{\rm bare}[\cC]$}

To regularize $\hat{L}_m^{\rm bare}[\cC]$, we will take a similar approach to the local operator. For this a picture may be useful. Consider the line operator along a curve $\cC$, which subtends a region $\cD$ denoted in grey in the following figure: 
\begin{equation*}
\begin{gathered}\includegraphics[height=3cm]{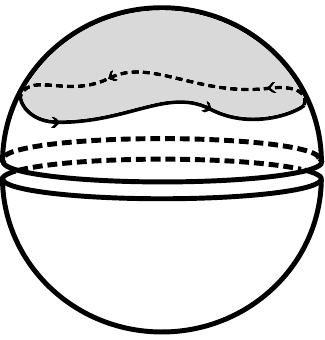}\put(-5,67){\footnotesize $\hat{L}_m^{\rm bare}[\cC]$}\end{gathered}\qquad\quad.
\end{equation*}
Using Stokes' theorem and \eqref{eq:fieldstrengthinstanton}, the line integral that appears inside the exponential in $\hat{L}_m^{\rm bare}[\cC]$ (in a $k$-instanton background) is: 
\begin{align}
    \oint_\cC \star J^{(1)}&=\oint_\cC \left(A^{(k)}_\mu + \frac{i }{2p}\epsilon_{\mu\nu}\partial^\nu B\right)\dd x^\mu = \int_{\mathcal{D}} \dd^2 x \sqrt{g} \left[\frac{1}{2}\epsilon^{\mu\nu}F_{\mu\nu}^{(k)}- \frac{i}{2p}\nabla^2 B(\mathbf{x})\right] \nonumber\\
    &= - \int_{S^2} \dd^2 x \sqrt{g}\,\Theta_{\cD}(\mathbf{x})\left[\frac{k}{2\ell^2}+ \nabla^2 \Phi(\mathbf{x})+\frac{i}{2p} \nabla^2 B(\mathbf{x}) \right]\, ,
\end{align}
where $\Theta_\cD$ is defined in \eqref{eq:Heavisidedef}. The issue that will arise when computing correlators of $\hat{L}_m^{\rm ba{}re}[\cC]$ is that the Heaviside function \eqref{eq:Heavisidedef} varies too sharply along the boundary $\cC$ of $\cD$. 

To remedy this, let us introduce a sequence of smooth functions $\cT_{\cD}^\delta(\mathbf{x})$ with the following properties: 
\begin{equation}\label{eq:Tproperties}
\int_{S^2} \dd^2 x \sqrt{g}\,\cT_{\cD}^\delta(\mathbf{x})=\cA_\cD~,\qquad\qquad \lim_{\delta\rightarrow 0}\cT_{\cD}^\delta(\mathbf{x})=\Theta_{\cD}(\mathbf{x})~,
\end{equation}
where $\cA_\cD$ is the area of the region $\cD$. We envision these functions $\cT^\delta_\cD$ to be almost constant over the bulk of $\cD$, varying smoothly to zero at the edge of $\cD$ in a width set by the small parameter $\delta$.

With these definitions in hand, we will define our renormalized line operators (in the gauge \eqref{Ak}) as follows: 
\begin{equation}\label{eq:Ldeltadef}
\hat{L}_m^\delta[\cC]=\cM_m^\delta \,e^{-\frac{{i m}k}{2\ell^2} \mathcal{A}_\mathcal{D} - {i m}\int_{S^2} \dd^2 x \sqrt{g}\,\cT_{\cD}^\delta(\mathbf{x})\left[ \nabla^2 \Phi(\mathbf{x})+\frac{i}{2p} \nabla^2 B(\mathbf{x}) \right]}~,
\end{equation}
with normalization 
\begin{equation}\label{RenormConst2}
\cM_m^\delta\equiv \exp\left[-\frac{\pi m^2}{2p^2}\int_{S^2} \dd^2 x \sqrt{g}\,\left(\partial_\mu\cT_{\cD}^\delta\right)\,\left(\partial^\mu\,\cT_{\cD}^\delta\right)\right]~.
\end{equation}
The exponent only receives  contributions  from the transition region near the edge of $\cD$ of width $\delta$.

\subsubsection{Local operators}

Now that we have expressions for the (regularized) topological operators in the bosonized language, we will discuss the various local operators available to us. Since the $p$-Schwinger model has non-trivial dynamics, these local operators will be sensitive to the background de Sitter spacetime and will exhibit interesting dynamics. 

\subsubsection*{The electric field}

The simplest local operator in this theory is the electric field. In two dimensions, the electric field is a scalar: 
\begin{equation}\label{eq:electricfielddef}
E(\mathbf{x})\equiv \frac{1}{2}\epsilon^{\mu\nu}F_{\mu\nu}(\mathbf{x})~.
\end{equation}
Given our gauge choice and using \eqref{eq:fieldstrengthinstanton}, the electric field in a $k$-instanton background will take the form:
\begin{equation}\label{eq:electricfieldkinst}
E^{(k)}(\mathbf{x})=-\left(\frac{k}{2\ell^2}+\nabla_{\mathbf{x}}^2\Phi\right)~.
\end{equation} 
We compute the two-point function of the electric field in the various vacua and pUniverses in \cref{ssub:electric_field_two_point_function}.

\subsubsection*{The meson vertex operators}

Naturally, the bosonized Schwinger model \eqref{eq: boson action} also admits local operators built out of $B(\mathbf{x})$. Technically, since $B(\mathbf{x})$ is a compact scalar, the only globally well-defined local operators built out of it are the vertex operators, as in \eqref{eq: BF top ops}:
\begin{equation}
\hat{V}_n^{\rm bare}(\mathbf{x})\equiv e^{in B(\mathbf{x})}~,\qquad n \in \bZ~,
\end{equation}
where we have included the superscript `bare' to indicate that this operator is ill-defined without a regularization scheme. Our scheme will parallel the regularization of the topological local operator $\hat{U}_n^{\rm bare}$ in \eqref{eq: renorm local top} with some minor differences which we will come to shortly. The question we want to address first is: what operator does this vertex operator correspond to in the fermionic picture? 

To answer this question it suffices to consider how $\hat{V}_n^{\rm bare}$ transforms under the unbroken axial  $\bZ_p^{(0)}$-global shift symmetry:
\begin{equation}\label{eq:Vcharge}
B(\mathbf{x})\rightarrow B(\mathbf{x})+\frac{2\pi}{p}~,\qquad \hat{V}_n^{\rm bare}(\mathbf{x})\rightarrow e^{\frac{2\pi in}{p}}\hat{V}_n^{\rm bare}(\mathbf{x})~.
\end{equation}
Our task is straightforward: we simply need to find the fermionic operators that carry the same axial $\bZ_p^{(0)}$ charges as the $\hat{V}_n^{\rm bare}$ above. We define a pair of projectors: 
\begin{equation}
P_L\equiv\frac{1+\gamma_*}{2}=\begin{pmatrix} 1 &\quad0\\0 &\quad0\end{pmatrix}~, \qquad \qquad P_R\equiv\frac{1-\gamma_*}{2}=\begin{pmatrix} 0 &\quad0\\0 &\quad1\end{pmatrix}~, 
\end{equation}
such that 
\begin{equation}
\bar{\Psi}\Psi= \bar{\Psi}P_L \Psi+\bar{\Psi}P_R\Psi~.
\end{equation}
The shift $B\rightarrow B+ \frac{2\pi}{p}$ in \eqref{eq:Vcharge} corresponds to an axial rotation of the fermions of the following form: 
\begin{equation}
\Psi(\mathbf{x})\rightarrow e^{\frac{i\pi}{p}\gamma_*}\Psi(\mathbf{x})~, \qquad\qquad\bar{\Psi}(\mathbf{x})\rightarrow \bar{\Psi}(\mathbf{x})e^{\frac{i\pi}{p}\gamma_*}~,
\end{equation}
under which
\begin{equation}\label{eq:Zpchargebilinear}
\bar{\Psi}P_L \Psi\rightarrow e^{\frac{2\pi i}{p}}\bar{\Psi}P_L \Psi~,\qquad\qquad\bar{\Psi}P_R\Psi\rightarrow e^{-\frac{2\pi i}{p}}\bar{\Psi}P_R \Psi~,
\end{equation}
which suggests the identifications:\footnote{The factors of $q$ are required by dimensional analysis. Locality dictates that we choose the UV scale $q$ over the IR scale $\tfrac{1}{\ell}$ for these purposes.} 
\begin{equation}
\left(\bar{\Psi}P_L \Psi\right)(\mathbf{x})\leftrightarrow q\,\hat{V}_1^{\rm bare}(\mathbf{x})~,\qquad\qquad\left(\bar{\Psi}P_R\Psi\right)(\mathbf{x})\leftrightarrow q\,\hat{V}_{-1}^{\rm bare}(\mathbf{x})~,
\label{BosMapBilinear}
\end{equation}
hence we see that the charge-one vertex operators capture the behavior of composite fermion bilinears, i.e. the mesons of the theory. Of course, we are dealing with composite operators on both sides of the identification above, so this map is schematic and only holds up to the cancellation of local divergences. We relegate the discussion of the specific regularization of $\hat{V}_n^{\rm bare}(\mathbf{x})$ to \cref{ssub:the_chiral_condensate}.

\subsection{Vacua and pUniverses in the \texorpdfstring{$p$}{p}-Schwinger model}\label{sub:vacua_and_puniverses_Schwinger}

\subsubsection*{First calculation: The Hartle-Hawking norm}
Just as we did in the topological BF theory, we will start with a calculation of the Hartle-Hawking norm of the bosonized $p$-Schwinger model, i.e. the bare path integral: 
\begin{equation}\label{eq:BDoverlapZdefSchwinger}
{\braket{0|0}}\propto \begin{gathered}\includegraphics[height=2.3cm]{Images/Overlaps.pdf}\end{gathered}= \int \frac{DB DA_\mu}{\textnormal{vol}\,\mathcal{G}} \,   e^{-S^{\textnormal{BS} }_E}\equiv \cZ_{\rm BS}~,
\end{equation}
with $S_E^{\rm BS}$ defined in \eqref{eq: boson action}. We will continue to decompose our gauge field in each $k$-instanton sector as (see equation \eqref{Ak}):
\begin{equation} 
A^{(k)}_\mu(\mathbf{x}) = k C_\mu(\mathbf{x}) +\epsilon_{\mu\nu} \partial^\nu\Phi(\mathbf{x}) + {i} h(\mathbf{x})^{-1} \partial_\mu h(\mathbf{x})~,
\end{equation}
where recall $C_\mu$ represents the single instanton background \eqref{eq:instantongaugefield}. The field strength for such a field configuration is: 
\be
F^{(k)}_{\mu\nu}(\mathbf{x})=-\epsilon_{\mu\nu}\left(\nabla^2_{\mathbf{x}}\Phi(\mathbf{x}) + \frac{k}{2\ell^2} \right) \, ,
\ee
as in \eqref{eq:fieldstrengthinstanton}.

 Taking all of this into account, the action of the theory \eqref{eq: boson action} in a $k$-instanton sector reduces to 
\begin{equation}
    S_E^{(k){\rm BS}} = \frac{\pi k^2}{2 q^2 \ell^2} + ip\, k\,b + \int_{S^2} \dd^2 x \sqrt{g} \left[ \frac{1}{8\pi} \partial_\mu B'\,\partial^\mu B' +\frac{1}{2q^2} \left( \nabla^2 \Phi\right)^2 + \frac{i p}{2\pi } B'\,\nabla^2 \Phi  \right] \, ,
    \label{eq:actionqschwinger}
\end{equation} 
where the first term is the on-shell action for the $k$-instanton background and we have preemptively split the field $B=b+B'(\mathbf{x})$ into its zero- and nonzero-modes. The last term will appear many times in the equations below, so we will give it a name:
\begin{equation}
S_E^{'\rm{BS}}[B',\Phi]\equiv \int_{S^2} \dd^2 x \sqrt{g} \left[ \frac{1}{8\pi} \partial_\mu B'\,\partial^\mu B' +\frac{1}{2q^2} \left( \nabla^2 \Phi\right)^2 + \frac{i p}{2\pi } B'\,\nabla^2 \Phi  \right] \, .
\end{equation}

Since the action \eqref{eq:actionqschwinger} is quadratic in $B'$ and $\Phi$, the Gaussian approximation around the saddle is exact. Integrating by parts and completing the square, we can write the above action as: 
\begin{multline}
    S_E^{(k){\rm BS}} = \frac{\pi k^2}{2 q^2 \ell^2} + ip\, k\,b \\+ \int_{S^2} \dd^2 x \sqrt{g} \left[ \frac{1}{8\pi}  B'\left(-\nabla^2+\frac{p^2q^2}{\pi}\right) B' +\frac{1}{2q^2} \left( -\nabla^2 \Phi-\frac{i p q^2}{2\pi} B'\right)^2  \right] \, .\label{eq:completesqure}
\end{multline} 

The equations of motion \eqref{eq:eomsbosonaction}, written in this parametrization amount to: 
\begin{equation}\label{eq:eombosonaction3}
 \nabla^2B'=2 i p\,\nabla^2 \Phi~,\qquad\qquad -\frac{1}{q^2}\nabla^2\left[-\nabla^2+\frac{p^2 q^2}{\pi}\right]\Phi=0 ~.
\end{equation}
And hence the full path integral is computed by the quadratic fluctuations around solutions to \eqref{eq:eombosonaction3}, which reduces to:
\begin{multline}
    \cZ_{\rm BS} = \sum_{k=-\infty}^\infty e^{-\frac{\pi k^2}{2q^2 \ell^2} } \frac{\ell}{\ell_{\rm UV}}\int_0^{2\pi}\dd b\, e^{-ip\,k\,b} \\
    \int \frac{D\delta B' D\delta\Phi Dh}{\textnormal{vol} \, \mathcal{G}} J_{\Phi,h} e^{-\int_{S^2} \dd^2 x \sqrt{g} \left[ \frac{1}{8\pi}  \delta B'\left(-\nabla^2\right)\delta B' +\frac{1}{2q^2}\delta\Phi\left(-\nabla^2\right) \left(-\nabla^2+\frac{p^2q^2}{\pi}\right)\delta\Phi   \right]}~.
\end{multline}
As before, the integral over the zero-mode $b$ collapses the instanton sum to the $k=0$ term.  Using the regularizations outlined in Appendix \ref{app:bosonregularization} as well as \eqref{eq:volgmeasure} we land on: 
\begin{equation}
\cZ^\epsilon_{\rm BS}=\left[\frac{\det_\epsilon'\left(-\nabla^2\right)}{\det_\epsilon'\left(-\frac{1}{q^2}\nabla^2\left(-\nabla^2+\frac{p^2q^2}{\pi}\right)\right)}\right]^{1/2}~.
\end{equation}
Using the identity $\det(AB)=\det(A)\det(B)$ reserved for finite operators,
this expression can be simplified to: 
\begin{equation}
\cZ^\epsilon_{\rm BS}=\left[\frac{{\det}'_\epsilon\left(q^2\right)}{{\det}'_\epsilon\left(-\nabla^2+\frac{p^2q^2}{\pi}\right)}\right]^{1/2}~.
\end{equation}
Finally, given our normalizations in \eqref{eq:BFmeasure}, we have that: 
\begin{equation}
\left[\det\left(q^2\right)\right]^{-1/2}=\left[\ell\Lambda_{\rm UV}\int_{-\infty}^{\infty}\dd c_0 e^{-\frac{q^2}{2}\left[\int_{S^2}\dd^2x \sqrt{g}\right]c_0^2}\right]\left[{\det}'\left(q^2\right)\right]^{-1/2}~,
\end{equation}
where we have pulled out the integral over the zero-mode from the determinant in the left hand side. Computing this integral, we find: 
\begin{equation}
\left[\det\left(q^2\right)\right]^{-1/2}=\frac{\Lambda_{\rm UV}}{\sqrt{2} q}\left[{\det}'\left(q^2\right)\right]^{-1/2}\implies \left[{\det}'\left(q^2\right)\right]^{1/2}=\frac{\Lambda_{\rm UV}}{\sqrt{2} q}\left[\det\left(q^2\right)\right]^{1/2}~.
\end{equation}
Putting these two computations together we arive at: 
\begin{equation}\label{ZBS}
\cZ^\epsilon_{\rm BS}=\frac{1}{q\ell_{\rm UV}\sqrt{2} }\left[\frac{{\det}_\epsilon\left(q^2\right)}{{\det}'_\epsilon\left(-\nabla^2+\frac{p^2q^2}{\pi}\right)}\right]^{1/2}~.
\end{equation}
The final step is to note that the quantity $\left[\det\left(q^2\right)\right]^{1/2}$ can be removed by adding a cosmological constant counterterm to the theory, so we are allowed to ignore this local divergence. The final answer is: 
\begin{equation}
\cZ^{\epsilon,{\rm renorm.}}_{\rm BS}=\frac{1}{q\ell_{\rm UV}\sqrt{2} }\left[{{\det}'_\epsilon\left(-\nabla^2+\frac{p^2q^2}{\pi}\right)}\right]^{-1/2}~.
\end{equation}
The $\det'$ of interest in the above equation is computed using heat kernel regularization in \cref{app:bosonregularizationExamples} (see equation \eqref{app:detprimemassive}) yielding:\footnote{The ultraviolet cutoff scale $\Lambda_{\rm UV}$ is related to the quantity $\epsilon$ in equation \eqref{eq:cutoffboson}.}
\begin{align}\label{Zboson}
\log\cZ^{\epsilon,{\rm renorm.}}_{\rm BS}=&\frac{2\ell^2}{\epsilon^2}+\left(\frac{1}{3}-\frac{p^2q^2\ell^2}{\pi}\right)\log(\ell\Lambda_{\rm UV})-\log\left(q\ell\sqrt{2}\right)+\left(\Delta-\frac{1}{2}\right)\log\frac{\Gamma(1+\bar{\Delta})}{\Gamma(1+\Delta)}\nonumber\\&+\left( \psi^{(-2)} (1+\Delta ) + \psi^{(-2)} (1+\bar{\Delta} ) \right)\nonumber\\&-\left(\psi^{(-2)}(1) + \psi^{(-2)}(2) +\frac{1}{4}-2\zeta'(-1)\right)+\cO(\epsilon)~,
\end{align}
where $\Delta$ is a solution to 
\begin{equation}\label{eq:deltadef}
\Delta(\Delta-1)=-\frac{p^2q^2\ell^2}{\pi}~.
\end{equation}
The finite part and the log diveregence of the above free energy for the bosonized $p$-Schwinger model on the sphere, computed in this scheme, matches precisely with the one computed for the $p=1$ Schwinger model in the original fermionic variables in \citep{Anninos:2024fty} using a similar scheme. The $\sim \tfrac{1}{\epsilon^{2}}$ divergence can be absorbed into the cosmological constant local counterterm, while the $\sim \log\epsilon$ logarithmic divergence into the Gauss-Bonnet local counterterm. Together, these constitute the entire scheme dependence of (\ref{Zboson}). 

\subsubsection*{Second calculation: The \texorpdfstring{$\mathbb{Z}_p^{(0)}$}{Zp0} `vacuum-to-vacuum' overlaps}
Now that we have an expression for $\mathcal{Z}_{\rm BS}$, we can move on to computing overlaps between states of different 0-form charge, as we did in BF theory in \eqref{eq:Zp0BFsetup}:
\begin{equation}
{\braket{m|n}}= \lim_{\delta\rightarrow0}\quad\begin{gathered}\includegraphics[height=2.3cm]{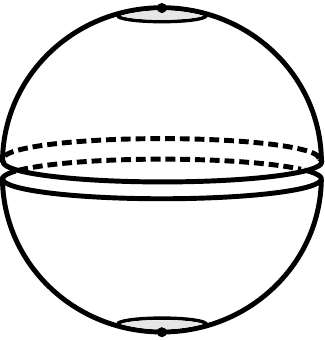}\put(-40,72){{\footnotesize $\hat{U}^\delta_n$}($\mathbf{x}$)}\put(-40,-12){{\footnotesize $\hat{U}_m^{\delta\dagger}$}($\mathbf{y}$)}\end{gathered}= \lim_{\delta\rightarrow0}\frac{1}{\cZ_{\rm BS}}\int \frac{DB DA_\mu}{\textnormal{vol}\mathcal{G}} \,\hat{U}_m^{\delta\dagger}(\mathbf{y})\hat{U}_n^\delta(\mathbf{x})   e^{-S^{\textnormal{BS} }_E}~.
\end{equation}
To perform this calculation, we insert the smeared operators \eqref{eq: renorm local top}, only taking the limit $\delta\rightarrow0$ at the end of the computation.
By combining \eqref{eq: J2} with \eqref{eq:fieldstrengthinstanton}, we find the following representation for the conserved current in a $k$-instanton background:
\begin{equation}
\star J^{(2)}= B-\frac{2\pi i}{pq^2}\left(\frac{k}{2\ell^2}+\nabla^2\Phi\right)~,
\end{equation}
and hence the smeared topological local operator can be expressed as follows:
\begin{equation}\label{eq:Undeltaexplicity}
\hat{U}_n^\delta(\mathbf{x}) =\mathcal{N}_n^\delta e^{\frac{\pi nk}{pq^2\ell^2}+inb+in\int_{S^2}\dd^2 x'\sqrt{g}f_\mathbf{x}^\delta(\mathbf{x'})\left(B'(\mathbf{x}')-\frac{2\pi i}{pq^2}\nabla^2\Phi(\mathbf{x'})\right)}~.
\end{equation}
To compute the state-overlap of interest, we must evaluate: 
\begin{align}\label{eq:vacuaoverlap1}
{\braket{m|n}}&=\lim_{\delta\rightarrow0}\frac{\mathcal{N}^\delta_n\mathcal{N}^\delta_m}{\cZ_{\rm BS}}\sum_{k=-\infty}^{\infty}e^{-\frac{\pi k}{pq^2 \ell^2}\left(\frac{kp}{2}+m-n\right) }\frac{\ell}{\ell_{\rm UV}}\int_{0}^ {2\pi}\dd b \, e^{-ib\left(p\, k+m-n\right)}\int \frac{ DB' D\Phi Dh}{\textnormal{vol}\,\mathcal{G}}J_{\Phi,h} \,  \nonumber\\&e^{-S^{'{\rm BS}}_E\left[B',\Phi\right]}\exp\left\lbrace
i\int_{S^2} \dd^2 x' \sqrt{g}\, \left(B'(\mathbf{x}')-\frac{2\pi i}{pq^2}\nabla^2\Phi(\mathbf{x'})\right) \left(n{f_\mathbf{x}^\delta(\mathbf{x}')}-m{f_\mathbf{y}^\delta(\mathbf{x}')}\right)   \right\rbrace~.
\end{align}
As in BF theory (see the manipulations around \eqref{eq:phishift}), the trick to simplify this expression is to shift the field $\Phi$ as follows
\begin{equation}\label{eq:phishiftBS}
\Phi(\mathbf{x}')\equiv\widetilde{\Phi}(\mathbf{x}')+\Phi_{\rm b}(\mathbf{x}')~,
\end{equation} 
where  $\Phi_{\rm b}$ is a fixed function that satisfies
\begin{equation}\label{eq:chidefBS}
 \nabla^2\Phi_{\rm b}(\mathbf{x}')=\frac{2\pi}{p}\left(nf_\mathbf{x}^\delta(\mathbf{x}')-m{f_\mathbf{y}^\delta(\mathbf{x}')}-\frac{n-m}{4\pi\ell^2}\right)~.
\end{equation}
Comparing this with \eqref{eq:chidef}, note that the only difference in the manipulations arises because the $p$-Schwinger model requires a specific regularization of ultraviolet divergences, which we were lucky to avoid in BF theory. 

Under this shift the path integral undergoes a remarkable simplification: 
\begin{align}
{\braket{m|n}}=&\lim_{\delta\rightarrow0}\frac{\mathcal{N}^\delta_n\mathcal{N}^\delta_m}{\cZ_{\rm BS}}\sum_{k=-\infty}^{\infty}2\pi\delta_{n,m+kp}e^{-\frac{\pi }{2p^2q^2 \ell^2}\left(kp+m-n\right)^2 }\frac{\ell}{\ell_{\rm UV}}\int \frac{ DB' D\widetilde{\Phi} Dh}{\textnormal{vol}\,\mathcal{G}}J_{\Phi,h} \,  \nonumber\\&e^{-S_E^{'{\rm BS}}\left[B',\widetilde{\Phi}\right]}\,\exp\left\lbrace\frac{2\pi^2}{p^2q^2}\int_{S^2} \dd^2 x' \sqrt{g}\,\left(n{f_\mathbf{x}^\delta(\mathbf{x}')}-m{f_\mathbf{y}^\delta(\mathbf{x}')}\right)^2\right\rbrace~,
\end{align}
allowing us to cancel a factor of $\cZ_{\rm BS}$ between the numerator and denominator, leaving 
\begin{align}
{\braket{m|n}}=\lim_{\delta\rightarrow0}\mathcal{N}^\delta_n\mathcal{N}^\delta_m\sum_{k=-\infty}^{\infty}\delta_{n,m+kp}\exp\left\lbrace\frac{2\pi^2}{p^2q^2}\int_{S^2} \dd^2 x' \sqrt{g}\,\left(n{f_\mathbf{x}^\delta(\mathbf{x}')}-m{f_\mathbf{y}^\delta(\mathbf{x}')}\right)^2\right\rbrace~.
\end{align}
If we assume that our smearing functions become tightly peaked around their central point as we take $\delta\to0$, namely:
\begin{equation}
\lim_{\delta\rightarrow0}\int_{S^2}\dd^2x\sqrt{g}f_\mathbf{x}^\delta(\mathbf{x'})f_\mathbf{y}^\delta(\mathbf{x'})=0~,
\end{equation}
then, using \eqref{eq: f def} and the precise form of the normalization factor \eqref{RenormConst1}, we obtain: 
\begin{equation}\label{eq:zp0statesschwinger}
\boxed{\braket{m|n}=\begin{cases}1~, & m=n \text{ mod } p\\
0~, & \text{otherwise} \end{cases}}~,
\end{equation}
precisely as in the BF theory.

\subsubsection*{Third calculation: Topological line inserted between \texorpdfstring{$\mathbb{Z}_p^{(0)}$}{Zp0} vacua}

The next calculation we will perform is the expectation value of a topological line operator sandwiched between states of definite $\mathbb{Z}_p^{(0)}$ charge: 
\begin{equation}\label{eq:lineopexpectationBS}
\bra{m}\hat{L}_j[\mathcal{C}]\ket{n}= \lim_{\beta,\delta\rightarrow0}\quad\begin{gathered}\includegraphics[height=3cm]{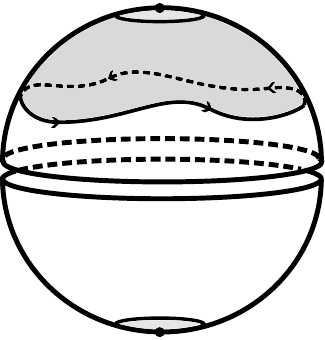}\put(-53,92){{\footnotesize $\hat{U}^\delta_n$}($\mathbf{x}$)}\put(-53,-12){{\footnotesize $\hat{U}_m^{\delta\dagger}$}($\mathbf{y}$)}\put(-5,67){\footnotesize $\hat{L}^\beta_j[\cC]$}\end{gathered}=\lim_{\beta,\delta\rightarrow0}\frac{1}{\cZ_{\rm BS}}\int \frac{DB DA_\mu}{\textnormal{vol}\mathcal{G}} \,\hat{U}_m^{\delta\dagger}(\mathbf{y})\hat{L}^\beta_j[\cC]\hat{U}_n^\delta(\mathbf{x})   e^{-S^{\textnormal{BS} }_E}~,\end{equation}
where $\cC$ is the curve traced out by the line operator $\hat{L}_j[\mathcal{C}]$. We will again label the smallest area subtended by the curve $\cC$ by $\cD$, which is represented by the grey shaded region in the above diagram. We will pull the expressions for the renormalized operators $\hat{U}_n^\delta$ and $\hat{L}^\beta_j[\cC]$ from \eqref{eq:Undeltaexplicity} and \eqref{eq:Ldeltadef}, respectively.

To proceed, we again split $B=b + B'$ between its zero mode and nonzero modes and write: 
\begin{align}
&\bra{m}\hat{L}_j[\mathcal{C}]\ket{n}=\lim_{\beta,\delta\rightarrow0}\frac{\mathcal{N}^\delta_n\cM^\beta_j\mathcal{N}^\delta_m}{\cZ_{\rm BS}}\sum_{k=-\infty}^{\infty}e^{-\frac{\pi k}{pq^2 \ell^2}\left(\frac{kp}{2}+m-n\right)-ijk\frac{\mathcal{A}_\mathcal{D}}{2\ell^2}  }\frac{\ell}{\ell_{\rm UV}}\int_{0}^ {2\pi}\dd b\, e^{-ib\left(p\, k+m-n\right)}\nonumber\\&\int \frac{ DB' D\Phi Dh}{\textnormal{vol}\,\mathcal{G}}J_{\Phi,h}e^{-S_E^{'\rm BS}[B',\Phi]} \exp\Bigg\lbrace i\int_{S^2} \dd^2 x' \sqrt{g}\, \left(B'(\mathbf{x}')-\frac{2\pi i}{pq^2}\nabla^2\Phi(\mathbf{x'})\right) \left(n{f_\mathbf{x}^\delta(\mathbf{x}')}-m{f_\mathbf{y}^\delta(\mathbf{x}')}\right)\nonumber\\
&\qquad\qquad\qquad\qquad\qquad\qquad\quad-i j\int_{S^2} \dd^2 x' \sqrt{g}\,\cT_\cD^\beta(\mathbf{x}')\left[ \nabla^2 \Phi(\mathbf{x}')+\frac{i}{2p} \nabla^2 B(\mathbf{x}') \right]\Bigg\rbrace~.
\end{align}
As the reader should anticipate, similarly to how we proceeded around \eqref{eq:bphishift}, we shift the fields as follows: 
\begin{equation}
B'(\mathbf{x}')\equiv \widetilde{B}'(\mathbf{x}')+\frac{2\pi j}{p}\left(\frac{\cA_\cD}{4\pi\ell^2}-\cT^\beta_\cD(\mathbf{x}')\right)~,\qquad
\Phi(\mathbf{x}')\equiv\widetilde{\Phi}(\mathbf{x}')+\Phi_{\rm b}(\mathbf{x}')~,
\end{equation} 
where $\Phi_{\rm b}$ is defined in \eqref{eq:chidefBS}.

Under this shift and integrating by parts where necessary, we again find a considerable simplification: 
\begin{align}
\bra{m}&\hat{L}_j[\mathcal{C}]\ket{n}=\lim_{\beta,\delta\rightarrow0}\frac{\mathcal{N}^\delta_n\cM^\beta_j\mathcal{N}^\delta_m}{\cZ_{\rm BS}}\sum_{k=-\infty}^{\infty}2\pi\delta_{n,m+kp}e^{-\frac{\pi }{2p^2q^2 \ell^2}\left(kp+m-n\right)^2-ij\left(kp+m-n\right)\frac{\mathcal{A}_\mathcal{D}}{2p\ell^2} }\nonumber\\&\frac{\ell}{\ell_{\rm UV}}\int \frac{ D\widetilde{B}' D\widetilde{\Phi} Dh}{\textnormal{vol}\,\mathcal{G}}J_{\Phi,h} \,e^{-S_E^{'\rm BS}\left[\widetilde{B}',\widetilde{\Phi}\right]+\frac{2\pi ij}{p}\int_{S^2} \dd^2 x' \sqrt{g}\,\left(m{f_\mathbf{y}^\delta(\mathbf{x}')}-n{f_\mathbf{x}^\delta(\mathbf{x}')}\right)\cT_\cD^\beta(\mathbf{x}')}  \nonumber\\&\exp\Bigg\lbrace\frac{2\pi^2}{p^2q^2}\int_{S^2} \dd^2 x' \sqrt{g}\,\left(n{f_\mathbf{x}^\delta(\mathbf{x}')}-m{f_\mathbf{y}^\delta(\mathbf{x}')}\right)^2+\frac{\pi j^2}{2p^2}\int_{S^2} \dd^2 x \sqrt{g}\,\partial_\mu\cT_{\cD}^\beta\,\partial^\mu\cT_{\cD}^\beta\Bigg\rbrace~,
\end{align}
which allows us to cancel a factor of $\cZ_{\rm BS}$ between the numerator and denominator, leaving 
\begin{multline}
\bra{m}\hat{L}_j[\mathcal{C}]\ket{n}=\lim_{\beta,\delta\rightarrow0}{\mathcal{N}^\delta_n\cM^\beta_j\mathcal{N}^\delta_m}\sum_{k=-\infty}^{\infty}\delta_{n,m+kp}e^{\frac{2\pi ij}{p}\int_{S^2} \dd^2 x' \sqrt{g}\,\left(m{f_\mathbf{y}^\delta(\mathbf{x}')}-n{f_\mathbf{x}^\delta(\mathbf{x}')}\right)\cT_\cD^\beta(\mathbf{x}')}\\\exp\left\lbrace\frac{2\pi^2}{p^2q^2}\int_{S^2} \dd^2 x' \sqrt{g}\,\left(n{f_\mathbf{x}^\delta(\mathbf{x}')}-m{f_\mathbf{y}^\delta(\mathbf{x}')}\right)^2+\frac{\pi j^2}{2p^2}\int_{S^2} \dd^2 x \sqrt{g}\,\partial_\mu\cT_{\cD}^\beta\partial^\mu\,\cT_{\cD}^\beta\right\rbrace~.
\end{multline}
The final step is to utilize the definitions of the normalizations \eqref{RenormConst1} and \eqref{RenormConst2} as well as the properties of the smearing functions $f_\mathbf{x}^\delta$ and $\cT_\cD^\beta$ given in \eqref{eq: f def}, \eqref{eq:fporperties},\eqref{eq:Tproperties}, and namely that the region $\cD$ overlaps with the insertion point of the toplogical local operator $\mathbf{x}$, leading to: 
\begin{equation}
\boxed{\bra{m}\hat{L}_j[\mathcal{C}]\ket{n}=\begin{cases}e^{-\frac{2\pi i }{p}n j}~, & m=n \text{ mod } p\\
0~, & \text{otherwise} \end{cases}}~,
\end{equation}
once more, just as in BF theory. 

From this we conclude that, once a consistent set of regularizations have been chosen, we can construct both the vacua and the pUniverses in the bosonized $p$-Schwinger model via the Euclidean path integral. The construction of the vacua follows from \eqref{eq:zp0statesschwinger}, while, to complete the construction of the pUniverses, we must show, using the exact same steps outlined above: 
\begin{equation}
\bra{m}\hat{L}_v[\mathcal{C}']\,\hat{L}_j[\mathcal{C}]\ket{n}= \quad\begin{gathered}\includegraphics[height=3cm]{Images/Overlaps7.pdf}\put(-53,92){{\footnotesize $\hat{U}_n$}($\mathbf{x}$)}\put(-53,-12){{\footnotesize $\hat{U}_m^\dagger$}($\mathbf{y}$)}\put(-5,67){\footnotesize $\hat{L}_j[\cC]$}\put(-5,9){\footnotesize $\hat{L}_v[\cC']$}\end{gathered}\qquad=\begin{cases}e^{\frac{2\pi i }{p}\left(vm-n j\right)}~, & m=n \text{ mod } p\\
0~, & \text{otherwise} \end{cases}~.
\end{equation} 
Given what we have already presented, it is uninformative for us to provide these final remaining steps. 
The conclusion remains: in the $p$-Schwinger model, we also have a path-integral derivation of the following: 
\begin{equation}
\boxed{\widetilde{\braket{m|n}}=\begin{cases}1~, & m=n \text{ mod } p\\
0~, & \text{otherwise} \end{cases}}~.
\end{equation}
We will now show that the vacua and pUniverses are de Sitter invariant and Hadamard.

\subsection{de Sitter invariance of the vacua and pUniverses}

We will now compute various correlation functions inside the multitude of vacua or pUniverses available to us. We will show that these correlation functions only exhibit coincident point singularities in the Euclidean section and are fully $SO(3)$ invariant. The $SO(3)$ invariance ensures that these correlation functions, upon analytic continuation, respect de Sitter invariance. 

Showing $SO(3)$ invariance is simple for two-point correlation functions. For this we simply need to demonstrate that the correlators are functions of the $SO(3)$-invariant distance between the two points, which we define now. For this, it will be useful to think of the $S^2$ as a hypersurface in $\bR^3$ satisfying the following constraint: 
\begin{equation}
\vec{\mathbf{r}}\cdot{\vec{\mathbf{r}}}=\ell^2~.
\end{equation} 
The $SO(3)$-invariant distance between two points can be expressed simply in terms of this embedding: 
\begin{equation}\label{invso3}
\cos\Theta_{xy}\equiv \frac{\vec{\mathbf{r}}_x\cdot\vec{\mathbf{r}}_y}{\ell^2}~.
\end{equation}
However, this notion of distance is not zero when the points are coincident, so instead we will often use the following related distance measure: 
\begin{equation}
u^E_{xy}\equiv\frac{\left(\vec{\mathbf{r}}_x-\vec{\mathbf{r}}_y\right)\cdot\left(\vec{\mathbf{r}}_x-\vec{\mathbf{r}}_y\right)}{2\ell^2}=2\sin^2\frac{\Theta_{xy}}{2}~, \label{eq:udef}
\end{equation}
where the superscript $E$ is a mnemonic indicating that this is a Euclidean distance measure.  These two distances are related by the following identity:  
\begin{equation}
\cos\Theta_{xy}=1-u^E_{xy}~.
\end{equation}
The distance $u_{xy}^E$ is naturally adapted to analytic continuation to dS$_2$, meaning that under analytic continuation of the coordinates to dS$_2$ we will find $u_{xy}^E\rightarrow u_{xy}^L$, where the superscript $L$ is there to remind the reader that this is a `Lorentzian' distance measure. 

\subsubsection{Electric field two-point function}\label{ssub:electric_field_two_point_function}

We will now compute the two-point correlator of the electric field scalar $E(\mathbf{x})$, defined in \eqref{eq:electricfielddef}. Our specific task is to demonstrate that the two-point function of the electric field in any of the vacua:  
\begin{equation}
\bra{m}E(\mathbf{v})E(\mathbf{w})\ket{n}=\lim_{\delta\to0}\bra{0}\hat{U}_m^{\delta\dagger}(\mathbf{y})E(\mathbf{v})E(\mathbf{w})\hat{U}_n^\delta(\mathbf{x})\ket{0}
\end{equation}
is de Sitter invariant. In Euclidean signature, this means that the correlation function only depends on $u_{vw}^E$ defined in \eqref{eq:udef}. The electric field two-point function in the pUniverse states will follow from the above calculation by considering linear superpositions of the vacua with phases appropriately chosen to match \eqref{eq:changeofbasiszp0zp1}. Combining \eqref{eq:vacuaoverlap1} with \eqref{eq:electricfieldkinst}, our task is to calculate: 
\begin{align}
\bra{m}E(\mathbf{v})&E(\mathbf{w})\ket{n}=\lim_{\delta\rightarrow0}\frac{\mathcal{N}^\delta_n\mathcal{N}^\delta_m}{\cZ_{\rm BS}}\sum_{k=-\infty}^{\infty}e^{-\frac{\pi k}{pq^2 \ell^2}\left(\frac{kp}{2}+m-n\right) }\frac{\ell}{\ell_{\rm UV}}\int_{0}^ {2\pi}\dd b \, e^{-ib\left(p\, k+m-n\right)}  \nonumber\\
&\int \frac{ DB' D\Phi Dh}{\textnormal{vol}\,\mathcal{G}}J_{\Phi,h} \,\left[\left(\nabla^2_{\mathbf{v}}\Phi(\mathbf{v}) + \frac{k}{2\ell^2} \right)\left(\nabla^2_{\mathbf{w}}\Phi(\mathbf{w}) + \frac{k}{2\ell^2} \right)\right]e^{-S_E^{'\rm BS}[B',\Phi]}\nonumber\\
&\times\exp\Bigg\lbrace+i\int_{S^2} \dd^2 x' \sqrt{g}\, \left(B'(\mathbf{x}')-\frac{2\pi i}{pq^2}\nabla^2\Phi(\mathbf{x'})\right) \left(n{f_\mathbf{x}^\delta(\mathbf{x}')}-m{f_\mathbf{y}^\delta(\mathbf{x}')}\right)   \Bigg\rbrace~.
\end{align}
The next step is to shift the $\Phi$ field as in \eqref{eq:phishiftBS} and \eqref{eq:chidefBS}, 
which will allow us to straightforwardly integrate the $B'$ field out. We are then left with the following Gaussian path integral to evaluate: 
\begin{align}
\bra{m}E(\mathbf{v})&E(\mathbf{w})\ket{n}=\lim_{\delta\rightarrow0}\mathcal{N}^\delta_n\mathcal{N}^\delta_m\sum_{k=-\infty}^{\infty}\delta_{n,m+kp}e^{+\frac{2\pi^2}{p^2q^2}\int_{S^2} \dd^2 x' \sqrt{g}\,\left(n{f_\mathbf{x}^\delta(\mathbf{x}')}-m{f_\mathbf{y}^\delta(\mathbf{x}')}\right)^2}~\nonumber\\
&\times\Bigg[\frac{4\pi^2}{p^2}\left(nf_\mathbf{x}^\delta(\mathbf{v})-m{f_\mathbf{y}^\delta(\mathbf{v})}\right)\left(nf_\mathbf{x}^\delta(\mathbf{w})-m{f_\mathbf{y}^\delta(\mathbf{w})}\right)\nonumber\\&+\nabla^2_{\mathbf{v}}\nabla^2_{\mathbf{w}}\left\lbrace\frac{1}{\cZ_{\widetilde{\Phi}}}\int  D\widetilde{\Phi}\,\left[\widetilde{\Phi}(\mathbf{v})\widetilde{\Phi}(\mathbf{w}) \right] e^{-\frac{1}{2q^2}\int_{S^2} \dd^2 x \sqrt{g} \, \widetilde{\Phi}\left(-\nabla^2\right)\left(-\nabla^2+\frac{p^2q^2}{\pi}\right)\widetilde{\Phi}   }\right\rbrace\Bigg]~.
\end{align}
Expressing $\widetilde{\Phi}$ in spherical harmonics, the last term in curly brackets is equivalent to
\begin{equation}\label{eq:Gdefsum}
 G_\Phi(\mathbf{v},\mathbf{w}) \equiv  \bra{0} \Phi(\mathbf{v})\Phi(\mathbf{w}) \ket{0} = q^2 \ell^2 \sum_{L=1}^\infty \sum_{M=-L}^L \frac{Y_{LM}(\mathbf{v})Y_{LM}^*(\mathbf{w})}{L (L+1) \left(L(L+1)+\frac{p^2q^2\ell^2}{\pi} \right)} ~,
\end{equation}
where $\ket{0}$ is the $n=0$ vacuum. The sum starts from $L=1$ because $\Phi$ and hence $\widetilde{\Phi}$ have no zero-mode.
This Green's function $G_\Phi$ is a solution to the following differential equation:
\begin{equation}
    -\frac{1}{q^2} \nabla^2 \left(-\nabla^2 + \frac{p^2 q^2}{\pi} \right) G_\Phi(\mathbf{x},\mathbf{y}) = \frac{\delta(\mathbf{x}-\mathbf{y})}{\sqrt{g}} - \frac{1}{4\pi \ell^2} \, ,
\end{equation}
where the constant term on the right hand side stems from the zero-mode removal. 
This sum can be performed exactly by making use of the following identity \citep{Anninos:2024fty} : 
\begin{equation}\label{eq:sumdiff}
\frac{q^2 \ell^2  }{L (L+1) \left(L(L+1)+\frac{p^2q^2\ell^2}{\pi}\right)}=\frac{\pi}{p^2}\left(\frac{1}{L(L+1)}-\frac{1  }{L(L+1)+\frac{p^2q^2\ell^2}{\pi}}\right)~,
\end{equation}
giving:
\begin{multline}
 G_\Phi(\mathbf{v},\mathbf{w}) =-\frac{1}{4p^2}\left(1-\frac{\pi}{p^2q^2\ell^2}\right)\\-\frac{1}{4p^2}\left[\log \frac{u_{vw}^E}{2}+{\Gamma(\Delta) \Gamma(1-\Delta)}\, {_2}F_1\left(\Delta,1-\Delta,1,1-\frac{u_{vw}^E}{2}\right)\right]~,
\label{GF}
\end{multline}
where $\Delta$ is as in \eqref{eq:deltadef}. The function $G_\Phi$ has a finite coincident-point limit as $u^E_{vw}\rightarrow0$: 
\begin{equation}\label{eq:G0}
    G_\Phi(0) =\frac{1}{4p^2} \left( \psi(1+\Delta) + \psi(2-\Delta)+2\gamma -1 \right)~,
\end{equation}
where $\psi(z)\equiv\frac{\Gamma'(z)}{\Gamma(z)}$ is the digamma function and $\gamma\approx0.5772$ is the Euler-Mascheroni constant.

The derivatives of interest may be evaluated directly using the summation representation \eqref{eq:Gdefsum}: 
\begin{equation}
\nabla^2_{\mathbf{v}}\nabla^2_{\mathbf{w}} G_\Phi(\mathbf{v},\mathbf{w})=q^2\frac{\delta(\mathbf{v}-\mathbf{w}) }{\sqrt{g}}-  \frac{p^2q^4}{\pi}\cG^\Delta(\mathbf{v},\mathbf{w})~,\label{chichi}
\end{equation}
where $\cG^\Delta$ is the two-point function of a massive scalar on the $S^2$:
\begin{equation}\label{eq:massivepropagatordef}
\cG^\Delta(\mathbf{x},\mathbf{y})\equiv \frac{\Gamma(\Delta)\Gamma(1-\Delta)}{4\pi} {_2}F_1\left(\Delta,1-\Delta; 1; 1-\frac{u^E_{xy}}{2} \right)~.
\end{equation}
 As is well known about the Schwinger model, the mass, captured by the quantity $\Delta$ in \eqref{eq:deltadef}, is dynamically generated as a result of the interaction between the gauge field and the charged fermion.

Finally putting everything together, we find that the electric field two-point function, after taking the smearing parameter $\delta\to0$, is: 
\begin{align}\label{4ptcontact}
&\bra{m}E(\mathbf{v})E(\mathbf{w})\ket{n}\nonumber\\&=\begin{cases}\frac{4\pi^2}{p^2}\left(n\frac{\delta(\mathbf{v}-\mathbf{x})}{\sqrt{g}}-m\frac{\delta(\mathbf{v}-\mathbf{y})}{\sqrt{g}}\right)\left(n\frac{\delta(\mathbf{w}-\mathbf{x})}{\sqrt{g}}-m\frac{\delta(\mathbf{w}-\mathbf{y})}{\sqrt{g}}\right)+\nabla^2_{\mathbf{v}}\nabla^2_{\mathbf{w}} G_\Phi(\mathbf{v},\mathbf{w})~, & m=n \text{ mod } p\\ 
0 ~,& \text{ otherwise} \end{cases}~,
\end{align}
where we remind the reader that $\mathbf{x}$ and $\mathbf{y}$ are the insertion locations of the toplogical operators $\hat{U}_{n/m}$ that prepare the vacua in the Euclidean path integral. These contact terms are almost undetectable, as they only have an effect when \emph{both} electric field operators are placed on top of one of the two state-preparation insertions. Nevertheless, they may have interesting consequences when we couple the theory to gravity, as discussed in \citep{Anninos:2023lin}.

Something interesting happens when we compute the electric field two-point function between pUniverses. Using \eqref{eq:changeofbasiszp0zp1}, we find: 
\begin{align}
&\widetilde{\bra{m}}E(\mathbf{v})E(\mathbf{w})\widetilde{\ket{n}}\nonumber\\&=\begin{cases}\frac{(4p-2)(p-1)\pi^2}{3p^2}\left(\frac{\delta(\mathbf{v}-\mathbf{x})}{\sqrt{g}}-\frac{\delta(\mathbf{v}-\mathbf{y})}{\sqrt{g}}\right)\left(\frac{\delta(\mathbf{w}-\mathbf{x})}{\sqrt{g}}-\frac{\delta(\mathbf{w}-\mathbf{y})}{\sqrt{g}}\right)+\nabla^2_{\mathbf{v}}\nabla^2_{\mathbf{w}} G_\Phi(\mathbf{v},\mathbf{w})~, & m=n \text{ mod } p\\ 
\frac{\pi^2}{p\sin^2\left(\frac{(m-n)\pi}{p}\right)}\left(e^{\frac{2\pi i(m-n)}{p}}+\frac{2-p}{p}\right)\left(\frac{\delta(\mathbf{v}-\mathbf{x})}{\sqrt{g}}-\frac{\delta(\mathbf{v}-\mathbf{y})}{\sqrt{g}}\right)\left(\frac{\delta(\mathbf{w}-\mathbf{x})}{\sqrt{g}}-\frac{\delta(\mathbf{w}-\mathbf{y})}{\sqrt{g}}\right) ~,& \text{ otherwise} \end{cases}~,
\end{align}
and hence the two point function of the electric field receives a non-diagonal pure-contact contribution across the pUniverses. Besides the contact terms, this two-point function is de Sitter invariant upon analytic continuation.\footnote{One should of course be cautious. A point on $S^2$ can map  to the entire static patch horizon, depending on how the Lorentzian continuation is performed. It would be interesting to carefully identify what part of the contact terms in the electric field two point function are scheme independent. A simpler avatar is a contact term that appears in the Abelian BF-model when computing the two-point function between $E(\mathbf{x})$ and $B(\mathbf{x})$. The BF-model contact term, which has a pure imaginary and quantised coefficient $\tfrac{2\pi i}{p}$, is tied to the symplectic structure of the theory and is scheme independent. It persists in the two point function between $E(\mathbf{x})$ and $e^{i nB(\mathbf{x})}$, where the coefficient picks up an additional factor of $n$. The contact term in (\ref{4ptcontact}) takes the form of the product of two such BF contact terms. The Lorentzian continuation of these contact terms is a spacetime contact term.}

\subsubsection{Wilson line expectation value}\label{ssub:wilson_line_expectation_value}

Now that we have shown that the electric field correlator is de Sitter invariant and Hadamard, we move on to compute the expectation value of the Wilson-loop operator $\hat{W_j}[\cC]$ defined in \eqref{eq: Wilson line}. As we've shown, the $\bZ_p^{(1)}$ one-form symmetry is spontaneously broken, so one may expect the expectation-value to obey a perimeter, rather than an area, law at asymptotically-large size \citep{Komargodski:2020mxz,Misumi:2019dwq}. However, since we will be computing these expectation values on the $S^2$, it is unclear how to test this hypothesis given that Wilson loops are bounded in size.

Unlike the topological local operator, we will not need to specify any regularization of the Wilson line, as we will now show. The calculation follows \eqref{eq:lineopexpectationBS}, namely we need to evaluate:
\begin{equation}
\bra{m}\hat{W}_j[\mathcal{C}]\ket{n}= \lim_{\delta\rightarrow0}\quad\begin{gathered}\includegraphics[height=3cm]{Images/Overlaps6_thickened.pdf}\put(-53,92){{\footnotesize $\hat{U}^\delta_n$}($\mathbf{x}$)}\put(-53,-12){{\footnotesize $\hat{U}_m^{\delta\dagger}$}($\mathbf{y}$)}\put(-5,67){\footnotesize $\hat{W}_j[\cC]$}\end{gathered}=\lim_{\delta\rightarrow0}\frac{1}{\cZ_{\rm BS}}\int \frac{DB DA_\mu}{\textnormal{vol}\mathcal{G}} \,\hat{U}_m^{\delta\dagger}(\mathbf{y})\hat{W}_j[\cC]\hat{U}_n^\delta(\mathbf{x})   e^{-S^{\textnormal{BS} }_E}~,\end{equation}
however this time we will be agnostic to whether the curve $\cC$ links with one of the topological local operators $\hat{U}_n$ or $\hat{U}_m$ creating the state. As we have drawn it, $\cC$ links with $\hat{U}_n$, but this need not be the case, and our calculations will reflect this. 

Using \eqref{eq:StokestheoremWilson}, we now want to compute:
\begin{align}
&\bra{m}\hat{W}_j[\mathcal{C}]\ket{n}=\lim_{\delta\rightarrow0}\frac{\mathcal{N}^\delta_n\mathcal{N}^\delta_m}{\cZ_{\rm BS}}\sum_{k=-\infty}^{\infty}e^{-\frac{\pi k}{pq^2 \ell^2}\left(\frac{kp}{2}+m-n\right)-ijk\frac{\mathcal{A}_\mathcal{D}}{2\ell^2}  }\frac{\ell}{\ell_{\rm UV}}\int_{0}^ {2\pi}\dd b\, e^{-ib\left(p\, k+m-n\right)}\nonumber\\&\int \frac{ DB' D\Phi Dh}{\textnormal{vol}\,\mathcal{G}}J_{\Phi,h}e^{-S_E^{'\rm BS}[B',\Phi]} \exp\Bigg\lbrace i\int_{S^2} \dd^2 x' \sqrt{g}\, \left(B'(\mathbf{x}')-\frac{2\pi i}{pq^2}\nabla^2\Phi(\mathbf{x'})\right) \left(n{f_\mathbf{x}^\delta(\mathbf{x}')}-m{f_\mathbf{y}^\delta(\mathbf{x}')}\right)\nonumber\\
&\qquad\qquad\qquad\qquad\qquad\qquad\quad-i j\int_{S^2} \dd^2 x' \sqrt{g}\,\Theta_\cD(\mathbf{x}')\nabla^2 \Phi(\mathbf{x}')\Bigg\rbrace~.
\end{align}
Following suit, the next step is to shift $\Phi\rightarrow \widetilde{\Phi}+\Phi_{\rm b}$ with $\Phi_{\rm b}$ satisfying \eqref{eq:chidefBS}, giving
\begin{align}
&\bra{m}\hat{W}_j[\mathcal{C}]\ket{n}=\sum_{k=-\infty}^{\infty}\delta_{n,m+kp}e^{+\frac{2\pi ij}{p}\int_{S^2} \dd^2 x' \sqrt{g}\,\left(m\frac{\delta(\mathbf{x}'-\mathbf{y})}{\sqrt{g}}-n\frac{\delta(\mathbf{x}'-\mathbf{x})}{\sqrt{g}}\right)\Theta_\cD(\mathbf{x}')}~\nonumber\\
&\times\frac{1}{\cZ_{\widetilde{\Phi}}}\int  D\widetilde{\Phi}\,e^{-\frac{1}{2q^2}\int_{S^2} \dd^2 x \sqrt{g} \left\lbrace \widetilde{\Phi}\left(-\nabla^2\right)\left(-\nabla^2+\frac{p^2q^2}{\pi}\right)\widetilde{\Phi}\right\rbrace-ij\int_{S^2} \dd^2 x \sqrt{g}\,\Theta_{\cD}\,\nabla^2\widetilde{\Phi}   }~.
\end{align}
where we have already taken the $\delta\to0$ limit and integrated out $B'$. Let us define the linking number $\cL$, which measures whether the curve $\cC$ links with any of the topological insertions: 
\begin{equation}
\cL(\mathbf{x},\cC)\equiv\int_{S^2} \dd^2 x' \sqrt{g}\,\frac{\delta(\mathbf{x}'-\mathbf{x})}{\sqrt{g}}\Theta_\cD(\mathbf{x}')~,
\end{equation}
where recall that $\cD$ is the `outward' region that subtends $\cC$. Then by completing the square, the Wilson line expectation value is:
\begin{multline}
\bra{m}\hat{W}_j[\mathcal{C}]\ket{n}=\\e^{\frac{2\pi i j}{p}\left(m\cL(\mathbf{y},\cC)-n\cL(\mathbf{x},\cC)\right)}\times\begin{cases}e^{-\frac{j^2}{2}\int_\cD\dd^2x\sqrt{g}\int_\cD\dd^2y\sqrt{g}\,\nabla_{\mathbf{x}}^2\nabla_{\mathbf{y}}^2G_\Phi(\mathbf{x},\mathbf{y})}~, & m=n \text{ mod } p\\
0~, & \text{otherwise} \end{cases}~.
\end{multline}
Using \eqref{chichi} we can write this as: 
\begin{multline}
\bra{m}\hat{W}_j[\mathcal{C}]\ket{n}=\\e^{\frac{2\pi i j}{p}\left(m\cL(\mathbf{y},\cC)-n\cL(\mathbf{x},\cC)\right)-\frac{j^2q^2}{2}\cA_\cD}\times\begin{cases}e^{\frac{p^2q^4j^2}{2\pi}\int_\cD\dd^2x\sqrt{g}\int_\cD\dd^2y\sqrt{g}\,\cG^\Delta(\mathbf{x},\mathbf{y})}~, & m=n \text{ mod } p\\
0~, & \text{otherwise} \end{cases}~.
\end{multline}
This result can be converted in to the pUniverse basis, where we notice something interesting. The phase that arises due to linking with the topological local operators gives us: 
\begin{multline}
\widetilde{\bra{m}}\hat{W}_j[\mathcal{C}]\widetilde{\ket{n}}=\\\begin{cases}e^{-\frac{j^2q^2}{2}\cA_\cD+\frac{p^2q^4j^2}{2\pi}\int_\cD\dd^2x\sqrt{g}\int_\cD\dd^2y\sqrt{g}\,\cG^\Delta(\mathbf{x},\mathbf{y})}~, & m=n \text{ mod } p \text{ and }\cC\text{ links no insertions }\\
0~, & \text{otherwise} \end{cases}~,
\end{multline}
meaning, the Wilson loop is only non-zero if it does not link with either of the topological operators (or both)! Besides this observation our result is standard and can be found in e.g. \citep{Sachs:1991en} for the model on the two-torus $\bT^2$. The fact that the Wilson line gives zero if it links one of the insertions is obvious upon further reflection: the Wilson lines, like the topological line operators $\hat{L}_n$ mediate transitions between pUniverses, making this expectation value vanish.

\subsubsection{The chiral condensate}\label{ssub:the_chiral_condensate}

We will now focus our efforts on computing expectation values of the vertex operator $\hat{V}^{\rm bare}_n(\mathbf{x})$. But recall that we have not yet addressed how we will regularize this operator. We turn to this discussion now. Before doing so, let us review some well-known facts. A theme of this paper is that, owing to the ABJ anomaly \eqref{eq:fermiactionrot}, the Schwinger model's axial symmetry is broken. For $p=1$ there is no residual unbroken symmetry, whereas for $p>1$ we are left with a residual $\bZ^{(0)}_p$ symmetry. 

In the $p=1$ theory, the order parameters that signal the breakdown of axial-symmetry are the following non-vanishing vacuum expectation values \citep{Lowenstein:1971fc,Nielsen:1976hs,Rothe:1978hx}: 
\begin{equation}
\left\langle\left(\bar{\Psi}P_L\Psi\right)(\mathbf{x})\right\rangle=\left\langle\left(\bar{\Psi}P_R\Psi\right)(\mathbf{x})\right\rangle\neq0~.
\end{equation}  
These vevs signal axial symmetry breaking in the $p=1$ theory because each of the fermion bilinears listed above are charged under the would-be $U(1)$-axial symmetry. In other words, if the symmetry were present, these vevs would vanish. But since the symmetry is not present, these operators do in fact obtain a vev and these non-vanishing vevs are often referred to as the \emph{chiral condensate}. 

For the $p=1$ theory on the $S^2$, the chiral condensates were first computed in \citep{Jayewardena:1988td} and later verified in \citep{Anninos:2024fty}. We quote the result here:
\begin{equation}\label{eq:chiralcondensatep=1}
\bra{0}\bar{\Psi}P_L\Psi\ket{0}=\bra{0}\bar{\Psi}P_R\Psi\ket{0}=-\frac{e^{\frac{1}{2}-\frac{\pi}{2q^2\ell^2}}}{4\pi\ell}e^{2G_\Phi(0)}~, \qquad p=1~,
\end{equation} 
with $G_\Phi(0)$ given in \eqref{eq:G0}, and where $\ket{0}$ should be understood as the as the Hartle-Hawking vacuum. The $\ell\rightarrow\infty$ limit of \eqref{eq:chiralcondensatep=1} matches the flat space-chiral condensates computed e.g. in \citep{Lowenstein:1971fc}. 

When $p>1$ the story changes \citep{Misumi:2019dwq}. Now that there is a residual unbroken $\bZ^{(0)}_p$ global symmetry under which the chiral condensates are charged, as per equation \eqref{eq:Zpchargebilinear}, the chiral condensates now \emph{must vanish} in any of the of the $p$ inequivalent vacua of the theory, precisely because the vacua also carry $\bZ^{(0)}_p$ charge. 

But let us interpret \eqref{eq:chiralcondensatep=1} slightly differently. For $p=1$, it is entirely equivalent to say that the result of \citep{Jayewardena:1988td,Anninos:2024fty} is:
\begin{equation}\label{eq:chiralcondensatepUniversep=1}
\widetilde{\bra{0}}\bar{\Psi}P_L\Psi\widetilde{\ket{0}}=\widetilde{\bra{0}}\bar{\Psi}P_R\Psi\widetilde{\ket{0}}=-\frac{e^{\frac{1}{2}-\frac{\pi}{2q^2\ell^2}}}{4\pi\ell}e^{2G_\Phi(0)}~, \qquad p=1~,
\end{equation}
because for $p=1$ there is no distinction between the vacuum and pUniverse of the theory. Taking inspiration from this, for $p\ge1$ we will normalize our operator such that: 
\begin{equation}\label{eq:chiralcondensatepuniverse}
\widetilde{\bra{0}}q\,\hat{V}_1\widetilde{\ket{0}}=\widetilde{\bra{0}}q\,\hat{V}_{-1}\widetilde{\ket{0}}=-\frac{e^{\frac{1}{2}-\frac{\pi}{2p^2q^2\ell^2}}}{4\pi\ell}e^{2p^2G_\Phi(0)}~, \qquad p\ge1~.
\end{equation}
We will not aim to compute this expression in the fermionic variables, hence the specific form for the chiral condensate written above is conjectural.  However it takes inspiration from the discussion in \citep{Misumi:2019dwq} whereby it was argued that for $p>1$ the chiral condensate's expectation value is suppressed by the action of a fractional instanton.\footnote{Note that the non-perturbative suppression in the gauge coupling is down by a factor of $p^{-2}$ compared to the one-instanton action \eqref{eq:actionqschwinger}.} We will see precisely how this happens in \eqref{eq:Bprimepathintegral} below. Furthermore, the above expression admits a finite flat-space limit if we take $\ell\rightarrow\infty$, for any $p$. When all is said and done, we will have normalized $\hat{V}_1$ such that it is consistent with \eqref{eq:chiralcondensatepuniverse}.

Let us now propose a regulated form of the vertex operator
\begin{equation}
    \hat{V}^\delta_n(\mathbf{x}) \equiv  \, \cV^\delta_n \,\, e^{ in \int_{S^2}\dd^2 z \sqrt{g} \,f_\mathbf{x}^\delta(\mathbf{z})\, B(\mathbf{z})}  \, ,
    \label{eq: renorm vertex}
\end{equation}
whose normalization $\cV^\delta_n$ we will fix to be consistent with \eqref{eq:chiralcondensatepuniverse} in the $\delta\to0$ limit. On our way to computing what we need, let us start by computing the following matrix elements:
\begin{equation}
\bra{m}\hat{V}_l(\mathbf{v})\ket{n}\equiv\lim_{\beta,\delta\to0}\bra{0}\hat{U}_m^{\delta\dagger}(\mathbf{y})\hat{V}_l^\beta(\mathbf{v})\hat{U}_n^\delta(\mathbf{x})\ket{0}~.
\end{equation}
Inserting the explicit expressions for the insertions, we find: 
\begin{align}
\bra{m}&\,\hat{V}_l(\mathbf{v})\ket{n}=\lim_{\beta,\delta\to0}\frac{\mathcal{N}^\delta_n\mathcal{V}^\beta_l\mathcal{N}^\delta_m}{\cZ_{\rm BS}}\sum_{k=-\infty}^{\infty}e^{-\frac{\pi k}{pq^2 \ell^2}\left(\frac{kp}{2}+m-n\right) }\frac{\ell}{\ell_{\rm UV}}\int_{0}^ {2\pi}\dd b \, e^{-ib\left(p\, k+m-l-n\right)}  \nonumber\\
&\int \frac{ DB' D\Phi Dh}{\textnormal{vol}\,\mathcal{G}}J_{\Phi,h}\,e^{-S_E^{'\rm BS}[B',\Phi]}
\exp\Bigg\lbrace+i\int_{S^2} \dd^2 x' \sqrt{g}\, B'(\mathbf{x}')\left(n{f_\mathbf{x}^\delta(\mathbf{x}')}+l{f_\mathbf{v}^\beta(\mathbf{x}')}-m{f_\mathbf{y}^\beta(\mathbf{x}')}\right)  \nonumber\\ 
&\qquad\quad+\int_{S^2} \dd^2 x' \sqrt{g}\, \frac{2\pi }{pq^2}\nabla^2\Phi(\mathbf{x'}) \left(n{f_\mathbf{x}^\delta(\mathbf{x}')}-m{f_\mathbf{y}^\delta(\mathbf{x}')}\right)\Bigg\rbrace~.
\end{align}
As we have done in all previous calculations so far, we again shift the path integral $\Phi\rightarrow\widetilde{\Phi}+\Phi_{\rm b}$ with $\Phi_{\rm b}$ defined in \eqref{eq:chidefBS}, which allows us to integrate out $\widetilde\Phi$ . We are then tasked with computing: 
\begin{align}\label{eq:Bprimepathintegral}
\bra{m}&\,\hat{V}_l(\mathbf{v})\ket{n}=\lim_{\beta\to0}\frac{\mathcal{V}^\beta_l}{\cZ_{ B'}}\sum_{k=-\infty}^{\infty}\delta_{n+l,m+kp}e^{-\frac{\pi l^2}{2p^2q^2 \ell^2} } \nonumber\\
&\int { DB' }\,
\exp\Bigg\lbrace -\int_{S^2}\dd^2 x' \sqrt{g}\,\left[\frac{1}{8\pi}  B'\left[-\nabla^2+\frac{p^2 q^2}{\pi}\right] B' -il\,B'(\mathbf{x}'){f_\mathbf{v}^\beta(\mathbf{x}')} \right]   \Bigg\rbrace~,
\end{align}
where we see the tell-tale suppression coming from the fractional instanton \citep{Misumi:2019dwq}.
We have skipped several steps in deriving the above expression. First we have preemptively taken the $\delta\to0$ limit using the specific form of the normalizations \eqref{RenormConst1}. Secondly we integrated out the field $\widetilde\Phi$ resulting in a mass term for the $B'$ field (see \eqref{eq:completesqure}). Recall that $B'$ is not a local field, since it is devoid of a zero-mode, hence its propagator will reflect this. The propagator for $B'$ satisfies: 
\begin{equation}\label{eq:Bprimepropagator}
\frac{1}{4\pi}\left(-\nabla_\mathbf{x}^2+\frac{p^2q^2}{\pi}\right)G^\Delta_{B'}(\mathbf{x},\mathbf{y})=\frac{\delta(\mathbf{x}-\mathbf{y})}{\sqrt{g}}-\frac{1}{4\pi\ell^2}
\end{equation}
whose solution is
\begin{equation}
G^\Delta_{B'}(\mathbf{x},\mathbf{y})=4\pi\sum_{L=1}^\infty \sum_{M=-L}^L \frac{Y_{LM}(\mathbf{x})Y^*_{LM}(\mathbf{y})}{L(L+1)+\frac{p^2q^2\ell^2}{\pi} } ~.
\end{equation}
The sum can be performed explicitly, yielding: 
\begin{equation}\label{eq:Bprimepropagator2}
G_{B'}^\Delta(\mathbf{x},\mathbf{y})=4\pi \cG^\Delta(\mathbf{x},\mathbf{y}) +\frac{1}{\Delta(\Delta-1)}~,
\end{equation}
with $\Delta$ defined in \eqref{eq:deltadef} and $\cG^\Delta$ is defined in \eqref{eq:massivepropagatordef}. Note that because of the zero-mode removal $G_{B'}^\Delta$ admits a smooth massless limit: 
\begin{equation}
G_{B'}^{\Delta=0}(\mathbf{x},\mathbf{y})\equiv \lim_{\Delta\rightarrow0}G_{B'}^\Delta(\mathbf{x},\mathbf{y})=-1-\log\frac{u_{xy}^E}{2}~.
\end{equation}
We can also relate these Green's functions with the $\Phi$ propagator $G_\Phi$ defined in \eqref{GF}: 
\begin{equation}\label{eq:propagatoridentity}
G_{B'}^{\Delta=0}(\mathbf{x},\mathbf{y})-G_{B'}^{\Delta}(\mathbf{x},\mathbf{y})=4p^2\,G_\Phi(\mathbf{x},\mathbf{y})~.
\end{equation} 

With this, evaluating \eqref{eq:Bprimepathintegral} is a simple exercise in completing squares. Note that the differential equation \eqref{eq:Bprimepropagator} ensures that 
\begin{equation}
\int_{S^2}\dd^2x\sqrt{g}\,G^\Delta_{B'}(\mathbf{x},\mathbf{y})=0~,
\end{equation}
which we make use of in the derivation. The final result is: 
\begin{multline}\label{eq:onepointvertexintermediate}
\bra{m}\hat{V}_l(\mathbf{v})\ket{n}=\lim_{\beta\to0}{\mathcal{V}^\beta_l}\sum_{k=-\infty}^{\infty}\delta_{n+l,m+kp}\\\times\exp\left[ -\frac{\pi l^2}{2p^2q^2 \ell^2}-\frac{l^2}{2} \int_{S^2} \dd^2 x\sqrt{g}\int_{S^2} \dd^2 y \sqrt{g} \, f_{\mathbf{v}}^\beta(\mathbf{x}) G_{B'}^\Delta(\mathbf{x},\mathbf{y}) f_{\mathbf{v}}^\beta(\mathbf{y}) \right]~.
\end{multline}
As expected the chiral condensate is nondiagonal in the basis of vacua for $p>1$. 

We now come to the normalization constant $\cV_l^\beta$, which is fixed by requiring that the above expectation value has a finite limit as $\beta\to0$. We will do this by ensuring that  $\mathcal{V}^\beta_l$ behaves as follows:
\begin{equation}
\mathcal{V}^\beta_l=v_l\exp\left[ +\frac{l^2}{2} \int_{S^2} \dd^2 x\sqrt{g}\int_{S^2} \dd^2 y \sqrt{g} \, f_{\mathbf{v}}^\beta(\mathbf{x}) G_{B'}^{\Delta=0}(\mathbf{x},\mathbf{y}) f_{\mathbf{v}}^\beta(\mathbf{y}) \right]~.
\end{equation}
In the above expression we are using $G^{\Delta=0}_{B'}$ rather than than the $G^{\Delta}_{B'}$ that appears in \eqref{eq:onepointvertexintermediate}.
To argue that this choice is the natural one, one notes that this choice ensures that the vertex operator is regulated in a way that doesn't depend on the coupling constant $q$. In other words, this choice of regularization holds whether we couple the compact scalar to Maxwell theory or not. It is entirely equivalent to the normal ordering or point-splitting prescription that one would follow in the free theory. We have also included an, as of yet, undetermined coefficient $v_l$ that is independent of the smearing parameter $\beta$. With this choice, using \eqref{eq:propagatoridentity}, we find: 
\begin{multline}
\bra{m}\hat{V}_l(\mathbf{v})\ket{n}=\lim_{\beta\to0}{v_l}\sum_{k=-\infty}^{\infty}\delta_{n+l,m+kp}\\\times\exp\left[ -\frac{\pi l^2}{2p^2q^2 \ell^2}+2p^2l^2 \int_{S^2} \dd^2 x\sqrt{g}\int_{S^2} \dd^2 y \sqrt{g} \, f_{\mathbf{v}}^\beta(\mathbf{x}) G_{\Phi}(\mathbf{x},\mathbf{y}) f_{\mathbf{v}}^\beta(\mathbf{y}) \right]~.
\end{multline}
Since $G_\Phi$ has a finite coincident-point limit, we can take the $\beta\to0$ limit and find: 
\begin{equation}
\bra{m}\hat{V}_l(\mathbf{v})\ket{n}=\begin{cases}v_l\, e^{-\frac{\pi l^2}{2p^2q^2 \ell^2}+2p^2l^2G_\Phi(0)}~, & m=n+l \text{ mod } p\\
0~, & \text{otherwise} \end{cases}~.
\end{equation}
with $G_\Phi(0)$ given in \eqref{eq:G0}.
Using \eqref{eq:changeofbasiszp0zp1} we can immediately compute the expectation value in the basis of pUniverses: 
\begin{equation}
\widetilde{\bra{m}}\hat{V}_l(\mathbf{v})\widetilde{\ket{n}}=\begin{cases}v_l\, e^{\frac{2\pi i}{p}nl-\frac{\pi l^2}{2p^2q^2 \ell^2}+2p^2l^2G_\Phi(0)}~, & m=n \text{ mod } p\\
0~, & \text{otherwise} \end{cases}~.
\end{equation}
We still need to match \eqref{eq:chiralcondensatepuniverse}, meaning we need to make a choice of $v_l$. The minimal choice that achieves this while having a finite $\ell\to\infty$ flat-space limit is: 
\begin{equation}
v_l\equiv\left(-{1}\right)^{ l}\left(\frac{e}{16\pi^2q^2\ell^2}\right)^{\frac{l^2}{2}}~.
\end{equation}
Unlike the discussion surrounding how we regulate $\hat{V}_l^\beta$, this choice for $v_l$ \emph{does} depend on the explicit value of the coupling. This should be understood as a choice of normalization for the operator rather than an intrinsic definition of the operator. Mainly, we make this choice of normalization to match the chiral condensate in the fermionic theory, bearing in mind that the vertex operator is  dimensionless. 
Putting everything together, we are choosing to normalize our operator such that: 
\begin{equation}
\boxed{\widetilde{\bra{m}}\hat{V}_l(\mathbf{v})\widetilde{\ket{n}}=\begin{cases}\frac{\left(-{1}\right)^{ l}}{\left({4\pi}q \ell \right)^{l^2}}\, e^{\frac{l^2}{2}\left(1-\frac{\pi }{p^2q^2 \ell^2}\right)+\frac{2\pi i}{p}nl+2p^2l^2G_\Phi(0)}~, & m=n \text{ mod } p\\
0~, & \text{otherwise} \end{cases}}~.
\end{equation}
If we take the flat space limit, we find: 
\begin{equation}
\lim_{\ell\to\infty}\widetilde{\bra{m}}\hat{V}_l(\mathbf{v})\widetilde{\ket{n}}=\begin{cases}\frac{\left(-1\right)^{l}}{(4\pi)^{l^2}}\,\left(\frac{p^2}{\pi}\right)^{\frac{l^2}{2}}\, e^{l^2\gamma+\frac{2\pi i}{p}nl}~, & m=n \text{ mod } p\\
0~, & \text{otherwise} \end{cases}~,
\end{equation}
which one can compare with the results of \citep{Lowenstein:1971fc,Sachs:1991en}.
An important realization is that the $\bZ_p$ phase of the chiral condensate is able to distinguish between the $p$ different pUniverses \citep{Misumi:2019dwq,Cherman:2022ecu}. 

Before computing correlation functions of vertex operators, let us write down once and for all the definition of our regulated vertex operator. It is given by \eqref{eq: renorm vertex} with 
\begin{equation}\label{eq:Vertexregulatorfinal}
\cV_n^\delta(\mathbf{v})\equiv\left(-1\right)^{ n} \left(\frac{e}{16\pi^2q^2\ell^2}\right)^{\frac{n^2}{2}} \exp\left[ +\frac{n^2}{2} \int_{S^2} \dd^2 x\sqrt{g}\int_{S^2} \dd^2 y \sqrt{g} \, f_{\mathbf{v}}^\delta(\mathbf{x}) G_{B'}^{\Delta=0}(\mathbf{x},\mathbf{y}) f_{\mathbf{v}}^\delta(\mathbf{y}) \right]~.
\end{equation}

To conclude, notice that higher powers of the fermion bilinear operators \eqref{BosMapBilinear} inserted at coincident points vanish identically due to fermionic statistics. From the perspective of the bosonic theory, this comes about as a consequence of the operator product expansion of vertex operators. In particular, one may easily verify that $\hat V_1(\mathbf{x})\hat V_1(\mathbf{y})\sim u^E_{xy} \hat V_2(\mathbf{y})$ (see \eqref{eq:vertexOPE} for a more detailed expression), hence vanishing in the coincident point limit $\mathbf{x}\to \mathbf{y}$.

This begs the question, how would one represent the higher vertex operators $\hat{V}_{n>1}(\mathbf{x})$ (and hence higher topological operators $\hat{U}_{n>1}(\mathbf{x})$) in the fermionic picture? This was discussed in \citep{Delmastro:2022prj} and \citep{Cherman:2022ecu} and involves strings of fermions and their covariant derivatives at coincident points. For example, by chiral charge counting, we should expect: 
\begin{equation}
    \hat{V}_2^{\rm bare}\propto \bar\Psi P_L \Psi (D_\mu^\dagger \bar\Psi)P_L (D^\mu \Psi)~, \qquad D_\mu \equiv\nabla_\mu+iA_\mu~,
\end{equation}
and with higher vertex operators constructed by introducing more fermions and more derivatives.\footnote{We thank Okasha Uddin for useful discussions around this point.}

\subsubsection{Meson \texorpdfstring{$N_I$}{n}-point function}\label{ssub:meson_n_point_function}

With our normalizations and regularizations in place, we can now provide a compact expression for the correlation function of an arbitrary number $N_I$ of meson vertex-operator insertions:
\begin{equation}
\bra{m}\left(\prod_{i=1}^{N_I}\hat{V}_{l_i}(\mathbf{x}_i)\right)\ket{n}\equiv\lim_{\beta,\delta\to0}\bra{0}\hat{U}_m^{\delta\dagger}(\mathbf{y})\left(\prod_{i=1}^{N_I}\hat{V}^\beta_{l_i}(\mathbf{x}_i)\right)\hat{U}_n^\delta(\mathbf{x})\ket{0}~.
\end{equation}
Our calculations have been very detailed so far, so we will allow ourselves to be a little bit less explicit here. The steps leading up to \eqref{eq:Bprimepathintegral} can be followed verbatim, leading to: 
\begin{align}
\bra{m}&\,\left(\prod_{i=1}^{N_I}\hat{V}_{l_i}(\mathbf{x}_i)\right)\ket{n}=\lim_{\beta\to0}\frac{\prod_{i}\mathcal{V}^\beta_{l_i}}{\cZ_{ B'}}\sum_{k=-\infty}^{\infty}\delta_{n+\sum_{i}l_i,m+kp}e^{-\frac{\pi \left(\sum_{i}l_i\right)^2}{2p^2q^2 \ell^2} } \nonumber\\
&\int { DB' }\,
\exp\Bigg\lbrace -\int_{S^2}\dd^2 x' \sqrt{g}\,\left[\frac{1}{8\pi}  B'\left[-\nabla^2+\frac{p^2 q^2}{\pi}\right] B' -i\sum_{i=1}^{N_I}l_i\,B'(\mathbf{x}'){f_{\mathbf{x}_i}^\beta(\mathbf{x}')} \right]   \Bigg\rbrace~.
\end{align}
The remainder of the calculation proceeds straightforwardly and we find: 
\begin{multline}
\bra{m}\left(\prod_{i=1}^{N_I}\hat{V}_{l_i}(\mathbf{x}_i)\right)\ket{n}=\lim_{\beta\to0}{\left(\prod_{i}\mathcal{V}^\beta_{l_i}\right)}\sum_{k=-\infty}^{\infty}\delta_{n+\sum_i l_i,m+kp}\\\times\exp\left[ -\frac{\pi \left(\sum_{i}l_i\right)^2}{2p^2q^2 \ell^2}-\sum_{i,j=1}^{N_I}\frac{l_i\,l_j}{2} \int_{S^2} \dd^2 v\sqrt{g}\int_{S^2} \dd^2 w \sqrt{g} \, f_{\mathbf{x}_i}^\beta(\mathbf{v}) G_{B'}^\Delta(\mathbf{v},\mathbf{w}) f_{\mathbf{x}_j}^\beta(\mathbf{w}) \right]~.\label{eq:Npointvertexintermediate}
\end{multline}
Subsituting the precise form of $\cV_{l_i}^\beta$ from \eqref{eq:Vertexregulatorfinal}, taking $\beta\to0$, and using: 
\begin{equation}
\left(\sum_{i=1}^{N_I}l_i\right)^2=\sum_{i=1}^{N_I}l_i^2+2\sum_{i<j}^{N_I}l_i \,l_j~,
\end{equation}
we find:
\begin{multline}
\bra{m}\left(\prod_{i=1}^{N_I}\hat{V}_{l_i}(\mathbf{x}_i)\right)\ket{n}=\left(\prod_{i=1}^{N_I}\frac{\left(-1\right)^{ l_i}e^{\frac{l_i^2}{2}\left(1-\frac{\pi }{p^2q^2 \ell^2}\right)+2p^2l_i^2G_\Phi(0)}}{\left(4\pi q\ell\right)^{l_i^2}}\right)\sum_{k=-\infty}^{\infty}\delta_{n+\sum_i l_i,m+kp}\\\times\exp\left[ -\sum_{i<j}^{N_I}{l_i\,l_j}\left(\frac{\pi }{p^2q^2 \ell^2}+G_{B'}^\Delta(\mathbf{x}_i,\mathbf{x}_j)\right)    \right]~.\label{eq:Npointvertexintermediate2}
\end{multline}
Now, it follows from \eqref{eq:Bprimepropagator2} combined with \eqref{eq:deltadef} that the quantity inside the exponential can be related to the massive propagator for a scalar field on the $S^2$ \eqref{eq:massivepropagatordef}: 
\begin{equation}
4\pi\cG^\Delta(\mathbf{x},\mathbf{y})=\left(\frac{\pi }{p^2q^2 \ell^2}+G_{B'}^\Delta(\mathbf{x},\mathbf{y})\right)~.
\end{equation}
Combining all these ingredients, we find
\begin{multline}
\bra{m}\left(\prod_{i=1}^{N_I}\hat{V}_{l_i}(\mathbf{x}_i)\right)\ket{n}=\left(\prod_{i=1}^{N_I}\frac{\left(-1\right)^{ l_i}e^{\frac{l_i^2}{2}\left(1-\frac{\pi }{p^2q^2 \ell^2}\right)+2p^2l_i^2G_\Phi(0)}}{\left(4\pi q\ell\right)^{l_i^2}}\right)\\\times\begin{cases}\prod_{i<j}^{N_I}e^{-4\pi l_i\,l_j\cG^\Delta(\mathbf{x}_i,\mathbf{x}_j)}~, & m=n+\sum_{i=1}^{N_I}l_i \text{ mod } p\\
0~, & \text{otherwise} \end{cases}~.\label{eq:matrixelementsverts}
\end{multline}
This result tells that that these correlation functions are not diagonal in the basis of vacua unless the $l_i$ sum to a multiple of $p$, reminscent of the zero-charge condition we must impose in the free-boson CFT \citep{DiFrancesco:1997nk}. But what happens in the basis of pUniverses? Interestingly, in the pUniverse basis we are no longer required to impose the zero-charge condition. Instead all correlators are diagonal in the pUniverse Hilbert space, as well as being degenerate up to a $\bZ_p$ phase: 
\begin{multline}
\widetilde{\bra{m}}\left(\prod_{i=1}^{N_I}\hat{V}_{l_i}(\mathbf{x}_i)\right)\widetilde{\ket{n}}=e^{\frac{2\pi i}{p}n\sum_{i=1}^{N_I}l_i }\left(\prod_{i=1}^{N_I}\frac{\left(-1\right)^{ l_i}e^{\frac{l_i^2}{2}\left(1-\frac{\pi }{p^2q^2 \ell^2}\right)+2p^2l_i^2G_\Phi(0)}}{\left(4\pi q\ell\right)^{l_i^2}}\right)\\\times\begin{cases}\prod_{i<j}^{N_I}e^{-4\pi l_i\,l_j\cG^\Delta(\mathbf{x}_i,\mathbf{x}_j)}~, & m=n \text{ mod } p\\
0~, & \text{otherwise} \end{cases}~.
\end{multline}
Finally, since these correlators depend only on the $SO(3)$-invariant distances between each pair of insertions, we have shown that upon analytic continuation to dS$_2$, these correlators respect the de Sitter symmetries. Hence both the vacua and pUniverses form a set of de Sitter invariant, Hadamard states at all values of the coupling $q$. 

We have derived the $N_I$-point functions of vertex operators of an interacting QFT on the $S^2$, fully non-perturbatively. We are now in a position to use these quantities for an in-depth study of the structure of a de Sitter invariant theory, via its analytic continuation to Lorentzian signature. We leave this  to future work. 

But, since we have computed all the matrix elements in \eqref{eq:matrixelementsverts} we can deduce the OPE structure of the vertex operators
\begin{equation}
\hat{V}_r(\mathbf{x})\hat{V}_s(\mathbf{y})\underset{\mathbf{x}\rightarrow \mathbf{y}}{=}\left((4\pi q \ell)^2\,\frac{u^E_{xy}}{2}\right)^{rs}\hat{V}_{r+s}(\mathbf{y})+\dots\label{eq:vertexOPE}
\end{equation}
which follows from our normalizations and from the short-distance structure of $\mathcal{G}^\Delta$ defined in \eqref{eq:massivepropagatordef}. 

\section{Coupling the \texorpdfstring{$\nf$}{Nf}-flavor \texorpdfstring{$p$}{p}-Schwinger model to 2d quantum gravity}\label{sec:gravity}

The purpose of this section is to demonstrate that a certain variant of the \texorpdfstring{$p$}{p}-Schwinger model can be coupled to two-dimensional gravity with $\Lambda>0$ in such a way that the combined theory admits a semiclassical de Sitter vacuum. At face value, this seems at odds with the topological nature of pure two-dimensional gravity, but the Schwinger model supplements the theory with locally propagating degrees of freedom that can yield non-trivial gravitational solutions. What is less clear is whether these solutions are semiclassically meaningful, or subject to large quantum fluctuations from the gravitational sector. To render such fluctuations small, as we shall see, we must promote the \texorpdfstring{$p$}{p}-Schwinger model to one admitting an $SU(\nf)$ flavor symmetry. Perhaps the $p$ locally indistinguishable  de Sitter-invariant vacua are simple examples of de Sitter horizon microstates in this theory. We will work in Euclidean signature throughout this section. 

\subsection{The \texorpdfstring{$\nf$}{Nf}-flavor \texorpdfstring{$p$}{p}-Schwinger model}

Our modified Schwinger action, on a general curved space, reads
\begin{equation}\label{eq:SschwNf}
    S_E^{\textnormal{Schwinger}} =   \int_{\mathcal{M}}  \textnormal{d}^2 x \,\sqrt{g}\left[\bar{\Psi}^I \gamma^\mu\left(\nabla_\mu  + i  p A_\mu \right) {\delta_I}^J\Psi_J+\frac{1}{4q^2} F^{\mu \nu} F_{\mu\nu}\right] ~ , 
\end{equation}
where now the fermion $\bar{\Psi}^I$, with $I=1,\ldots,\nf$. In section 2 of \citep{Misumi:2019dwq}, a careful and  complete treatment of the symmetry structure for this model is presented on a Minkowski background.  The $\bar{\Psi}^I$ transform in the anti-fundamental representation of a global vector-$U(\nf)$ symmetry. The vector-$U(1)$ subgroup of the $U(\nf)$ is gauged in the model, so in reality the vector-like global symmetry group is $SU(\nf)$. The vector-$SU(\nf)$ conserved currents are given by
\begin{equation}
j_a^\mu = \bar{\Psi}^I \gamma^\mu {(T^a)_I}^J\Psi_J + \text{h.c.}~,\quad\quad a=1,\ldots, \nf^2-1~,
\end{equation}
where the $T^a$ are the generators of $SU(\nf)$.  There will also be an axial-$U(\nf)$ global symmetry whose axial-$U(1)$ subgroup is anomalously broken down to a $\mathbb{Z}^{(0)}_{\nf p}$. Finally, there will be a $\mathbb{Z}_p^{(1)}$ 1-form global symmetry associated to $p$ unbreakable Wilson loop operators. The model exhibits spontaneous symmetry breaking in its discrete symmetry sector, with the 1-form symmetry being completely spontaneously broken, and  the $\mathbb{Z}^{(0)}_{\nf p}$ spontaneously broken down to a $\mathbb{Z}^{(0)}_{\nf}$ subgroup. The model exhibits $p$-Universes due to the mixed 't Hooft anomaly structure between the 1-form and 0-form symmetries.  More details of this model have been studied on a Minkowski background in \citep{Coleman:1976uz, Gepner:1984au,Affleck:1985wa,Misumi:2019dwq,Dempsey:2023gib},\footnote{See also \citep{Delmastro:2021otj} for an alternative approach based on GKO coset chiral algebras.} but nothing obstructs its analysis in curved space.  The multiflavour Schwinger model, and close variants of it exhibiting the same pattern of symmetries and anomalies, have been proposed as effective descriptions at the worldvolume of domain walls in the high temperature phase of four dimensional non-Abelian gauge theories \citep{Anber:2018jdf,Anber:2018xek}, as well as certain brane configurations in string theory \citep{Armoni:2018bga}.
 
The theory (\ref{eq:SschwNf}) permits a bosonized picture. The $SU(\nf)$ fermionic currents map to those of a decoupled level-one Wess-Zumino-Witten model, whilst the vector-$U(1)$ current maps to a compact boson as in \ref{jB}. The bosonized action reads
\begin{equation}\label{eq:SschwNfB}
    S_E^{\textnormal{BS}_\tn{f}} =   \int_{\mathcal{M}}  \dd^2 {x} \,\sqrt{g}\left[\frac{1}{8\pi}\partial_\mu B_{\tn{f}} \, \partial^\mu B_{\tn{f}}+\frac{1}{4q^2} F^{\mu \nu} F_{\mu\nu} - \frac{i p \sqrt{\nf}}{4\pi} B_{\tn{f}}\epsilon^{\mu\nu}F_{\mu\nu}\right] + S_{\text{WZW}}~ , 
\end{equation}
where now $B_{\tn{f}}\cong B_{\tn{f}}+2\pi\sqrt{\nf}$, and $S_{\text{WZW}}$ is a level-one $SU(\nf)$ Wess-Zumino-Witten (WZW) conformal field theory \citep{Wess:1971yu,Witten:1983tw,DiFrancesco:1997nk}, itself endowed with two $SU(\nf)$ current algebras. The conformal anomaly of the WZW theory at hand is given by $c_{\text{WZW}} = \nf-1$, and this follows from standard treatments \citep{DiFrancesco:1997nk}. Following the same steps outlined in section \cref{sub:vacua_and_puniverses_Schwinger} we can integrate out the vector-$U(1)$ gauge field, we note that the mass generated is now given by 
\begin{equation} 
m_{\tn{f}}^2 \equiv \frac{p^2 q^2\nf}{\pi}~.
\end{equation}

\subsection{Gravitational saddle}

Having defined the \texorpdfstring{$\nf$}{Nf}-flavor \texorpdfstring{$p$}{p}-Schwinger model, we are now ready to couple it to a dynamical metric. At least from the perspective of low energy effective field theory, nothing precludes us from doing so. Our model can be placed on an arbitrary curved space and the quantum field theory exhibits no diffeomorphism anomaly. Our gravitational action will be endowed with a (positive-)cosmological constant term  such that the combined action is
\begin{equation}\label{NfSTot}
S_{\text{tot}} = -\frac{\vartheta_b}{4\pi} \int_{\mathcal{M}} \dd^2{x} \,\sqrt{g} \,R + \Lambda_b\int_{\mathcal{M}}  \dd^2{x} \,\sqrt{g} + S_E^{\textnormal{BS}_{\tn{f}}}~.
\end{equation}
We would like to argue that the above theory exhibits a round-$S^2$ saddle. Even more, we will show that in the large-$\nf$ limit, whilst keeping $m_{\tn{f}}$ fixed, the saddle is subject to small metric fluctuations. 
Throughout our discussion we restrict to a closed two-manifold $\mathcal{M}$ with $S^2$ topology, such that our first term evaluates to $-2\vartheta_b$ irrespective of $g_{\mu\nu}$. The gravitational couplings $\vartheta_b$, and $\Lambda_b$ are viewed here as bare couplings that can absorb any ultraviolet divergences that arise from the path integral over the various fields. The physical couplings are labelled by $\Lambda$ and $\vartheta$. We can suppress non-trivial topology by going to the regime of parameterically large $\vartheta$.

To proceed, we will first integrate out the matter fields to obtain an induced action that is purely gravitational. Since the WZW model is decoupled from the compact scalar $B_{\rm{f}}$, we can write down its path integral directly. Furthermore, since two-dimensional gravity is invariant under the two-dimensional diffeomorphism group we can fix the metric, at least in a small neighborhood around any point, to the Weyl gauge 
\begin{equation}\label{Weyl}
    \dd s^2 = e^{2\omega(\mathbf{x})} \tilde{g}_{\mu \nu} \dd x^\mu \dd x^\nu \, ,
\end{equation}
where $\omega(\mathbf{x})$ is a real valued function that encodes the Weyl factor of the physical metric. The fiducial metric $\tilde{g}_{\mu \nu}$ is taken to be the unit round metric on $S^2$. Up to local UV divergencies, that can be absorbed into the bare couplings of \eqref{NfSTot}, the effective action of the WZW sector of \eqref{eq:SschwNfB} is fixed by the conformal anomaly 
\begin{equation}\label{ZWZW}
    \log \cZ_{\tn{WZW}}[\omega,\tilde{g}] = \frac{\nf-1}{24\pi} \int_{S^2} \dd^2 x \sqrt{\tilde{g}} \left( \tilde{g}^{\mu \nu} \partial_\mu \omega \partial_\nu \omega + \omega \tilde{R} \right) \, , 
\end{equation}
where $\tilde{R}$ is the Ricci scalar associated to $\tilde{g}_{\mu \nu}$, and taking a variational derivative with respect to $\omega$ yields the standard trace anomaly.\footnote{The action \eqref{ZWZW} can also be written as a non-local functional of the metric 
\begin{equation*}\label{ZWZWP}
\log \mathcal{Z}_{\text{WZW}}[g_{\mu\nu}] = \frac{c_{\text{WZW}}}{96\pi} \int_{\mathcal{M}} \textnormal{d}^2 x \textnormal{d}^2y \sqrt{g(\mathbf{x})}\sqrt{g(\mathbf{y})} R(\mathbf{x}) \, (\nabla^{-2}_g)_{\mathbf{x},\mathbf{y}} R(\mathbf{y)}~,
\end{equation*}
known as the Polyakov action \citep{Polyakov:1987zb}. However, when the Laplacian has zero modes the treatment is subtle and thus it is more convenient to work instead with \eqref{ZWZW}. 
}

We also need to consider the path integral over $B_{\tn{f}}(\mathbf{x})$ and $A_\mu(\mathbf{x})$ which, due to the presence of the dimensionful coupling $q$, is not a CFT. Yet, under the same regularization scheme such path integral yields a modified version of \eqref{ZBS} 
\begin{equation}\label{ZBSg}
\mathcal{Z}_{\text{BS}_{\tn{f}}}^\epsilon[g_{\mu\nu}] = \frac{\sqrt{\nf}}{q \ell_{\text{UV}}\sqrt{2}}\left[{\det}'_\epsilon\left( -\nabla_g^2+ \frac{p^2 q^2 \nf}{\pi} \right) \right]^{-1/2}~,
\end{equation}
where the overall factor of $\sqrt{\nf}$ stems from the modified radius of the compact scalar $B_{\tn{f}}(\mathbf{x})$. 

In general, the variation with respect to $g_{\mu \nu}$ of the effective actions \eqref{ZWZW}, \eqref{ZBSg} will yield the expectation value of the stress-energy tensor on a curved space. For \eqref{ZWZW} this can be treated effectively  as in \citep{Anninos:2024iwf}, while for \eqref{ZBSg} the functional dependence on $g_{\mu \nu}$ is more obscure. Namely, before taking the variation with respect to the metric, one needs to compute the functional determinant in \eqref{ZBSg} for an arbitrary Weyl factor $\omega(\mathbf{x})$, such analytic treatment is not known to us.

Nonetheless, in the large-$\nf$ limit, the effect of $\cZ_{\tn{BS}_{\tn{f}}}\left[ g_{\mu \nu} \right]$ will be subleading. To see this it is convenient to define 
\begin{equation}
    q_{\tn{f}} = q \sqrt{\nf} \, ,
\end{equation}
in such a way that we take $\nf \to \infty$ while keeping $q_f$ fixed. Effectively 
\begin{equation}
    m_{\tn{f}}^2 = \frac{q_{\tn{f}}^2 p^2}{\pi} \, ,
    \label{NfMass}
\end{equation}
remains fixed in the large $\nf$ limit of interest. The partition function \eqref{ZBSg} is now given by 
\begin{equation}
    \cZ_{\textnormal{BS}_{\tn{f}}}^\epsilon\left[ g_{\mu \nu} \right] = \frac{\nf}{q_{\tn{f}}\, \ell_{\textnormal{UV}}} \left[{\det}'_\epsilon\left( -\nabla_g^2+ \frac{p^2 q_{\tn{f}}^2 }{\pi} \right) \right]^{-1/2}~ .
\end{equation}
The pre-factor can be absorbed in a renormalization of the bare coupling $\vartheta_b$ in  \eqref{NfSTot}. Thus, the leading metric dependence in the large $\nf$ and fixed $q_{\tn{f}}$ limit stems solely from \eqref{ZWZW}. Effectively, to leading order in the large-$\nf$ fixed $q_{\tn{f}}$ approximation, we have a theory of gravity coupled to a two-dimensional conformal field theory at large central charge. The sphere path integral for such a system was analyzed extensively in \citep{Anninos:2021ene,Muhlmann:2021clm}, where it was indeed shown that the theory admits a semiclassical round two-sphere saddle point geometry whose metric, $g^{(\text{cl})}_{\mu\nu}$, is given by
\begin{equation}\label{saddleg}
\textnormal{d}s^2 = \frac{\nf}{24\pi\Lambda} \left(\textnormal{d}\vartheta^2 + \sin^2\vartheta d\varphi^2\right)~,
\end{equation}
to leading order in our variables. Thus, the size of this world scales with $\sqrt\nf$ and can become macroscopically large. We have a classical de Sitter length scale $\ell_\text{cl} \equiv \sqrt{\tfrac{\nf}{24\pi\Lambda}}$.  

A physical interpretation  of why the above saddle exists is that the positive energy from the cosmological constant balances the negative quantum mechanical Casimir energy from the conformal field theory \citep{Anninos:2021ene}. Happily, in this case, the background two-dimensional de Sitter geometry is not put in by hand, but emerges dynamically in the theory of gravity plus matter.

\subsection{Fluctuation theory}

On top of the saddle point geometry, one will have small fluctuations from the matter fields and the metric. To study such fluctuations, we need to effectively fix the path-integration measure over the gravitational field. In the Weyl gauge, the combination of the gravitational measure and Polyakov action stemming from the matter CFT \eqref{ZWZW} have been conjectured in \citep{Distler:1988jt,David:1988hj} to yield a Liouville theory. This has been tested extensively in the matrix model literature, see \citep{Ginsparg:1993is,Anninos:2020ccj} and references therein. 

As customary in the Liouville literature, its convenient to work with the following Weyl mode $   \varphi(\mathbf{x}) \equiv \beta^{-1} \omega(\mathbf{x})$ in \eqref{Weyl}. The field $\varphi(\mathbf{x})$ captures the fluctuations of the dynamical Weyl factor on top of the classical saddle (\ref{saddleg}), 
\begin{equation}\label{phicl}
\varphi_{\text{cl}} = \frac{1}{2\beta}\log \frac{\nf}{24\pi\Lambda}~,
\end{equation}
the parameter $\beta$ is introduced for later convenience and fixed by the central charge. In addition to the Weyl mode and the matter fields of the theory, one must also introduce the $\mathfrak{bc}-$ghost system to properly fix the gauge. 

As discussed in the previous section, to leading order in the large-$\nf$ limit, the theory \eqref{eq:SschwNfB} is dominated by the WZW conformal field theory, with large positive central charge. When one  couples a conformal field theory of central charge $c>26$ to two-dimensional gravity, and integrates out the conformal fields,  the resulting effective action is given by  a timelike Liouville conformal field theory, with action
\begin{equation}\label{tL}
S_{\text{tL}} = \frac{1}{4\pi} \int \textnormal{d}^2x \sqrt{\tilde{g}}\left( - \tilde{g}^{\mu\nu}\partial_\mu \varphi \partial_\nu \varphi - Q \tilde{R} \varphi + 4\pi \Lambda e^{2\beta\varphi}\right)~,
\end{equation}
with $Q =\beta^{-1} -\beta$. The central charge of timelike Liouville is $c_{\tn{L}}=1-6Q^2$. Consistency of the theory, viewed as a theory of gravity coupled to conformal matter, requires the cancellation of the anomaly via $c_{\tn{L}}-26+c=0$, see \citep{Anninos:2021ene} for further details. In our case, the large-$\nf$ limit is equivalent to the small-$\beta$ limit, which renders the theory perturbative. To leading order $\nf \approx 6\beta^{-2}$. The wrong sign of the kinetic term in (\ref{tL}) is a two-dimensional counterpart to the conformal mode problem of Euclidean gravity \citep{Gibbons:1978ac}. It is a feature, not a bug, which has recently attracted a rigorous treatment \citep{Chatterjee:2025yzo}. Solving the equations of motion of (\ref{tL}) yields (\ref{phicl}) in the leading large-$\nf$ regime.

In addition to the gravity plus WZW sector, we must also compute the contribution to the action governing the Weyl mode that stems from the effective action $S_{\text{BS}_{\tn{f}}} \equiv-\log\mathcal{Z}_{\text{BS}_{\tn{f}}}^\epsilon$. The divergences stemming from ${Z}_{\text{BS}_{\tn{f}}}$ are absorbed into the bare couplings $\vartheta_b$ and $\Lambda_b$ such that our gravitational theory, unlike the bare quantum field theory, is void of ultraviolet divergences. They get traded for finite valued physical couplings. We would like to expand $S_{\text{BS}_{\tn{f}}}$ to quadratic order in the fluctuation $\delta\varphi = \varphi-\varphi_{\text{cl}}$. This yields the non-local functional 
\begin{equation}\label{fluct}
S_{\text{BS}_{\tn{f}}}[\delta\varphi(\mathbf{x})] = -\frac{\beta^2}{2} \int \dd^2 x \dd^2 {y} \sqrt{{g}^{\text{(cl)}}(\mathbf{x})}\sqrt{{g^{\text{(cl)}}}(\mathbf{y})} \delta\varphi(\mathbf{x}) \, _{{g}^{(\text{cl})}}\langle 0| T(\mathbf{x})T(\mathbf{y})|0\rangle_{{g}^{(\text{cl})}} \,\delta\varphi(\mathbf{y})~.
\end{equation}

In the above, $T(\mathbf{x})$ is the trace of the stress tensor of the \eqref{ZBSg} theory in the classical sphere background. Recall that the BF term is topological and thus the only contributions to $T(\mathbf{x})$ stem from the compact scalar $B_{\tn{f}}(\mathbf{x})$ and the gauge field $A_\mu(\mathbf{x})$. One has
\begin{equation}
    T(\mathbf{x})\,= \frac{1}{q^2} \nabla^2_\mathbf{x} \nabla^2_\mathbf{x} :\Phi(\mathbf{x})^2: \, .
\end{equation}
Additional contributions to the local action of $\delta\varphi(\mathbf{x})$ stemming from \eqref{ZBSg} are subleading in the large-$\nf$ expansion. 

In fact, $T(\mathbf{x})$ will itself have a non-vanishing expectation value that depends on the background metric and $m_{\tn{f}}$. This will lead to  a small $\mathcal{O}(\beta^2)$ shift in the saddle. Also, let us comment on the non-integral in (\ref{fluct}) near coincident points. The two-point function computed in the classical saddle \eqref{saddleg} is
\begin{equation}
_{{g}^{(\text{cl})}}\langle 0| T(\mathbf{x})T(\mathbf{y})|0\rangle_{{g}^{(\text{cl})}} = 2 \, m_{\tn{f}}^4\,\cG^{\Delta_{\tn{f}}}(\mathbf{x},\mathbf{y})^2~,
\end{equation}
where $\cG^{\Delta_{\tn{f}}}(\mathbf{x},\mathbf{y})$ is given in \eqref{eq:massivepropagatordef}, and $\Delta_{\rm f}$ satisfies
\begin{equation}
    \Delta_{\tn{f}}(1-\Delta_{\tn{f}}) = m^2_{\tn{f}} \ell_{\tn{cl}}^2 \, ,
\end{equation}
with $m^2_{\tn{f}}$ defined in \eqref{NfMass}. In the coincident point limit, $\cG^{\Delta_{\tn{f}}}(\mathbf{x},\mathbf{y})$ grows as a logarithm, whose square is thus integrable.

All in all, to leading order in the small-$\beta$ expansion, the quadratic fluctuations of $\varphi$ are governed by the action
\begin{equation}
\begin{split}
S^{(2)}[\delta\varphi(\mathbf{x})] = &\frac{1}{4\pi} \int \textnormal{d}^2 x \sqrt{\tilde{g}}\left( - \tilde{g}^{\mu\nu}\partial_\mu \delta\varphi \partial_\nu \delta\varphi +2  \delta\varphi^2 \right) \\ - & \beta^2 m_{\tn{f}}^4 \,\ell^4_{\text{cl}}\int \textnormal{d}^2 x  \textnormal{d}^2 y\sqrt{\tilde{g}(\mathbf{x})}\sqrt{\tilde{g}(\mathbf{y})} \delta\varphi(\mathbf{x}) \,\cG^{\Delta_{\tn{f}}}(\mathbf{x},\mathbf{y})^2 \,\delta\varphi(\mathbf{y})~,
\end{split}
\end{equation}
accompanied by the $\mathfrak{bc}$-ghost action. We have kept the leading local and non-local contributions in the small-$\beta$ expansion. At higher orders, we will have non-trivial interactions. This seems like an interesting extension of two-dimensional quantum gravity theories  that is worthy of study (see also \citep{Allameh:2025gsa}).

\subsection{Gravitational observables}

In a theory of quantum gravity, physical observables must be invariant under diffeomorphisms. Generically, to express such operators in a gauge-invariant way would require integrating the insertion points over the entire manifold $\cM$. After introducing the ghost system, and fixing the Weyl gauge, there is a residual gauge redundancy given by $\varphi(\mathbf{x}) \to \varphi(\mathbf{x}) + \tilde{\omega}(\mathbf{x})$ with $\tilde{g}_{\mu\nu} \to e^{-2\beta\tilde{\omega}(\mathbf{x})}\tilde{g}_{\mu\nu}$ and thus, it is generally argued that the combined timelike Liouville theory plus the matter system should be a Weyl invariant theory with respect to the background metric $\tilde{g}_{\mu \nu}$ \citep{Ginsparg:1993is}. This implies then, that the operators of the theory have to be gravitationally dressed by the timelike Liouville ones in such a way that the dressed operators have scaling dimension $\Delta=\bar{\Delta}=1$. If the matter sector is a conformal theory itself, the conformal primaries with conformal weight $(\Delta,\bar{\Delta})$ can be consistently coupled to the gravitational theory provided they have $\Delta = \bar{\Delta}$ and are dressed by a Liouville primary of the form $\cO_\alpha=e^{2\alpha \varphi}$ with a suitably-chosen $\alpha$ \citep{Bautista:2020obj, Anninos:2021ene}. On the other hand, if the matter theory is not a CFT, it is generally a difficult problem to construct such dressed operators. 

Having said this, it is interesting to note that the $p$-Schwinger model admits topological operators $\hat{U}_m(\mathbf{x})$ that might yield slightly more local observables in the gravitational theory, at least in the low-energy effective field theory regime. Also interesting are the line operators $\hat{L}_m[\mathcal{C}]$ which are defined along a closed curve $\mathcal{C}$ whose dependence is again topological. If we view these from a Lorentzian perspective they could either wrap the spatial cycle, in which case they cut across the de Sitter horizon, or reside along one (or multiple) timelike curves connecting past and future infinities. In any case, $\hat{U}_m(\mathbf{x})$ and $\hat{L}_m[\mathcal{C}]$ may constitute candidate observables in the gravitational theory. Similar comments may apply when coupling higher dimensional quantum field theories with higher-form symmetries, and hence topological operators, to general relativity.\footnote{We should note here that the Standard Model (and presumable any of its GUT completions) does not admit topological line or point operators. (It does have topological volume, {\it i.e.} co-dimension one, operators.) On the other hand,  as noted in the introduction, one can in principle construct topological line operators in a container filled with (say a litre of) superfluid helium-4. The principal role of this topological line operator is to register quantum vortices in the superfuid. To do so practically is not a straightforward affair, as one would have to exponentiate the derivative of the condensate field $\Phi$. Nonetheless, it is curious that we can in principle construct diffeomorphism invariant operators localised on a timelike curve in a low energy effective field theory of general relativity coupled to a superfluid.}

\subsection{Symmetries?}

As a final brief comment, it is interesting to note that when we restrict ourselves to only consider the gravitational theory on a spherical topology, then the topological operators, and consequently the symmetries they generate, persist. It appears, then, that we have an ultraviolet-finite theory of quantum gravity with exact global symmetries. It has been argued \citep{Banks:2010zn,Polchinski:2003bq}, however, that these global symmetries are not compatible with quantum gravity. 

Perhaps the resolution is that we are in a two-dimensional setting. In higher dimensions, the argument proceeds by considering appropriately-charged black holes that would potentially destroy the global symmetries by providing end points for otherwise unbreakable Wilson loops. In two dimensions, no such black hole solutions exist. However, the theory does have a cosmological horizon, which could play a similar role at least from the perspective of a local observer confined within a single static patch. From this perspective, a globally closed topological line operator is perceived as an open Wilson line ending on the two points constituting the dS$_2$ horizon, giving rise to a type of edge mode physics at the de Sitter horizon \citep{Anninos:2021ihe,Law:2025ktz,Law:2026tuk}. (We can also arrange the topological line operator in the form of a pair of entangled worldlines traversing each static patch of dS$_2$ with opposite orientation.)
\newline\newline
Alternatively, it could be that the global symmetries no longer persist once we include additional topologies in the gravitational path integral (see for example \citep{Heckman:2024obe}). This effect can be parametrically suppressed by driving $\vartheta$ to be large. 

Consider, for example, a Euclidean geometry with a small handle. By a simple pictorial argument, 
\begin{equation*}
\begin{gathered}\includegraphics[height=4.5cm]{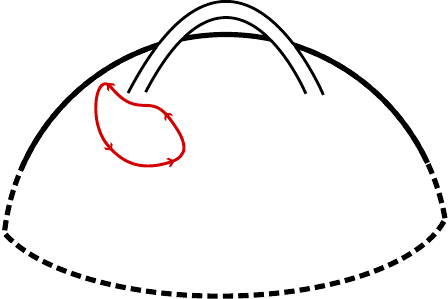}\put(-110,55){{\small $\hat{L}_n$}}\end{gathered}\qquad=\qquad\begin{gathered}\includegraphics[height=4.5cm]{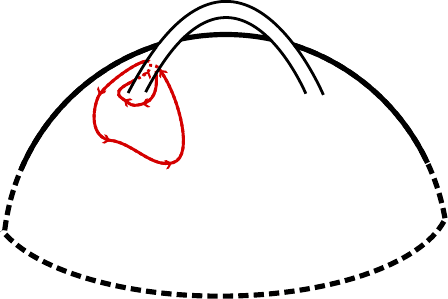}\put(-110,55){{\small $\hat{L}_n$}}\end{gathered}~,
\end{equation*}
we can convince ourselves that a wormhole mouth carries $\bZ_p^{(0)}$ charge.\footnote{The authors owe a great debt of gratitude to Nabil Iqbal for providing the arguments found here.} In BF theory the argument proceeds as follows: BF theory on an $S^2$ with a small handle is, for all intents and purposes, the same as BF theory on the two-torus $\bT^2$, because the two geometries are topologically equivalent. Now let us interpret the path integral of BF theory on a $\bT^2$ as a trace over the Hilbert space of the theory quantized on a spatial $S^1$ and we will let that $S^1$ be the wormhole mouth. Then, the trace will count contributions from all the vacua of the theory \eqref{eq: BF fixed holonomy basis}, and this will be reflected if we imagine passing a topological line operator around the wormhole mouth, as in the figure above. For example, if we were to cut the Euclidean $\bT^2$ on the wormhole mouth, then we would obtain the maximally-mixed density matrix of all the vacua. 

We can then argue similarly to \citep{Coleman:1988cy,Klebanov:1988eh,Hawking:1991vs}  that, in the dilute-wormhole approximation,  we can approximate the contribution from the wormholes as: 
\begin{equation}
    \cZ_{\rm wormhole}=\int\frac{DB DA_\mu}{\textnormal{vol}\,\mathcal{G}} \,   e^{-S^{\textnormal{BS} }_E}\int \dd^2 \alpha\,  e^{-\frac{|\alpha|^2}{2}-\int\dd^2x\sqrt{g}\left(\alpha \sum_{n}\hat{U}_n(\mathbf{x})+\bar{\alpha}\sum_{n}\hat{U}_{-n}(\mathbf{x})\right)}~.
\end{equation}
In the BF theory, this would be akin to adding a series Sine-Gordon potentials for $B$ which explicitly break the $\bZ_p^{(0)}$ global symmetry, with the $\alpha$-parameter being induced by higher-genus contributions from the gravity path integral. 

And since no global symmetries can survive the coupling to gravity, we must also find a mechanism that breaks $\bZ_p^{(1)}$. To achieve this, we would like to argue that wormholes may also act as endpoints for topological lines, in a somewhat vague sense. From the Euclidean theory, a configuration such as: 
\begin{equation*}
\begin{gathered}\includegraphics[height=4cm]{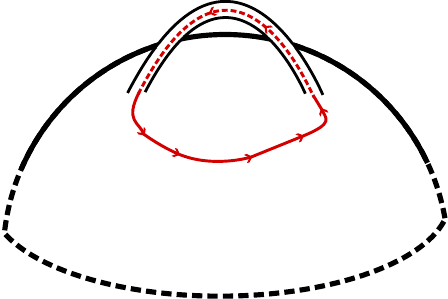}\put(-80,41){{\small $\hat{W}_n$}}\end{gathered}~,
\end{equation*}
may appear as a broken Wilson line to a Euclidean observer that does not have access to the internal dynamics of the wormhole. This would imply that the wormhole induces charged matter in the effective action, resulting in a broken one-form symmetry as well. 

It goes without saying that the arguments presented in this brief section are heuristic and deserving of a more in-depth analysis.

\section*{Acknowledgments}
The authors would like to thank  Jackson Fliss, Loukas Grimanellis, Diego Hofman, Manolo Loparco, Beatrix M\"uhlmann, Priyadarshi Paul, Guilherme Pimentel, Rajath Radhakrishnan, Edgar Shaghoulian, Luigi Tizzano, Okasha Uddin, Stathis Vitouladithis, and especially Nabil Iqbal for insightful discussions.  We are also especially grateful to Jackson Fliss and Stathis Vitouladitis for clarifying discussions on  BF theory.  We thank Magdalena Zalewska for help with designing the graphics. DA is funded by the Royal Society under the grant “Concrete Calculables in Quantum de Sitter,” the STFC consolidated grant ST/X000753/1, and the KU Leuven grant C16/25/010. 
TA and ARF are supported by UKRI Future Leaders Fellowship ``The materials approach to quantum spacetime'' under reference MR/X034453/1. JG is funded under the EPSRC Grant 2895509 ``Matrix quantum mechanics and emergent spacetime.'' The work of JAD is funded by the Spanish MCIN/AEI/10.13039/501100011033 grant PID2022-126224NB-C21. 

\appendix 
\section{Regularization and normalization of bosonic path integrals}\label{app:bosonregularization}

\subsection{Definitions}
In presenting detailed calculations about the BF model and the Schwinger model, we implicitly need to compute (and regularize) certain bosonic path integrals. There are several subtleties involved in this endeavor which we aim to clarify in this appendix. Our conventions are chosen to match those of appendix B of \citep{Anninos:2024fty}. 

Let us start by considering the Euclidean path integral for a standard massive free boson: 
 \begin{equation}\label{eq:massivescalar}
\cZ_\phi=\int D\phi\, e^{-\frac{1}{2}\int_{S^2} \phi \left(-\nabla^2+m^2\right)\phi}~.
\end{equation}
Being a local field composed of an infinite set of modes, the above path integral is not formally defined and needs to be regulated. A standard regularization procedure is via the \emph{heat kernel} (see \citep{vassilevich} for a standard reference on this procedure) whereby one instead computes: 
\begin{equation}\label{eq:massivescalarreg}
\cZ_\phi\rightarrow \cZ_\phi^\epsilon\equiv\exp\left[-\frac{1}{2}\log\det{}_\epsilon\left(-\nabla^2+m^2\right)\right]
\end{equation} 
where for any positive operator $-D^2$ we define the regularized determinant: 
\begin{equation}\label{eq:heatkernelharishboson}
    \log {\det}_\epsilon\, \left(-D^2\right)\equiv -\int_0^\infty\frac{d\tau}{\tau} e^{-\frac{\epsilon^2}{4\tau}}\text{Tr}~e^{+D^2\tau}~,
\end{equation}
and where the function $e^{-\frac{\epsilon^2}{4\tau}}$ serves to cut off the high-frequency modes that contribute to the above sum at small $\tau$ \citep{Anninos:2020hfj}. Note that in these expressions, the operator $-D^2$ has units of $[\text{length}]^{-2}$, hence $\epsilon$ has units of $[\text{length}]$, and as we will soon see, this procedure introduces an emergent UV cutoff scale: 
\begin{equation}\label{eq:cutoffboson}
    \ell_{\rm UV} = \frac{\epsilon e^{\gamma}}{2} \, , \qquad \Lambda_{\rm UV} = \frac{1}{ \ell_{\rm UV}} \, ,
\end{equation}
associated with this regularization scheme, where $\gamma\approx 0.5772$ is the Euler-Mascheroni constant.

In the main text we often find ourselves concerned with computations involving $\det'_{\epsilon}\left(-D^2\right)$ rather than $\det_{\epsilon}\left(-D^2\right)$ (for various choices of $-D^2$) where the prime $'$ means removal of the $L=0$ mode on the $S^2$. So, it will be useful for us to consider how these two quantities are related. 

Let us start with the massive scalar \eqref{eq:massivescalarreg}, and which we will compute by separating out the $L=0$ mode on the $S^2$:
\begin{equation}
    \log {\det}_\epsilon\, \left(-\nabla^2+m^2\right)=-\int_0^\infty\frac{d\tau}{\tau} e^{-\frac{\epsilon^2}{4\tau}}e^{-m^2\tau} -\int_0^\infty\frac{d\tau}{\tau} e^{-\frac{\epsilon^2}{4\tau}}\text{Tr}'~e^{-\left(-\nabla^2+m^2\right)\tau}~.
\end{equation}
In this expression, $\text{Tr}'$ sums over all $L\neq0$ families of modes. We can perform this integral explicitly, yielding: 
\begin{equation}
    \log {\det}_\epsilon\, \left(-\nabla^2+m^2\right)=-2K_0(\epsilon m)+ \log {\det}'_\epsilon\, \left(-\nabla^2+m^2\right)~,
\end{equation}
where $K_0$ is the modified Bessel function. Plugging this back to \eqref{eq:massivescalarreg}, we find
\begin{equation}
\cZ_\phi^\epsilon=e^{K_0(\epsilon m)}\left[{\det}'_\epsilon\, \left(-\nabla^2+m^2\right)\right]^{-1/2}~.
\end{equation}
This equation is useful in that it lets us study the $m\rightarrow 0$ limit of the above expression. This must invariably lead to a divergence, since the integral over the $L=0$ mode is no longer Gaussian-suppressed in this limit. Nevertheless, we can now study precisely how this divergence appears: 
\begin{equation}\label{eq:masslessdet}
\cZ_\phi^\epsilon\underset{m\rightarrow0}{\approx}\frac{\Lambda_{\rm UV}}{m}\left[{\det}'_\epsilon\, \left(-\nabla^2\right)\right]^{-1/2}+\cO(m)+\dots
\end{equation}
In the above expression notice the appearance of the emergent UV cutoff scale introduced in \eqref{eq:cutoffboson}, as well as the `IR-cutoff' scale $m$. In fact, we can go further: The factor of $\Lambda_{\rm UV}$ is dictated by locality, and indicates that the regulated path integral we computed is that of a local quantum field. 

\subsubsection*{BF theory measure}
In the BF theory of \cref{sec: BF} and in the bosonized $p$-Schwinger model of \cref{sec: boson Schwinger}, we must compute similar determinants for a scalar field $B$ whose zero-mode $b$ is compact: $b\cong b+2\pi$. Because of the compactness of the zero-mode, we often find ourselves needing to treat the zero-mode separately in the path integral, that is, we often write the field $B=b+B'(\mathbf{x})$ where $B'(\mathbf{x})$ contains all the non-zero modes. Given the above analysis, this suggests the following treatment of the measure of the path integral: 
\begin{equation}\label{eq:BFmeasure}
\int DB=\ell\Lambda_{\rm UV}\int_{0}^{2\pi}{\dd b}\int DB'
\end{equation}
where we envision that the integral $\int DB'$ will be regulated into an quantity containing $\det'_\epsilon$~. The zero-mode path integral contains a factor of $\Lambda_{\rm UV}$ as dictated by the analysis leading up to \eqref{eq:masslessdet}, and we have replaced the IR-cutoff scale $m$ by the natural IR scale on the $S^2$, namely $1/\ell$.

\subsubsection*{Gauge theory measure}

Concerning gauge theory path integrals, we also must divide by the infinite volume of gauge orbits. This results in equation \eqref{eq: Dh int}, whose derivation we give here. Precisely, we must specify how the following quantity: 
\begin{equation}\label{eq: Dh int appendix}
\int \frac{ Dh}{\textnormal{vol}\,\mathcal{G}}
\end{equation}
is calculated. In this expression, the gauge parameter $h\in U(1)$, thus we can represent $h(\mathbf{x})\equiv e^{i\alpha(\mathbf{x})}$ with $\alpha(\mathbf{x})$ a compact scalar, meaning $\alpha(\mathbf{x})\cong\alpha(\mathbf{x})+2\pi$. Since constant $\alpha$ do not affect the gauge orbit, the measure $Dh$ excludes the zero-mode, which, as we just stated, is compact. 

In this sense we may use the result \eqref{eq:BFmeasure} to calculate \eqref{eq: Dh int appendix}, that is: 
\begin{equation}\label{eq:volgmeasure}
\int \frac{ Dh}{\textnormal{vol}\,\mathcal{G}}\equiv\frac{\int DB'}{\int D B}=\frac{1}{2\pi\ell \Lambda_{\rm UV}}=\frac{\ell_{\rm UV}}{2\pi \ell}~.
\end{equation}
To recap: the volume of the gauge group $\cG$ includes the zero mode, whereas the integral over $h$ does not, hence the ratio is given precisely by $(2\pi)^{-1}$ times a regulator-dependent dimensionless quantity.

\subsection{Examples}\label{app:bosonregularizationExamples}
An example that will appear in the main text quite often is the following determinant:
\begin{equation}\label{eq:detprimedef}
    \log {\det}_\epsilon'\, \left(-\nabla^2+m^2\right)= -\int_0^\infty\frac{d\tau}{\tau} e^{-\frac{\epsilon^2}{4\tau}}\text{Tr}'~e^{-\left(-\nabla^2+m^2\right)\tau}~.
\end{equation}
We refer the reader to appendix D of \citep{Anninos:2024fty}, where this quantity was computed explicitly. The answer can be expressed as a single integral: 
\begin{equation}
    \log {\det}'_{{\epsilon}}\left(-{\nabla}^2+ m^2\right) = -2\int_{0}^\infty \frac{\dd v}{\sqrt{v^2 + \frac{{\epsilon}^2}{ \ell^{2}}}}\left[ \frac{1+e^{-\sqrt{v^2 + \frac{{\epsilon}^2}{ \ell^{2}}}}}{\left(1-e^{-\sqrt{v^2 + \frac{{\epsilon}^2}{ \ell^{2}}}}\right)^2}-1\right] e^{-\frac{1}{2}\sqrt{v^2 + \frac{{\epsilon}^2}{ \ell^{2}}}}\cos\left(\nu \, v\right) \, ,
\end{equation}
where $\nu\equiv\sqrt{m^2\ell^2-\frac{1}{4}}$ and the massless limit therefore corresponds to $\nu=i/2$. We are only interested in the functional form of this integral in the small $\epsilon$ limit. We can obtain this information by direct methods (see \citep{Anninos:2020hfj,Anninos:2024fty}), and the answer is: 
\begin{align}\label{app:detprimemassive}
    \log {\det}'_{{\epsilon}}\left(-{\nabla}^2+ m^2\right) = -2\Bigg[&\frac{2\ell^2}{\epsilon^2}+\left(-\frac{2}{3}+\Delta(\Delta-1)\right)\log(\ell\Lambda_{\rm UV})+\left(\Delta-\frac{1}{2}\right)\log\frac{\Gamma(1+\bar{\Delta})}{\Gamma(1+\Delta)}\nonumber\\&+\left( \psi^{(-2)} (1+\Delta ) + \psi^{(-2)} (1+\bar{\Delta} ) \right)\nonumber\\&-\left(\psi^{(-2)}(1) + \psi^{(-2)}(2) +\frac{1}{4}-2\zeta'(-1)\right)\Bigg]+\cO(\epsilon)~,
\end{align}
where we have defined 
\begin{equation}
\Delta=\frac{1}{2}+i\nu~,\qquad\qquad \bar{\Delta}=1-\Delta~,\qquad\qquad \Delta(\Delta-1)=-m^2 \ell^2~ .
\end{equation}
We can take the massless limit of the above expression by setting $\Delta=0$:
\begin{equation}
    \log {\det}'_{{\epsilon}}\left(-{\nabla}^2\right) = -2\bigg[\frac{2\ell^2}{\epsilon^2}-\frac{2}{3}\log(\ell\Lambda_{\rm UV})-\frac{1}{4}+2\zeta'(-1)\bigg]+\cO(\epsilon)~.\label{eq:masslessdetprime}
\end{equation}

\section{Background fields and topological operators}\label{app: background}

A common theme across the various computations presented in the main text is that, in order to obtain physically sensible results, the smeared topological operators must be equipped with local counterterms in order to cancel all divergences as the regulator is taken to zero. In this section we expose how this counterterms can be systematically determined by exploiting the relation among topological operators and background fields.

Indeed, a complementary perspective on global symmetries defines their action in terms of background gauge connections \citep{Gaiotto:2014kfa}. More precisely, through Poincare duality, networks of topological defects implementing a $p-$form symmetry $G^{(p)}$, extended over closed submanifolds $\Sigma_{d-p-1}$ and classified by $ H_{d-p-1}(\cM_d,G)$, are in one-to-one correspondence with configurations of $G$-valued flat connections, namely elements belonging to equivalence classes in $H^{p+1}(\cM_d,G)$. 
Under this map, the group element $g\in G$ associated to $\cU_g(\Sigma_{d-p-1})$ is mapped to discontinuities on the flat background connection $B^{(p+1)}$ along submanifolds $\Sigma_{p+1}$ intersecting $\Sigma_{d-p-1}$. When the Euclidean spacetime $\cM_d$ is a Riemannian manifold, equipped with a Riemannian metric, this relation   can be made explicit through the delta function form, with support over a closed manifold $\Sigma_{d-p-1}$, namely
\be\label{eq: delta form}
\delta(\Sigma_{d-p-1})_{\mu_1\ldots \mu_{p+1}}=\frac{1}{(d-p-1)!}\frac{\tilde{\epsilon}_{\nu_1\ldots \nu_{d-p-1} \mu_1\ldots \mu_{p+1}}}{\sqrt{g}}\int \dd X^{\nu_1}\wedge\ldots \wedge \dd X^{\nu_n}\delta^{(d)}\left(x-X(\sigma)\right)
\ee
and $X^\mu=X^\mu(\sigma)$ denotes the embedding of $\Sigma_{d-p-1}$ in spacetime, in terms of some arbitrary worldvolume coordinates $\{\sigma^1,\ldots,\sigma^{d-p-1}\}$.\footnote{With this definition, the holonomy of a 1-form $\omega=\omega_\mu dx^\mu$ over a curve $\cC$ reads 
$$
\oint \omega = \oint d\sigma \dot X^\mu(\sigma) \omega_\mu(X(\sigma))= \int \omega\wedge \delta(\cC)
$$
}

We refer the reader to the specialized literature for a careful treatment (see for instance \citep{Bhardwaj:2023kri,Schafer-Nameki:2023jdn} for reviews and references therein), whereas here we will limit ourselves to illustrate these notions applied to the examples relevant for this paper. 
We therefore restrict the discussion to $d=2$, and let us begin describing the case of a 1-form symmetry with group $\bZ_p^{(1)}$. Group elements are then labeled by $m\in\bZ_p$.
Background fields for this symmetry comprise 2-form gauge fields $\cB$. Collectively denoting the fields in our theory by $\phi$ and the corresponding action by $S[\phi]$, then a network of topological local operators with support on a set of points $\{\mathbf{x}_1,\ldots, \mathbf{x}_l\}$ is mapped to
\be\label{eq: U network}
 \langle\hat{U}_{m_1}(\mathbf{x}_1)\ldots \hat{U}_{m_l}(\mathbf{x}_l)\ldots \rangle \longmapsto \int\mathcal{D}\phi e^{-S\left[\phi, \cB \right]}\ldots
\ee
where $S[\phi,\cB]$ denotes the gauge invariant action coupled to the background field and $\ldots$ denotes further insertions of, not necessarily topological, operators. Poincare duality thus prescribes the following configuration for the background 2-form field 
\be \label{eq: U to B}
\cB_{\mu \nu}= \frac{2\pi}{p} \sum_{i=1}^l m_i \frac{\delta(\mathbf{x}-\mathbf{x}_i)}{\sqrt{g}} \epsilon_{\mu \nu} \, .
\ee

The story is similar for the case of a 0-form $\bZ_p$ global symmetry, associated to a background 1-form field $\cA$. Now a network of topological line operators with support along closed curves $\cC_i$ corresponds to coupling the theory to the following 1-form background field configuration  
\be\label{eq: L to A}
\hat L_{n_1}[\cC_1]\ldots \hat L_{n_l}[\cC_l] \,\, \longmapsto \,\,  \cA_\mu = \frac{2\pi}{p}\sum_{i=1}^l n_i \delta(\cC_i)_\mu \, .
\ee

Quite importantly, the background field configurations just introduced are singular, hence the relations \eqref{eq: U to B} and \eqref{eq: L to A} are valid only formally. In practice, one needs to resolve the delta function singularities. As explained in the main text, we achieve this by appealing to the smearing functions introduced in \eqref{eq: f def}, \eqref{eq:Tproperties}, in terms of which the regularized background field configurations now read
\be \label{eq: reg backgrounds}
\cB_{\mu \nu}= \frac{2\pi}{p} \sum_{i=1}^l m_i \, f^\delta_{\mathbf{x}_i}(\mathbf{x}) \epsilon_{\mu \nu}\quad , \quad 
 \cA_\mu = \frac{2\pi}{p}\sum_{i=1}^l n_i \, \partial_\mu\cT^\delta_{\cD_i}
\ee  
where $\cD_i$ is the region enclosed by the curve $\cC_i$, that is $\cC_i=\partial\cD_i$. We refer the reader to the main text for a precise account of the properties satisfied by these smearing functions.

Now we are in place to apply this construction to the particular case of the bosonic $p$-Schwinger model. The coupling to the background fields is achieved by a simple covariantization, namely
\be
F_{\mu\nu}\to F_{\mu\nu}+\cB_{\mu\nu} \quad , \quad \partial_\mu B\to \partial_\mu B+\cA_\mu 
\ee

For the case of the background 2-form field configuration in \eqref{eq: reg backgrounds} describing the (smeared) network \eqref{eq: U network}, and specializing to the case of $l=1$, the action yields 
\begin{equation}
    \begin{split}
        -S^{\rm BS}_E[\cB] &= -\int \dd^2 x \sqrt{g} \left[ \frac{1}{4q^2}(F_{\mu\nu}+\cB_{\mu\nu})(F^{\mu\nu}+\cB^{\mu\nu})+\frac{1}{8\pi}\partial_\mu B\partial^\mu B-i\frac{p }{4\pi}B\epsilon^{\mu\nu}(F_{\mu\nu}+\cB_{\mu\nu})\right] 
        \\
        &= -S^{\rm BS}_E +i\frac{2\pi m}{p} \int\dd^2 x \sqrt{g} f^\delta_{{\mathbf x}_i}(\mathbf{x}) \left[\frac{i}{2q}\epsilon^{\mu \nu} F_{\mu \nu} + \frac{p}{2\pi}B \right] \\&\quad -\frac{2\pi^2 m^2}{p^2q^2} \int \dd^2 x \sqrt{g} \,  f^\delta_{{\mathbf x}_i}(\mathbf{x}) f^\delta_{{\mathbf x}_j}(\mathbf{x})  \, .
    \end{split}
\end{equation}
In the expression above, we recognize the smeared 2-form current together with the appropriate counterterm \eqref{RenormConst1}, in the second and third lines respectively.

Passing to the insertion of a topological line $\hat L_n[\cC]$, we need to couple the action of the model to the background field configuration for $\cA$ in \eqref{eq: reg backgrounds}. For $l=1$ we therefore get
\begin{equation}
\begin{split}
    -S^{\rm BS}_E[\cA]&= -\int\dd^2 x \sqrt{g}\left[\frac{1}{4q^2}F_{\mu\nu}F^{\mu\nu} +\frac{1}{8\pi}(\partial_\mu B+\cA_\mu)(\partial^\mu B+\cA^\mu) +i\frac{p}{2\pi}  \epsilon^{\mu\nu}(\partial_\mu B+\cA_\mu)A_{\nu}\right]\\ 
    &= -S^{\rm BS}_E + i \frac{2\pi n}{p} \int\dd^2 x \sqrt{g} \,  \epsilon^{\mu \nu} \left[ \frac{i}{4\pi} \epsilon_{\mu \rho} \partial^\rho B + \frac{p }{2\pi}A_\mu \right]\partial_\nu \cT^\delta_\cD \\
    & \qquad - \frac{\pi n^2}{2p^2}  \int \dd^2 x \sqrt{g} \,\left( \partial_\mu \cT^\delta_\cD\right)\left(\partial^\mu \cT^\delta_\cD \right)\, .
\end{split}
\end{equation}
hence prescribing the counterterm \eqref{RenormConst2} listed in the main text.

\bibliographystyle{utphys2}
\bibliography{refs.bib}

\end{document}